\documentclass[10pt,journal,compsoc]{IEEEtran}

\usepackage{graphicx}
\usepackage{balance}  
\usepackage{multirow}
\usepackage{algorithmicx}

\usepackage{hyperref}

\usepackage{caption}
\usepackage{subfigure}

\usepackage{algpseudocode}

\usepackage[linesnumbered, ruled, vlined]{algorithm2e}

\SetKwRepeat{Do}{do}{while}

\usepackage{tabu}
\usepackage[export]{adjustbox}
\usepackage{amsmath}
\usepackage{MnSymbol}%
\usepackage{wasysym}%

\usepackage{ulem}

\usepackage{float}

\usepackage{color}
\definecolor{Xiang}{rgb}{1,0,0}
\definecolor{Weilong}{rgb}{0,0,1}

\newcommand{\nop}[1]{}
\newtheorem{definition}{Definition}
\newtheorem{theorem}{Theorem}[section]
\newtheorem{lemma}{Lemma}[section]
\newtheorem{example}{Example}
\newcommand{\ttlit}[1]{#1}

\begin{document}

\title{{\ttlit $kt$}-Safety: Graph Release via {\ttlit $k$}-Anonymity and {\ttlit $t$}-Closeness (Technical Report)}

\author{Weilong Ren, Kambiz Ghazinour, and Xiang Lian
\IEEEcompsocitemizethanks{
\IEEEcompsocthanksitem W. Ren is with Shenzhen Institute of Computing Sciences, Shenzhen, China.
E-mail: renweilong@sics.ac.cn
\IEEEcompsocthanksitem K. Ghazinour is with the State University of New York, Canton, NY, USA. 
E-mail: ghazinourk@canton.edu
\IEEEcompsocthanksitem X. Lian is with the Department
of Computer Science, Kent State University, Kent, Ohio, USA. 
E-mail: xlian@kent.edu
}
}

\IEEEtitleabstractindextext{
\begin{abstract}
In a wide spectrum of real-world applications, it is very important to analyze and mine graph data such as social networks, communication networks, citation networks, and so on. However, the release of such graph data often raises privacy issue, and the graph privacy preservation has recently drawn much attention from the database community. While prior works on graph privacy preservation mainly focused on protecting the privacy of either the graph structure only or vertex attributes only, in this paper, we propose a novel mechanism for graph privacy preservation by considering attacks from both graph structures and vertex attributes, which transforms the original graph to a so-called $kt$-safe graph, via $k$-anonymity and $t$-closeness. We prove that the generation of a $kt$-safe graph is NP-hard, therefore, we propose a feasible framework for effectively and efficiently anonymizing a graph with low anonymization cost. In particular, we design a cost-model-based graph partitioning approach to enable our proposed divide-and-conquer strategy for the graph anonymization, and propose effective optimization techniques such as pruning method and a tree synopsis to improve the anonymization efficiency over large-scale graphs. Extensive experiments have been conducted to verify the efficiency and effectiveness of our proposed $kt$-safe graph generation approach on both real and synthetic data sets.
\end{abstract}

}


\maketitle

\section{Introduction}
The rapid development of social-network applications such as Twitter, Facebook, MySpace, and Friendster has brought a great opportunity for researchers and industries to better explore and understand social behaviors hidden in the world of human beings. To do so, some organizations or companies (e.g., Facebook) may need to release social-network data to the public, where the released data may contain users' privacy information (e.g., salary and relationships with other users). To protect such privacy information from leakage, some critical identifiers (e.g., name and social security numbers) that can uniquely identify users are usually removed before publishing. However, simply removing these identifiers cannot prevent attackers from locating a user (node) in social networks, since a user can also be identified by one's quasi-identifier (e.g., a combination of some non-sensitive attribute values) \cite{sweeney2002k} and/or one's neighborhood information in the graph \cite{zou2009k,cheng2010k}.

\begin{figure}[t!]
\centering
\hspace{-1ex}\includegraphics[scale=0.15]{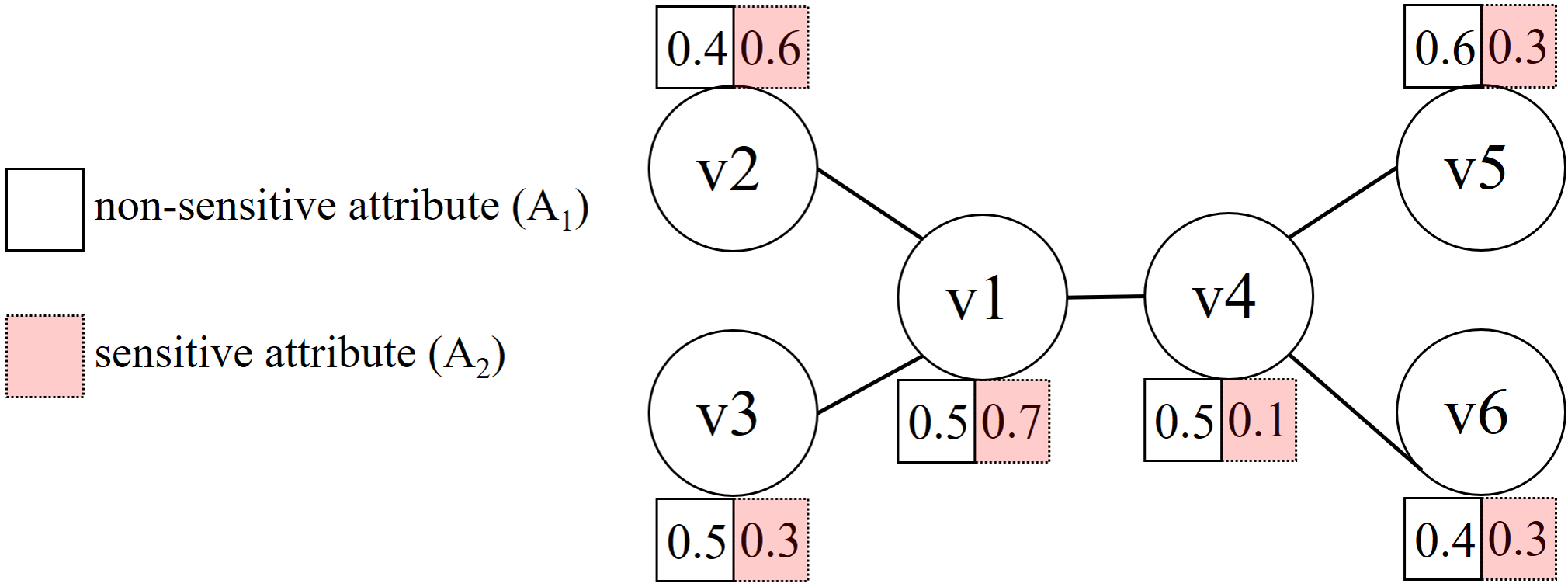}
\caption{\small An example of social-network graph $G$ before releasing.}
\label{fig:motivation1}
\end{figure}

\begin{example}{\bf (Social-Network)}
Figure \ref{fig:motivation1} is an example of a social-network graph before the anonymization and release. Each vertex in the graph corresponds to a user with two attributes, one non-sensitive attribute $A_1$ and one sensitive attribute $A_2$. Here, for sensitive attribute $A_2$ (e.g., salary), we assume that its value range is sensitive, if $A_2$ is less than 0.2 (i.e., low salary). Each edge represents the friendship between two users. 
\label{example_1}
\end{example}

From Figure \ref{fig:motivation1}, the release of such a graph may greatly threaten the privacy of users, for example, vertex $v_4$ with low salary. Thus, it is very important to study how to release the most graph information (i.e., attribute values and graph structure), without revealing the users' privacy.

\setlength{\textfloatsep}{1pt}

In order to publish such social-network data while protecting the users' privacy, one possible direction is to separately release the anonymized graph structure and users' attribute values. For example, \cite{zou2009k,cheng2010k} considered the graph structure attack only, and protected the users' identities via any structure attack. As an example in Figure \ref{fig:motivation1}, by removing all attribute values at vertices, we can obtain a 2-automorphism anonymized graph \cite{zou2009k}, where each node cannot be distinguished from at least 1 ($=2-1$) other symmetric vertex via structural information. A better graph releasing strategy is to release both the graph structure and attribute values together. Yuan et al. \cite{yuan2011protecting} proposed a graph anonymization mechanism, called $k$-degree-$l$-diversity, which guarantees that the candidate set of any vertex has at least $k$ candidate vertices (i.e., $k$-anonymity \cite{sweeney2002k}) with the same degree and contains at least $l$ different values on sensitive attribute (i.e., $l$-diversity \cite{machanavajjhala2006diversity}). 

\begin{example}{\bf (Privacy Disclosure)}
In Figure \ref{fig:motivation1}, vertices $v_1$ and $v_4$ satisfy the 2-degree-2-diversity requirement, where they both have 2 similar vertices (i.e., $v_1$ and $v_4$) and contain 2 different values (i.e., 0.1 and 0.7) on the sensitive attribute. However, this is actually not safe for vertices $v_1$ and $v_4$ in Figure \ref{fig:motivation1}, if we consider their neighbors' information. For instance, the neighbors of vertices $v_1$ and $v_4$ are $\{v_2, v_3, v_4\}$ and $\{v_1, v_5, v_6\}$, with non-sensitive attribute values $\{0.4, 0.5, 0.5\}$ and $\{0.5, 0.6, 0.4\}$, respectively. If an attacker knows neighbors of $v_1$ and $v_4$ (including their non-sensitive attributes), this attacker can uniquely identify $v_1$ and $v_4$, which reveals the privacy information of $v_1$ and $v_4$. 
\label{example_2}
\end{example}

Therefore, most existing works (e.g., \cite{tai2013identity, yuan2011protecting, zou2009k,cheng2010k}) usually protected the privacy of either the graph structure only or vertex attributes only. Their proposed techniques cannot be directly applied to the problem of privacy preservation of graph data, since graph data not only have sensitive attribute values, but also contain sensitive relationships (i.e., edges) among vertices in the graph structure. Meanwhile, although there are a few works (e.g., \cite{song2012sensitive}) that consider both vertex attributes and the graph structure, they assumed that the attacker only knows direct neighbors of vertices (i.e., 1-hop neighbors), which cannot resist from attackers with $n$-hop ($n>1$) neighbor information.

In this paper, we will explore how to release both the graph structure and attributes of vertices, while preventing the users' identities and sensitive attribute values from the leakage. To achieve this goal, there are two requirements on the released graph. First, to avoid the graph structure attack, we need to take into account the structure of \textit{$n$-hop neighbors} for any vertex in the graph. Moreover, we need to consider the intersection attack, raised by the possible combinations of \textit{$n$-hop neighbor graph structure} and vertex attributes (from either the target vertex or its neighbors). One straightforward solution for the intersection attack is to separately publish the anonymized graph structure and vertex attribute values, in which vertex IDs are not consistent in both anonymized parts. However, this method will directly destroy the relationship between the graph structure and vertex attributes. Instead, in this paper, we aim to publish both the graph structure and attribute values of each vertex, under the premise of protecting the graph privacy via either $n$-hop neighborhood attack or attribute-related attack (e.g., quasi-identifier). 

Specifically, in this paper, we will adopt $k$-anonymity \cite{sweeney2002k} and $t$-closeness \cite{li2007t} (a stronger privacy mechanism than $l$-diversity \cite{machanavajjhala2006diversity}) to guarantee the privacy preservation for vertex attributes in graph data. Furthermore, for the graph structure, we will protect the vertex identity from $n$-hop neighbor attack by inserting fake nodes, edges, and/or attributes.

\begin{figure}[t!]
\centering
\hspace{0ex}\includegraphics[scale=0.14]{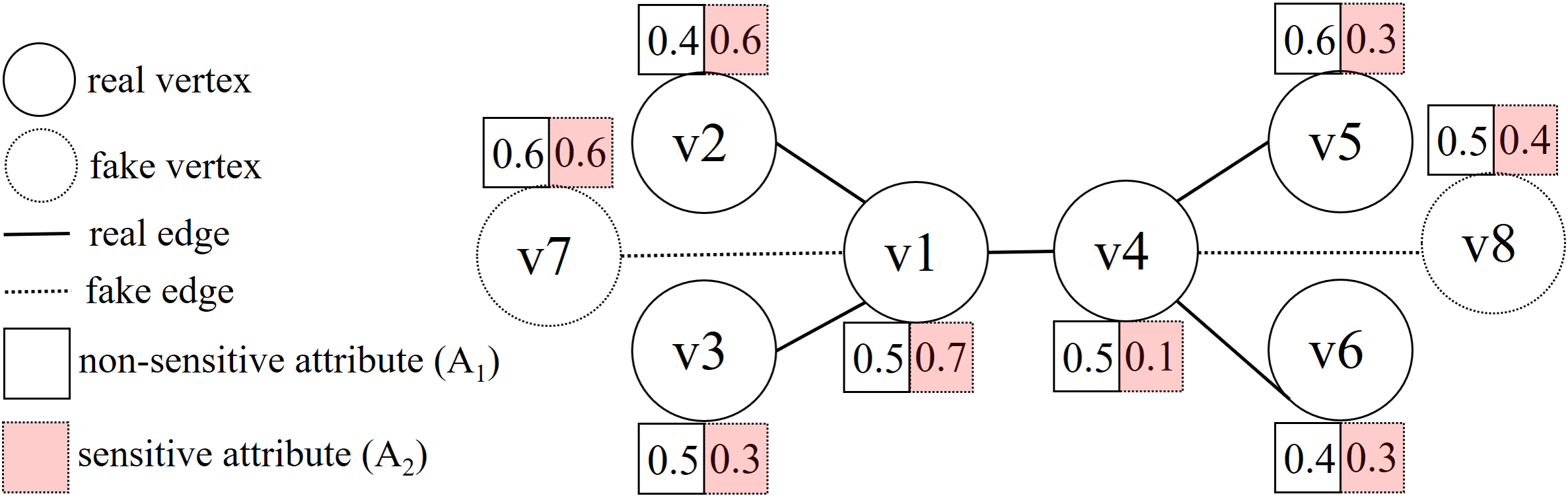}
\caption{\small An anonymized graph $G'$ from graph $G$ in Figure \ref{fig:motivation1}.}
\label{fig:kt_safe_example}
\end{figure}

\begin{example}{\bf (An Anonymized Graph)}
Figure \ref{fig:kt_safe_example} illustrates an anonymized graph of the social-network graph in Figure \ref{fig:motivation1} by adding two fake vertices, $v_7$ and $v_8$, with non-sensitive attribute values 0.6 and 0.5, as the new neighbors of vertices $v_1$ and $v_4$, respectively. We can see that vertices $v_1$ and $v_4$ cannot be identified by an attacker through their neighbors' information, since either $v_1$ or $v_2$ has 4 neighbors with the same set of non-sensitive attributes. 
\label{example_3}
\end{example}

To tackle the graph privacy preservation problem, in this paper, we will formalize a novel graph privacy preservation mechanism, namely $kt$-safe graph, which can efficiently anonymize the graph by letting their $n$-hop neighbors contain the same or similar information (e.g., same/similar non-sensitive values). We prove that the $kt$-safe graph generation is NP-hard and not tractable. Therefore, we propose a feasible framework for the $kt$-safe graph anonymization, and optimize the framework by designing a cost-model-based graph partitioning strategy, proposing an effective pruning method, and devising a tree synopsis for facilitating efficient anonymization process (especially for a large-scale graph).  


Note that, unlike previous works that release graph statistics only \cite{day2016publishing,dwork2011differential} or synthetically generated (anonymized) graphs \cite{jorgensen2016publishing}, in this paper, we aim to publish a ``real'' anonymized graph.
Moreover, even if we use the widely adopted \textit{differential privacy} mechanism \cite{hay2009accurate,dwork2011differential} to release a real anonymized graph (rather than graph statistics \cite{dwork2011differential}), it may introduce too many nodes/edges in the real attributed graph due to its strong privacy guarantee, which leads to poor utility of the anonymized graph.
We would like to leave interesting topics of releasing real attributed anonymized graphs via the \textit{differential privacy} as our future work.

\noindent {\bf Contributions.} In this paper, we make the following contributions:

\begin{enumerate}
\item We formalize a novel and important problem, namely \textit{kt-safety graph anonymization}, based on our newly proposed graph privacy-preservation mechanism, \textit{kt-safety}, in Section \ref{sec:problemDefinition}.

\item We prove the NP-hardness of the \textit{kt-safety graph anonymization} problem, and propose a feasible $kt$-safe graph generation framework in Section \ref{sec:kt_safe_graph_generation}.
\item We propose the detailed algorithms, under our $kt$-safe graph generation framework in Section \ref{sec:alg_details}.
\item We design a cost-model-based graph partitioning strategy, a pruning method, and a tree synopsis to optimize the $kt$-safe graph anonymization process in Section \ref{sec:optimization_for_kt_safe}.
\item We evaluate through extensive experiments the performance of our $kt$-safe graph generation approach on real/synthetic data in Section \ref{sec:experimentation_evaluation}. 
\end{enumerate}
In addition, Section \ref{sec:related_work} reviews related works on privacy preservation over tabular and graph data. Section \ref{sec:conclusions} concludes this paper. 

\section{Problem Definition}
\label{sec:problemDefinition}
\subsection{Models}
In this subsection, we present the graph model considered in this paper, and the model of the attacker's knowledge used for identifying some target vertices.

\noindent {\bf Graph Model:} First, we give the definition of graph data (e.g., social networks) as follows.

\begin{definition} {\textbf{(Graph, $G$)}} A \textit{graph} $G$ is represented by a triple $(V(G), E(G), A(G))$. Here, $V(G)$ is a set of vertices (nodes) $v_i$ (for $1 \le i \le |V(G)|$), $E(G)$ is a set of unlabelled edges $e_j$ (for $1 \le j \le |E(G)|$), and $A(G) = (A_1, ..., A_{d-1}, A_d)$ is an attribute set at each vertex $v_i\in V(G)$, where the first $(d-1)$ attributes $A_j$ (for $1\le j \le d-1$) are non-sensitive attributes and the $d$-th attribute $A_d$ is a sensitive attribute. \qquad $\blacksquare$
\label{def:lg}
\end{definition}

In Definition \ref{def:lg}, we use $v_i.A_j$ to denote the value of attribute $A_j$ at vertex $v_i$. When an attribute value $v_i.A_j$ is not available (or missing), we denote it as $v_i.A_j = $``-''. The first $(d-1)$ attributes, $(v_i.A_1, ..., v_i.A_{d-1})$, of each vertex $v_i$ can be used as a \textit{quasi-identifier}, denoted as $v_i.QI$, which partially discloses the identity of vertex $v_i$. Moreover, the $d$-th attribute, $v_i.A_d$, contains a sensitive value that the user (vertex) $v_i$ is not willing to disclose (e.g., user's age, salary, home address, or mobile number).  

For simplicity, in this paper, we assume that there is only one sensitive attribute for each vertex. Nonetheless, our graph model can be easily extended to multiple sensitive attributes to be anonymized, by treating these sensitive attributes as a composite attribute. 


\noindent {\bf Attacker Model:} An attacker who wants to uniquely identify a vertex in the graph (e.g., some user in a social-network graph) may have a priori knowledge about features of this vertex, which is defined as follows.

\begin{definition} {\textbf{(Attacker Model)}} We assume that an \textit{attacker} knows the quasi-identifier, $v_i.QI$, of a target vertex $v_i$ and its $n$-hop neighbor information, denoted as $HN(v_i,n)$, where $HN(v_i,n)$ is a subgraph containing all neighbor vertices (with their non-sensitive attributes) that are less than or equal to $n$ hops away from $v_i$. \qquad $\blacksquare$
\label{def:ak}
\end{definition}
In Definition \ref{def:ak}, an attacker can have two types of background information about a target vertex $v_i$, which are the target's quasi-identifier $v_i.QI$ and its $n$-hop neighbor information $HN(v_i,n)$ (i.e., a subgraph structure with $n$-hop neighbors of $v_i$, containing non-sensitive attributes). In a special case where $n=0$, the attacker only knows the quasi-identifier of the target vertex $v_i$. 

\subsection{$kt$-Safe Graph via Privacy Mechanisms}
\label{subsec:kt-safety-model}


\noindent {\bf Preliminary:} Prior privacy mechanisms such as $k$-anonymity \cite{sweeney2002k} and $t$-closeness \cite{li2007t} are usually proposed for tabular data, which produce an anonymized version of the original tabular data. As a result, even if an attacker has the quasi-identifier of a target record, he/she can only obtain a set of candidate records, but still cannot uniquely identify the target from these records. Specifically, $k$-anonymity requires the size of this candidate set not be smaller than $k$, whereas $t$-closeness requires the distribution of sensitive attribute values within this set be similar to that of the entire released data (i.e., distances between the two distributions should not be larger than a threshold $t$). 

Different from tabular data, the privacy protection over graph data is much more complex, since it involves the relationships (i.e., edges in the graph structure) among vertices in the graph (e.g., users in social networks). 

\noindent {\bf The Protection Set of a Vertex:} Specifically, for each vertex $v_i$ (associated with a quasi-identifier $v_i.QI$ and its $n$-hop neighbor information $HN(v_i,n)$) in the privacy-preserved graph, we aim to guarantee that there exists a set, $v_i.PS$, of (at least $k$) candidate vertices $v_m$ with the same quasi-identifier $v_m.QI$ as $v_i.QI$ and $n$-hop neighbor information $HN(v_m,n)$ similar to $HN(v_i,n)$. 

Below, we define the protection set, $v_i.PS$, of vertex $v_i$.

\begin{definition} {\textbf{(Protection Set, $v_i.PS$})} Given a parameter $t$, a graph edit distance threshold $\epsilon$, and a vertex $v_i$ with its quasi-identifier $v_i.QI$ and $n$-hop neighbor information $HN(v_i,n)$, a protection set, $v_i.PS$, of vertex $v_i$ is a set of vertices $v_m$ that satisfy the following three conditions:
\begin{enumerate}
 \item $v_i.QI = v_m.QI$;
 \item $ged(HN(v_i,n), HN(v_m,n)) \leq \epsilon$; and
 \item $att\_dist(HN(v_i,L), HN(v_m,L), A_j) \leq t$ (for $1\le j\le d-1$ and $1\le L\le n$);
\end{enumerate}
where $ged(g_1, g_2)$ is the graph edit distance between subgraphs $g_1$ and $g_2$ by considering vertex/edge insertion operators, $HN(v_i,L)$ is a subgraph that contains all neighbor vertices that are less than or equal to $L$ hops away from $v_i$, and $att\_dist(g_1, g_2, A_j)$ is a distance function between attribute value distributions from subgraphs $g_1$ and $g_2$, for each non-sensitive attribute $A_j$ ($1\leq j\leq d-1$). \qquad $\blacksquare$
\label{def:anonymization_set}
\end{definition}

Intuitively, in Definition \ref{def:anonymization_set}, the protection set $v_i.PS$ contains a set of vertices $v_m$ with (1) the same quasi-identifiers as $v_i.QI$, (2) graph structures of $n$-hop neighbors similar to that of $HN(v_i,n)$ (restricted via parameters $n$ and $\epsilon$), and (3) non-sensitive attributes $A_j$ in subgraphs $HN(v_m,L)$ similar to that of $HN(v_i,L)$ (restrained via parameter $t$). Note that, 
we do not want to identify vertex $v_i$ from the distributions of non-sensitive attributes of its $L$-hop neighbors $HN(v_i,L)$. Thus, 
in Definition \ref{def:anonymization_set}, we apply $t$-closeness \cite{li2007t} to the distributions of non-sensitive attributes $A_j$, which can indirectly protect sensitive attribute values of graph nodes. 

To preserve the privacy of graph data, in this paper, we consider graph edit operators such as node and edge insertions. Therefore, the graph edit distance, $ged(HN(v_i,n),$ $HN(v_m,n))$, is given by considering vertex or edge insertion operators (i.e., changes of the graph structure). 

To insert new attribute values, we will introduce new fake nodes with attributes, such that $att\_dist(HN(v_i,L), HN(v_m,L), A_j) \leq t$ holds for each attribute $A_j$ (for $1\le j\le d-1$) and with $(\leq L)$-hop neighbors (for $1\leq L\leq n$). 

Following \cite{li2007t}, we adopt \textit{Earth Mover's Distance} (EMD) as our distance function $att\_dist(g_1, g_2, A_j)$. To compute the \textit{Earth Mover's Distance} for category data \cite{li2007t}, function $att\_dist(HN(v_i,L), HN(v_m,L), A_j)$ is given by half of the $L_1$-norm distance between two normalized (attribute frequency) distributions from subgraphs $HN(v_i,L)$ and $HN(v_m,L)$.
\begin{eqnarray}
\hspace{-4ex}&&att\_dist(HN(v_i,L), HN(v_m,L), A_j) \label{eq:EMD} \\
\hspace{-2ex}&=& \hspace{-2ex}\frac{1}{2}\cdot \hspace{-1ex}\sum_{\forall a_j\in dom(A_j)} \hspace{-2ex}|pdf(a_j \in HN(v_i,L)) - pdf(a_j \in HN(v_m,L))|, \notag
\end{eqnarray}
where $dom(A_j)$ is the value domain of attribute $A_j$, function $pdf(a_j\in HN(v,L))$ = $\frac{freq(a_j\in HN(v,L))}{\sum_{\forall a_x\in dom(A_j)} freq(a_x\in HN(v,L))}$ is the frequency fraction of value $a_j\in dom(A_j)$ appearing in vertices of subgraph $HN(v,L)$ (i.e., nodes within $L$ hops away from vertex $v$), and $freq(a_j\in HN(v,L))$ is the frequency (count) of value $a_j\in dom(A_j)$ in subgraph $HN(v,L)$. We can extend the distance function $att\_dist(g_1, g_2, A_j)$ from EMD to other distances like $L_1$-norm distance of two histograms (i.e., Hamming distance), by removing the parameter $\frac{1}{2}$ from Eq. (\ref{eq:EMD}).

\noindent {\bf Goals of Graph Privacy Preservation:} We want to release an anonymized (privacy-preserved) graph that prevents the privacy of any vertex $v_i$ from leakage, with the following two goals: 
\begin{enumerate}
  \item the attacker cannot uniquely link (identify) any vertex $v_i$ in graph $G$ to some vertices via quasi-identifier $v_i.QI$ and $n$-hop neighbor information $HN(v_i,n)$ with high confidences, and; 
  \item the attacker cannot deduce a vertex's sensitive attributes to precise values via $v_i.QI$ and $HN(v_i,n)$ with high confidences. 
\end{enumerate}

We define a safe (privacy-preserved) vertex as follows.

\begin{definition} {\textbf{($kt$-Safe Vertex)}} 
Given thresholds $k$, $t$, $\alpha$, and $\epsilon$, a vertex $v_i$ is $kt$-safe, if it satisfies the two conditions below:
\begin{enumerate}
 \item $|v_i.PS| \ge k$, and;
 \item $P_{sens}(v_i.PS, A_d) \le \alpha$,
\end{enumerate}
where $|\cdot|$ is the size of $\cdot$, and $P_{sens}(v_i.PS, A_d)$ is the fraction of vertices $v_m \in v_i.PS$ that have sensitive values for sensitive attribute $A_d$. \qquad $\blacksquare$
\label{def:KT_Safe_Vertex}
\end{definition}

Intuitively, in Definition \ref{def:KT_Safe_Vertex}, a vertex $v_i$ is $kt$-safe, if (1) $v_i$ has at least $k$ vertices, $v_m$, in the protection set $v_i.PS$, and (2) less than $(100\cdot \alpha)\%$ of vertices $v_m \in v_i.PS$ have sensitive values for attribute $A_d$. Note that, here we consider attribute $A_d$ having either sensitive or non-sensitive values. For example, for the disease attribute, its values like ``AIDS'' and ``cancer'' are sensitive, whereas attribute values such as ``flu'' and ``cold'' are not so sensitive. With the two conditions above, we cannot distinguish $v_i$ from other vertices in $v_i.PS$ and disclose sensitive attribute values. Thus, we call $v_i$ a $kt$-safe vertex.

\begin{definition} {\textbf{($kt$-Safe Graph)}} 
Given a graph $G$ and thresholds $k$, $t$, $\alpha$, and $\epsilon$, we call the anonymized graph $G'$ of $G$ a $kt$-safe graph, if any vertex $v_i \in V(G')$ is $kt$-safe.\qquad $\blacksquare$
\label{def:KT_Safe_Graph}
\end{definition}

In Definition \ref{def:KT_Safe_Graph}, given a graph $G$, the anonymized graph $G'$ is $kt$-safe, if all vertices $v_i \in V(G')$ are $kt$-safe (i.e., the privacy of every vertex in $G'$ is well preserved).


\noindent {\bf Discussions on the $kt$-Safety Model}: Our $kt$-safety model adopts the $k$-anonymity \cite{sweeney2002k,doka2015k} for the protection set (as given by Definition \ref{def:KT_Safe_Vertex}), and the $t$-closeness for non-sensitive attributes in the protection set (as given by Definition \ref{def:anonymization_set}). Essentially, our goal is to protect the identity and sensitive attribute values of each node $v_i$ in the graph.

Note that, our $kt$-safety model can be easily extended to protecting both the identity and link safety, by setting neighbor-related parameters $\epsilon$ and $t$ (given in Definition \ref{def:anonymization_set}) to 0. Moreover, we can easily extend our $kt$-safety model to other privacy mechanisms, for example, updating $t$-closeness with $\beta$-likeness \cite{cao2012publishing}, by revising the distance function from $att\_dist(HN(v_i,L), HN(v_m,L),$ $A_j)$ to $att\_dist(HN(v_i,L), HN(v_m,L), a_j)$, where $a_j\in dom(A_j)$ are the values on non-sensitive attributes $A_j$. 

\subsection{$kt$-Safe Graph Anonymization Problem}

\noindent {\bf The Anonymization Cost:} We define the cost of the graph anonymization as follows.

\begin{definition} {\textbf{(The Anonymization Cost)}} Given a graph $G$ and its anonymized graph $G'$, the anonymization cost, $Cost(G,G')$, from $G$ to $G'$ is given by:
\begin{eqnarray}
\hspace{-2ex}Cost(G, G')  = |(V(G') \cup V(G)) - (V(G') \cap V(G))|\nonumber\\
+|(E(G') \cup E(G)) - (E(G') \cap E(G))|, \label{eq:eq1}
\end{eqnarray}
\noindent where $V(G)$ and $E(G)$ are the sets of nodes and edges in graph $G$, respectively, and $|S|$ is the size of set $S$.\qquad $\blacksquare$
\label{def:anony_cost}
\end{definition}

In Definition \ref{def:anony_cost}, the anonymization cost, $Cost(G, G')$, is defined as the total number of graph edit operators from $G$ to $G'$ (as given by Eq.~(\ref{eq:eq1})). 

In particular, the first term in Eq.~(\ref{eq:eq1}) (i.e., $|(V(G') \cup V(G)) - (V(G') \cap V(G))|$) is the cost of vertex edits from $G$ to $G'$, while the second term in Eq.~(\ref{eq:eq1}) (i.e., $|(E(G') \cup E(G)) - (E(G') \cap E(G))|$) corresponds to the cost of edge edits from $G$ to $G'$. The anonymization cost $Cost(G, G')$ sums up the two terms above.

\noindent {\bf Graph Release via $k$-Anonymity and $t$-Closeness:} After defining the anonymization cost for graph privacy preservation, we provide the definition of our $kt$-safe graph anonymization problem below.

\begin{definition} {\textbf{($kt$-Safe Graph Anonymization Problem)}} Given a graph $G$, and thresholds $k$, $t$, $\alpha$, and $\epsilon$, the problem of \textit{$kt$-safe graph anonymization} is to produce an anonymized $kt$-safe graph $G'$ from $G$, with minimal anonymization cost $Cost(G, G')$ (given in Eq.~(\ref{eq:eq1})).\qquad $\blacksquare$
\label{def:kt_graph_problem}
\end{definition}

In Definition \ref{def:kt_graph_problem}, given a graph $G$, the $kt$-safe graph anonymization problem aims to generate and release an anonymized $kt$-safe graph $G'$ from graph $G$ with the minimum anonymization cost in Definition \ref{def:anony_cost}. 

\noindent {\bf Challenges:} The major challenges to tackle the $kt$-safe graph anonymization problem are twofold. First, previous works assumed that an attacker knows the quasi-identifier and degree of a target only \cite{yuan2011protecting}, or the graph structure of the target's $n$-hop neighbors only \cite{zou2009k,cheng2010k}. However, this assumption may not always hold in practice, since the attacker may know both quasi-identifier and $n$-hop neighbors (for  $n \ge 1$, as given in Definition \ref{def:ak}). It is non-trivial how to protect the identity and sensitive attribute values of each vertex against attacker's knowledge in Definition \ref{def:ak}. Thus, we need to propose an effective solution to transforming each vertex in $G$ to a $kt$-safe vertex.

Second, it is very challenging to efficiently anonymizing vertices for the large-scale graph. The $kt$-safe graph anonymization problem is NP-hard (as will be proved in Section~\ref{subsec:NP_hardness}) and thus intractable. We need to design efficient heuristic approaches to improve the efficiency of the graph anonymization.

Table \ref{symbols_and_descriptions} depicts the commonly used symbols and their descriptions in this paper.

\begin{table}\centering
{\small\scriptsize
    \caption{\small Symbols and descriptions.}
    \label{symbols_and_descriptions}
    \begin{tabular}{l|l} \hline
    {\bf Symbol} & \qquad\qquad\qquad\qquad{\bf Description} \\ \hline \hline
    $G$  & a graph \\ \hline
    $G'$  & a $kt$-safe graph of $G$ (an anonymized graph)\\ \hline    
    $\mathcal{G}$  & a set of disjoint subgraphs from graph $G$\\ \hline        
    $v_i.QI$ & the quasi-identifier of vertex $v_i$ \\ \hline
    $HN(v_i,n)$ & the $n$-hop neighbor information of vertex $v_i$ \\ \hline
    $v_i.PS$ & a set of vertices for protecting the privacy of vertex $v_i$ \\ \hline
    $v.CS$ & a candidate set of vertices with quasi-identifiers and \\ 
    & $n$-hop graph structures similar to that of a target vertex $v$ \\ \hline
    \end{tabular}
}
\end{table}

\section{$kt$-safe graph Generation}
\label{sec:kt_safe_graph_generation}

\subsection{NP-hardness of the $kt$-Safe Graph Anonymization Problem}
\label{subsec:NP_hardness}

\begin{theorem} {\textbf{(NP-hardness)}} 
The structural-attack-resistance-only version of $kt$-safe graph anonymization problem in Definition \ref{def:kt_graph_problem} is NP-hard.
\label{theo:NP_hardness}
\end{theorem}
\noindent {\it Proof:} Please refer to Appendix~\ref{subsec:proof_of_NPhard} in our supplemental material. \qquad $\square$

\normalem

\subsection{$kt$-Safe Graph Generation Framework}

Since the $kt$-safe graph anonymization problem is NP-hard and not tractable, we will propose alternative heuristics-based algorithms. Specifically, Algorithm \ref{alg:kt_safe_frame} illustrates a heuristic-based $kt$-safe graph generation framework, which produces a $kt$-safe graph $G'$ from a given graph $G$, where we only consider graph adding operations (i.e., inserting vertices/edges) for simplicity. We leave as our future work for extending Algorithm \ref{alg:kt_safe_frame} by embedding graph deletion operations (i.e., removing vertices/edges). 

This framework consists of three phases: \textit{preprocessing}, \textit{$kt$-safe graph generation}, and \textit{$kt$-safe graph merging} phases. In the \textit{preprocessing} phase, we first partition graph $G$ into $s$ disjoint subgraphs, and store these partitioned subgraphs in a set $\mathcal{G}$ via function $partition\_graph(G, \gamma, s)$ (line 1). 

In the \textit{$kt$-safe graph generation} phase, we make each partitioned subgraph $g\in \mathcal{G}$ a $kt$-safe subgraph. For each partitioned subgraph $g$ in $\mathcal{G}$, we sort vertices in $g$ in a queue $Q_g$ by the counts of their $(\leq n)$-hop neighbors in ascending order (line 4), based on which we anonymize vertices. Intuitively, the vertices with less neighbors should be considered first, since graph edit operators (i.e., insertion operators over vertex/edge/attribute) over these vertices may have less effect on the neighbor information of other unprocessed vertices. Next, for each vertex $v_i\in Q_g$ (assuming that $v_i$ is the target vertex), we invoke function $initial\_candidate(v_i, g)$, and retrieve an initial candidate set $v_i.CS$ for $v_i\in g$ (lines 5-6), which can be used for quickly obtaining the protection set $v_i.PS$ of $v_i$.

Then, for each vertex $v_i\in g$ in the queue $Q_g$, we make $v_i$ a $kt$-safe vertex via function $kt\_safety\_vertex(v_i, v_i.CS, g)$ (line 7).  Note that, in the anonymization process of $v_i$, if graph edit operators (i.e., adding edges/attributes) over a vertex $v_l\in g$ affects the privacy of previously anonymized vertices (i.e., identity or sensitive attribute values), then we will create a duplicate $v_l'$ of $v_l$, and perform graph edits on duplicate vertex $v_l'$ (instead of $v_l$). This way, we can obtain a $kt$-safe graph $g'$ from subgraph $g\in \mathcal{G}$, and add $g'$ to the set $\mathcal{G'}$ (lines 8-9).

In the \textit{$kt$-safe graph merging} phase, we merge vertices in $kt$-safe subgraphs $g' \in \mathcal{G'}$ and their duplicates in other $kt$-safe subgraphs in $\mathcal{G'}$ via function $merge\_subgraphs(\mathcal{G'})$ (line 10). Finally, we return the merged graph $G'$ as the $kt$-safe graph (line 11).

Note that, Algorithm \ref{alg:kt_safe_frame} can work for producing a $kt$-safe anonymized graph without restricting the setup of $n$-hop neighbors (i.e., $n\geq 1$). However, since exact GED calculation between two $n$-hop neighbor subgraphs $HN(v,n)$ is an NP-hard problem for $n>1$, the cost of GED computation for $n>1$ is expected to be higher than that for $n=1$. Moreover, we have an attack resistance discussion on the anonymized graph via Algorithm \ref{alg:kt_safe_frame} (e.g., against minimality attack \cite{wong2007minimality}). Please refer to Appendix \ref{subsec:attack_resistance_discussion} in the supplemental material for details.

\begin{algorithm}[t!]\small
\KwIn{a graph $G$, and thresholds $k$, $t$, $\epsilon$, $\alpha$, $\gamma$, and $s$}
\KwOut{an anonymized $kt$-safe graph $G'$}
\tcp{\small Preprocessing Phase}
$\mathcal{G} \leftarrow partition\_graph(G, \gamma, s)$ \tcp{\small partition graph into disjoint subgraphs (Section \ref{subsec:prep_phase})}

\tcp{\small $kt$-safe graph Generation Phase}
$\mathcal{G'} \leftarrow null$

\For{each subgraph $g\in \mathcal{G}$}{
    $Q_g \leftarrow$ vertices $v_i\in g$ sorted by the count of their $(\leq n)$-hop neighbors in ascending order
    
    \For{$v_i \in Q_g$}{
    
    \tcp{\small retrieve initial candidate vertices $v_m$ of $v_i$ in $g$ (Algorithm \ref{alg:init_cand} in Section \ref{subsubsec:cand_set_detec})} 
    $v_i.CS \leftarrow initial\_candidate(v_i, g)$

    \tcp{\small make vertex $v_i$ a $kt$-safe vertex (Section \ref{subsubsec:kt_safe_vert_gener})}
    invoke function $kt\_safety\_vertex(v_i, v_i.CS, g)$
    
    }
    
    obtain a $kt$-safe subgraph $g'$ from $g$
    
    $\mathcal{G'} \leftarrow \mathcal{G'} \cup \{g'\}$
}

\tcp{\small kt-Safe Graph Merging Phase}
$G^{'} \leftarrow merge\_subgraphs(\mathcal{G'})$ \tcp{\small merging $|\mathcal{G'}|$ subgraphs $g' \in \mathcal{G'}$ by resolving conflicts between vertices and their duplicates (Section \ref{subsec:kt_safe_graph_refine})}

return a $kt$-safe graph $G'$
\caption{$kt$-Safe Graph Generation Framework}
\label{alg:kt_safe_frame}
\end{algorithm}

\section{Algorithms for $kt$-safe graph Anonymization}
\label{sec:alg_details}

Please refer to Appendix~\ref{subsec:demo} in our supplemental material for a running example with the 3 phases of anonymizing the graph in Figure~\ref{fig:motivation1} via our $kt$-safe graph generation approach.

\subsection{Preprocessing Phase}
\label{subsec:prep_phase}

\noindent {\bf Graph Partitioning:} We use the divide-and-conquer strategy to divide $G$ into subgraphs with similar sizes, which reduces the time cost to anonymize each vertex to be $kt$-safe by making each subgraph a $kt$-safe graph.

Given a graph $G$, the maximum size, $\gamma$, of subgraphs, and the number, $s$, of partitions, we have an algorithm to produce a set, $\mathcal{G}$, of partitioned subgraphs. Please refer to Algorithm \ref{alg:graph_partition} in Appendix \ref{subsec:graph_partition} in our supplemental material for details.

Note that, after we divide $G$ into subgraphs $g_i$, we will expand each subgraph $g_i$ by including duplicated $n$-hop neighbors of border vertices $v\in g_i$. We will conduct the anonymization process on each expanded subgraph, and then merge the anonymized subgraphs.

\noindent \textbf{Discussions on Parameters $\gamma$ and $s$.} Please refer to Appendix \ref{subsec:gamma_s_appendix} in our supplemental material for the discussions on parameters $\gamma$ and $s$.

\subsection{$kt$-Safe Graph Generation Phase}
\label{subsec:kt_safe_graph_gene}


\subsubsection{Candidate Set Detection}
\label{subsubsec:cand_set_detec}

For any vertex $v_i$, in order to obtain its protection set $v_i.PS$ (as given in Definition \ref{def:anonymization_set}), we will first retrieve its initial candidate set, $v_i.CS$, of vertices $v_m$ that satisfy the first two conditions (w.r.t. quasi-identifier and graph structure) in Definition \ref{def:anonymization_set}, which we can further anonymize to obtain $v_i.PS$ (satisfying the third condition). Please refer to Algorithm \ref{alg:init_cand} in Appendix \ref{subsec:cand_set_detect} in our supplemental material for details.

Note that, if many vertices $v_m\in V(G)$ have the same quasi-identifier (i.e., $v_m.QI$) as $v_i$, we will obtain many candidates in $v_i.CS$, which leads to a lower anonymization cost to anonymize $v_i$ as a $kt$-safe vertex later in Section \ref{subsubsec:kt_safe_vert_gener}.

\subsubsection{$kt$-Safe Vertex Generation}
\label{subsubsec:kt_safe_vert_gener}

Algorithm \ref{alg:kt_safe_vertex} illustrates the details of the $kt$-safe graph generation phase, which obtains a protection set $v_i.PS$ of each vertex $v_i$, based on the candidate set $v_i.CS$, and protects the identity and sensitive attribute values of $v_i$, leveraging $v_i.PS$.

\noindent {\bf Initialization of the Protection Set $v_i.PS$:} In Algorithm \ref{alg:kt_safe_vertex}, we first compute an initial protection set $v_i.PS$ of vertex $v_i$ (based on the candidate set $v_i.CS$) that satisfies the three conditions in Definition \ref{def:anonymization_set} (lines 1-11). Specifically, candidate vertices $v_m$ in $v_i.CS$ satisfy the first two conditions of Definition \ref{def:anonymization_set}. Thus, 
we only need to further check whether vertices $v_m$ also satisfy the third condition (i.e., similar quasi-identifiers in $L$-hop neighbors of vertex $v_i$). 

Specifically, for each vertex $v_m\in v_i.CS$, we check the similarity of value distributions on attribute $A_j$ (for $1\le j\le d-1$) between $L$-hop neighbors of $v_i$ and $v_m$ (for $1\le L\le n$; lines 1-9). If it holds that $att\_dist(HN(v_i,L), HN(v_m,L), A_j)\leq t$ for all non-sensitive attributes $A_j$ ($1\le j\le d-1$) and all $L$-hop neighbors (i.e., flag $att\_safe=true$ holds), then we will add $v_m$ to the protection set $v_i.PS$ (lines 10-11).

\begin{algorithm}[t!]\small
\KwIn{a graph $G$, a vertex $v_i$,  a candidate set $v_i.CS$, and thresholds $k$, $t$, $\epsilon$, and $\alpha$}
\KwOut{a $kt$-safe vertex $v_i$}

\tcp{\small retrieve an initial protection set $v_i.PS$ of $v_i$}

\For{each $v_m\in v_i.CS$}{
    $att\_safe \leftarrow true$
    
    \For{$L=1$ to $n$}{
        \For{$j = 1$ to $d-1$}{
            \If{$att\_dist(HN(v_i,L), HN(v_m,L), A_j) > t$}{
                $att\_safe \leftarrow false$\\
                break;
            }
        }
       \If{$att\_safe = false$}{
            break;
        }

    }
    
    \If{$att\_safe = true$}{
        $v_i.PS \leftarrow v_i.PS \cup \{v_m\}$
    }
}

\tcp{\small make $v_i$ satisfy $|v_i.PS| \ge k$}
\If{$|v_i.PS| \in [0, k)$}{
    $N_{needed} \leftarrow k - |v_i.PS|$
    
    \For{each $v_m\in (v_i.CS - v_i.PS)$}{
        \If{$N_{needed} = 0$}{
            break;
        }
        
        \If{$v_m$ can be a member of $v_i.PS$ via adding fake nodes and edges}{
            $v_i.PS \leftarrow v_i.PS \cup \{v_m\}$

            \For{each vertex $v_l\in Q_G$ prior to $v_i$ in $Q_G$}{
                \If{graph edit operators over vertex $v_a$ affect the privacy of $v_l$}{
                    make a duplicate $v_a'$ of $v_a$
                    
                    proceed graph edit operators over $v_a'$
                }
            }

            $N_{needed} \leftarrow N_{needed} - 1$
        }
        }

    \If{$N_{needed} > 0$}{
        add $N_{needed}$ fake duplicate vertices $v_i'$ of existing vertices $v_m \in v_i.PS$ with non-sensitive value on sensitive attribute $A_d$, via fake vertex/edge/attribute insertion
        
        $v_i.PS \leftarrow v_i.PS \cup \{v_i'\}$
    }
}

\tcp{\small make $v_i$ satisfy $P_{sens}(v_i.PS, A_d) \le \alpha$}
\If{$P_{sens}(v_i.PS, A_d) > \alpha$}{
    add $x$ ($=\left\lceil\frac{N_{sens}}{\alpha} - |v_i.PS|\right\rceil$) fake duplicate vertices $v_i'$ of existing vertices $v_m \in v_i.PS$ with non-sensitive value on sensitive attribute $A_d$, via fake vertex/edge/attribute insertion
    
    $v_i.PS \leftarrow v_i.PS \cup \{v_i'\}$
}

\caption{$kt\_safety\_vertex(v_i, v_i.CS, G)$}
\label{alg:kt_safe_vertex}
\end{algorithm}

\noindent {\bf Identity Protection of $v_i$:} After obtaining an initial protection set $v_i.PS$, we next aim to protect the identity of vertex $v_i$, by letting $|v_i.PS| \ge k$ hold (i.e., the first condition of the $kt$-safe vertex in Definition \ref{def:KT_Safe_Vertex}). 

Specifically, if the initial protection set $v_i.PS$ satisfies the condition $|v_i.PS| \ge k$, then we do nothing; otherwise, we will let the remaining candidates in set $(v_i.CS-v_i.PS)$ become protection vertices for $v_i$ (lines 12-26). That is, when $|v_i.PS|$ is less than $k$, we will obtain $N_{needed}$ ($=k-|v_i.PS|)$ more vertices as protection vertices, either from $(v_i.CS-v_i.PS)$ (lines 14-23) and/or by adding fake duplicate nodes (lines 24-26).

For the remaining vertices $v_m \in (v_i.CS-v_i.PS)$ that do not satisfy the third condition (w.r.t. attribute distributions) in Definition \ref{def:anonymization_set}, we will attempt to add new nodes (and edges as well) to their $L$-hop neighbors to make them the members of $v_i.PS$ (i.e., $v_m \in v_i.PS$). Specifically, we will do a rehearsal to add some fake nodes and edges to make $v_m$ satisfy the third condition of Definition~\ref{def:anonymization_set}, where the addition operations are finally executed only if they will not make $v_m$ violate the first two conditions of Definition~\ref{def:anonymization_set}. If we can transform $v_m$ to a member of $v_i.PS$, then we will add it to $v_i.PS$ and update those vertices $v_l$ affected by node/edge insertions for $v_m$, via vertex/edge duplication (lines 17-23). In particular, the anonymization operations of $v_m \in (v_i.CS-v_i.PS)$ may affect the privacy (i.e., identity and sensitive attribute values) of those $kt$-safe vertices $v_l$, prior to $v_i$ in the queue $Q_G$. Thus, in the case that graph edit operators on a vertex $v_a$ affect $v_l$, we will duplicate $v_a$ and obtain $v_a'$ to proceed the operators (lines 19-22). That is, the anonymization operations of a vertex do not affect the privacy of previously anonymized vertices. When we include a new vertex $v_m \in v_i.PS$, we will decrease the variable $N_{needed}$ by 1 (line 23). The iteration process terminates, when $N_{needed} = 0$ holds (lines 15-16) or there are no more vertices in $(v_i.CS - v_i.PS)$ (line 14).

After checking all vertices in $v_i.CS$, if we still need more vertices in the protection set $v_i.PS$ (i.e., $N_{needed} >0$), then we have to add $N_{needed}$ fake duplicate nodes $v_i'$ of $v_i$ with non-sensitive values on attribute $A_d$ and with the same associated edges as $v_i$ (lines 24-25). Then, we include duplicate nodes $v_i'$ in the protection set $v_i.PS$ (line 26). 

\setlength{\textfloatsep}{1pt}
\begin{algorithm}[t!]\small
\KwIn{a set, $\mathcal{G'}$, of anonymized subgraphs $g^{'}$}
\KwOut{a refined $kt$-safe graph $G'$}
\For{any two expanded subgraphs, $g_1$ and $g_2$, from $\mathcal{G'}$}{
    \For{each common vertex $v$ or its duplicate $v'$ between $g_1$ and $g_2$}{
        \If{$HN(v,n)$ and $HN(v',n)$ in $g_1$ and $g_2$ are not changed w.r.t. that in initial graph $G$}{
            merge $v$ and $v'$ into one vertex
        }
        \Else{
            keep both $v$ and $v'$ in $g_1$ and $g_2$, respectively
        }
    }
    obtain a merged graph $g$ from $g_1$ and $g_2$
    $\mathcal{G'} \leftarrow \mathcal{G'} \cup \{g\} -\{g_1, g_2\}$
}

return $G' \in \mathcal{G'}$
\caption{$merge\_subgraphs(\mathcal{G'})$}
\label{alg:graph_merge}
\end{algorithm}

\noindent {\bf Sensitive Attribute Value Protection of $v_i$:} Next, we will let vertex $v_i$ satisfy the second condition of the $kt$-safe vertex in Definition \ref{def:KT_Safe_Vertex} (i.e., $P_{sens}(v_i.PS, A_d) \le \alpha$). Specifically, if it holds that $P_{sens}(v_i.PS, A_d) > \alpha$, we will add to graph $G$ $x$ fake duplicates of existing vertices $v_m \in v_i.PS$, but with different non-sensitive values on attribute $A_d$, where $x$ is given by $\left\lceil\frac{N_{sens}}{\alpha} - |v_i.PS|\right\rceil$ and $N_{sens}$ is the count of vertices $v_m\in v_i.PS$ with sensitive values on attribute $A_d$ (line 28). Finally, we include these vertices $v_i'$ in the protection set $v_i.PS$ (line 29). This way, we can obtain a protection set $v_i.PS$ for vertex $v_i$, such that $v_i$ is a $kt$-safe vertex (satisfying the two conditions in Definition \ref{def:KT_Safe_Vertex}).

\noindent {\bf Discussions on Parameters $k$, $t$, $\epsilon$ and $\alpha$.} Please refer to Appendix \ref{subsec:k_t_epsilon_alpha_appendix} in our supplemental material for the discussions on parameters $k$, $t$, $\epsilon$ and $\alpha$.

\subsection{$kt$-Safe Graph Merging Phase}
\label{subsec:kt_safe_graph_refine}
In the $kt$-safe graph merging phase, Algorithm \ref{alg:graph_merge} merges the anonymized $kt$-safe subgraphs from $\mathcal{G'}$ back to one single $kt$-safe graph $G'$. In particular, for any two (expanded) $kt$-safe subgraphs, $g_1$ and $g_2$, from $\mathcal{G'}$, we need to merge the original vertices $v$ and their duplicates $v'$ (lines 1-6). That is, if $HN(v,n)$ and $HN(v',n)$ are not modified (compared with the original graph $G$), then we can merge $v$ and $v'$ into the same vertex; otherwise, we will keep both versions (lines 3-6). This way, we can obtain a merged graph $g$ from $g_1$ and $g_2$, and update the set $\mathcal{G'}$ (lines 7-8). Finally, there is only one graph $G'$ left in $\mathcal{G}'$, which will be returned as the refined $kt$-safe graph (line 9).

\subsection{Complexity Analysis}
\label{subsec:anal_and_disc}

In this subsection, we first provide the time complexity of $kt$-safe graph anonymization algorithms. The graph partitioning (Algorithm \ref{alg:graph_partition} in Appendix \ref{subsec:graph_partition} in our supplemental material) needs $O(|V(G)|\cdot log(|V(G)|))$ time complexity, which includes the cost to load vertices $O(|V(G)|)$ and that to assign vertices to $s$ subgraphs $O(|V(G)|\cdot log(|V(G)|))$. The candidate set detection (Algorithm \ref{alg:init_cand} in Appendix \ref{subsec:cand_set_detect} in our supplemental material) needs $O(|V(G)|\cdot C_{n\text{-}hop})$ time complexity in the worst case (i.e., all vertices share the same quasi-identifiers), where $C_{n\text{-}hop}$ is the average time cost to calculate the graph edit distance between $HN(v_i,n)$ and $HN(v_m,n)$. The $kt$-safe vertex anonymization (Algorithm \ref{alg:kt_safe_vertex}) takes $O(|v_i.CS|\cdot |V(G)|\cdot (|V(G)|+E(G)|)+\frac{|v_i.CS|}{\alpha}\cdot (|V(G)|+|E(G)|))$ time complexity in the worst case, which contains 1) the cost to initialize the protection set $v_i.PS$ of $v_i$ $O(|v_i.CS|\cdot n\cdot (d-1))$ when the neighbors $HN(v_m,n)$ of all candidate vertices $v_m\in v_i.CS$ of $v_i$ have the similar value distribution on all non-sensitive attributes (lines 1-11), 2) the cost to make each vertex $v_m\in (v_i.CS-v_i.PS)$ a member of $v_i.PS$ $O(|v_i.CS|\cdot |V(G)|\cdot (|V(G)|+E(G)|))$ when the anonymization of $v_m\in v_i.CS$ affects the privacy of all previous anonymized $kt$-safe vertices ($O(|V(G)|)$) and requires graph operations to be conducted over duplicated vertices (needing $O(|V(G)|+E(G)|)$) (lines 14-23), 3) the cost to add fake vertices and make them as the members of $v_i.PS$ $O(k\cdot(|V(G)|+|E(G)|))$ when the vertices $v_m\in v_i.CS$ cannot be anonymized as members of $v_i.PS$ (lines 24-26), and 4) the cost to add extra fake vertices to meet the privacy requirement $P_{sens}(v_i.PS, A_d) \leq \alpha$ and anonymize these fake vertices as members of $v_i.PS$ $O(\frac{|v_i.CS|}{\alpha}\cdot (|V(G)|+|E(G)|))$ when all vertices $v_m\in v_i.CS$ can be anonymized as members of $v_i.PS$ but all $v_m$ hold sensitive values on attribute $A_d$. The subgraph merging (Algorithm \ref{alg:graph_merge}) requires $O(|\mathcal{G'}|^2 \cdot |g^{'}|)$ time complexity, where $|\mathcal{G'}|$ is the number of anonymized subgraphs $g^{'}$ in set $\mathcal{G'}$, and $|g^{'}|$ is the size of vertices in a subgraph $g^{'}$.

For the anonymization cost (as given in Definition \ref{def:anony_cost}), our proposed $kt$-safety anonymization algorithms need to add $O((k-1)\cdot (|V(G)|+|E(G)|)\cdot \prod_{j=1}^{d-1}|dom(A_j)|)$ nodes and/or edges in the worst case, when each node has a unique value on each non-sensitive attribute $A_j$ ($1\leq j\leq d-1$), where $|dom(A_j)|$ is the domian size of attribute $A_j$. However, via the empirical evaluation of our anonymization algorithm over real/synthetic graphs, this anonymization cost is usually smaller than $O((k-1)\cdot (|V(G)|+|E(G)|)$. Please refer to Section \ref{subsec:effi_scal_eva} and Appendix \ref{subsec:more_anony_cost_eva} in our supplemental material for the anonymization cost evaluation.

\section{Optimizations for the $kt$-safe graph Anonymization}
\label{sec:optimization_for_kt_safe}
In this section, we propose optimization techniques for improving the effectiveness and efficiency of $kt$-safe graph anonymization algorithms. Specifically, we design a cost model to guide the graph partitioning in Section \ref{subsec:cost_model}, propose effective pruning strategies to accelerate the candidate set retrieval for each vertex in Section \ref{subsec:pruning_strategy}, and devise a tree synopsis to facilitate efficient anonymization of a graph $G$ in Section \ref{subsec:kt-tree}. 

\begin{algorithm}[t!]\small
\KwIn{a graph $G$, thresholds $\gamma$ and $s$, and iterations $ite$}
\KwOut{a good graph partitioning set $\mathcal{G}$}

\For{$i=1$ to $ite$}{
    \If{$i=1$}{
        $cts \leftarrow$ randomly and recursively select $|\mathcal{G}|$ centers (vertices) from $G$
        
        $\mathcal{G} \leftarrow$ obtain $|\mathcal{G}|$ subgraphs centered at $cts$ via clustering
        
        $cost \leftarrow$ estimate anonymization cost of set $\mathcal{G}$ via Eq. (\ref{eq:cost_model_for_gp})
    }
    \Else{
        $CTs \leftarrow$ randomly replacing a vertex in $cts$
        
        $\mathcal{G}_{new} \leftarrow$ obtain $|\mathcal{G}|$ subgraphs centered at $CTs$ via clustering
        
        $COST \leftarrow$ estimate the anonymization cost of set $\mathcal{G}_{new}$ via Eq. (\ref{eq:cost_model_for_gp})
        
        \If{$COST < cost$}{
            $cts \leftarrow CTs$
            
            $cost \leftarrow COST$
        
            $\mathcal{G} \leftarrow \mathcal{G}_{new}$
        }
    }
    
}

return $\mathcal{G}$
\caption{Graph Partitioning Set $\mathcal{G}$ Selection}
\label{alg:partition_set}
\end{algorithm}

\subsection{Cost Model for Graph Partitioning}
\label{subsec:cost_model}

In the preprocessing phase in Section \ref{subsec:prep_phase}, we need to partition graph $G$ into disjoint subgraphs of similar sizes. Since the anonymization cost highly depends on the graph partitioning strategy, in this subsection, we will propose a cost model to estimate the anonymizaiton cost of a graph partitioning strategy $\mathcal{G'}$, and guide the graph partitioning process. 

The basic idea of our cost model is as follows. We retrieve a (small) sample graph $S$ from $G$, and anonymize $S$ as a $kt$-safe graph $S^{'}$ via our proposed approach (Algorithm \ref{alg:kt_safe_frame}). The total cost of anonymizing $S$ comes from two parts, anonymization and merging. For anonymization cost, it is the sum of anonymization cost $u.C_{mkt}$ for making each vertex $u\in S$ to be $kt$-safe. For merging cost, it equals to the sum of cost $u^{'}.C_{mer}$ for merging each border vertex $u^{'}\in S$ and its duplicates. Then, given a graph partitioning strategy, we estimate the anonymization cost $Cost(G, G')$ of $G$ in Eq. (\ref{eq:eq1}), by summing up the approximate cost of each vertex $v\in \mathcal{G}$ to be $kt$-safe and that of each border vertex $v^{'}\in \mathcal{G}$ to be merged based on the sample $S$. Specifically, we estimate the anonymization cost, $Cost(G, G')$, below.
\begin{eqnarray}
 && Cost(G, G') \label{eq:cost_model_for_gp}\\ 
 &=& Cost_{anonymization} + Cost_{merging} \notag\\
 &=& \sum_{\forall g_i \in \mathcal{G}}  \sum_{\forall v\in g_i}\{u.C_{mkt}|arg \min\limits_{u} \sum_{\forall u\in S} \vartheta(u, v, n)\}  \notag \\ 
&& + \hspace{-5ex}\sum_{\forall g_i\in\mathcal{G}, g_h\in\mathcal{G}(i\ne h)}\sum_{\forall v^{'}\in g_i\cup g_h}\hspace{-2ex} \{u^{'}.C_{mer}|arg \min\limits_{u^{'}} \sum_{\forall u^{'}\in S} \vartheta(u^{'}, v^{'}, n)\}, \notag
\end{eqnarray}
\noindent where $S$ is a sample graph from $G$ containing vertices $u$, $u.C_{mkt}$ is the anonymization cost of sample $u$ to be $kt$-safe, $u^{'}.C_{mer}$ is the cost of border vertices $u^{'}$ and their duplicates to be merged, and function $\vartheta(u,v,n)$$=$$|V(HN(u,n))|-|V(HN(v,n))|$ calculates the difference of vertex numbers from subgraphs $HN(u,n)$ and $HN(v,n)$.

For each subgraph $g_i\subseteq \mathcal{G}$, based on the processing order $Q_{g_i}$, we estimate the $kt$-safety cost of each vertex $v\in g_i$ as the cost $u.C_{mkt}$ of sample $u\in S$ to be $kt$-safe, where $u$ is the most similar sample to $v$ among all samples in $S$ w.r.t. their $n$-hop neighbor size (i.e., $u=arg \min\limits_{u} \sum_{\forall u\in S}\vartheta(u, v, n)$). Similarly, for each border vertex $v^{'}$ in any two subgraphs $g_i$ and $g_h$, we estimate its merging cost via the cost $u^{'}.C_{mer}$ of the most similar border sample $u^{'}$ to $v^{'}$ in $S$ to be merged.

Therefore, given a graph partitioning set $\mathcal{G}$, we can use Eq.~(\ref{eq:cost_model_for_gp}) to estimate the anonymization cost (i.e., Eq. (\ref{eq:eq1})). Intuitively, a good partitioning set will have a low estimated anonymization cost via Eq.~(\ref{eq:cost_model_for_gp}). We can also adopt existing graph partitioning approaches (e.g., \cite{wang2014partition}) to accelerate the graph anonymization, as long as the estimated anonymization cost via Eq.~(\ref{eq:cost_model_for_gp}) remains low. 

\noindent {\bf Cost-Model-Based Graph Partitioning.} Given a graph $G$, thresholds $\gamma$ and $s$, and an iteration $ite$, we have Algorithm \ref{alg:partition_set} to select a good graph partitioning set $\mathcal{G}$, based on the cost model in Eq.~(\ref{eq:cost_model_for_gp}) (lines 1-14). Specifically, in the first iteration (i.e., $i=1$), we randomly and recursively select $|\mathcal{G}|$ (determined by parameters $\gamma$ and $s$) centers, $cts$, from $G$, and we obtain $|\mathcal{G}|$ subgraphs via clustering, in which cluster nodes in $G$ have the shortest path to a center in $cts$ (lines 2-4). Note that, if the size of a subgraph reaches $\gamma$, we will not assign more vertices to this subgraph. Then, we use Eq.~(\ref{eq:cost_model_for_gp}) to estimate the anonymization cost for anonymizing and merging subgraphs in $\mathcal{G}$ (line 5). In the remaining iterations (i.e., $i\ne 1$), we obtain a new set, $CTs$, of centers (i.e., vertices) by randomly replacing a vertex in $cts$, obtain a new subgraph set $\mathcal{G}_{new}$ based on centers in $CTs$, and estimate the anonymization cost of the set $\mathcal{G}_{new}$ via Eq. (\ref{eq:cost_model_for_gp}) (lines 6-9). We choose the partitioning set (either $\mathcal{G}$ or $\mathcal{G}_{new}$) with lower cost (lines 10-13). Finally, we return the set $\mathcal{G}$ with the minimal cost from $ite$ iterations (line 14).

\subsection{Pruning Strategy}
\label{subsec:pruning_strategy}

In the candidate set detection phase in Section \ref{subsubsec:cand_set_detec} (i.e., Algorithm \ref{alg:init_cand} in Appendix \ref{subsec:cand_set_detect} of our supplemental material), given a vertex $v_i\in V(G)$, in order to obtain its candidate set $v_i.CS$, we need to compute the graph edit distance $ged(HN(v_i,n), HN(v_m,n))$ between $n$-hop neighbors of $v_i$ and each vertex $v_m\in V(G)$, which is rather costly. Assume that there is a vertex $v_p$ whose $n$-hop neighbors $HN(v_p,n)$ can be used as a pivot subgraph. Then, we can reduce the distance computation cost above via the triangle inequality by utilizing the pivot subgraph $HN(v_p,n)$ to filter out those vertices $v_m$ that cannot be in the candidate set $v_i.CS$.

\noindent {\bf The Pruning via Pivots.} We have the pruning lemma below to rule out vertices $v_m\in G$ whose $n$-hop neighbor subgraphs are dissimilar to $HN(v_i,n)$. 

\begin{lemma}{\bf (Pivot Pruning)}
Given two vertex $n$-hop neighbor subgraphs $HN(v_i,n)$ and $HN(v_m,n)$, and a pivot $n$-hop neighbor subgraph $HN(v_p,n)$, vertex $v_m$ can be safely pruned (i.e., $v_m \notin v_i.CS$), if it holds that $|ged(HN(v_i,n), HN(v_p,n)) - ged(HN(v_m,n), HN(v_p,n))| > \epsilon$.
\label{lem:lem1}
\end{lemma}
\noindent {\it Proof:} Please refer to Appendix \ref{subsec:proof_lem1} in the supplemental material.
\qquad $\square$

In practice, we can select a set, $PVTs$, of pivots $v_p\in G$ whose $n$-hop neighbor subgraphs $HN(v_p,n)$ can be adopted to enable the pivot pruning (as discussed in Lemma \ref{lem:lem1}). If $v_m$ can be pruned via any pivot $v_p\in PVTs$, then $v_m$ is not a candidate of the set $v_i.CS$. Please refer to Appendix \ref{subsec:pivot_set_appendix} in our supplemental material for the details of our pivot set selection strategy.

\subsection{$kt$-Tree}
\label{subsec:kt-tree}

Based on the selected pivot set $PVTs$ (via Algorithm \ref{alg:pivot_set} in Appendix \ref{subsec:pivot_set_appendix} in our supplemental material), we propose a tree synopsis, $kt$-tree ($ktT$), to further accelerate the retrieval process of candidates in $v_i.CS$ for each vertex $v_i\in G$.

\noindent {\bf $kt$-Tree Synopsis.} The $kt$-tree, $ktT$, is a tree index built over graph $G$, where each non-leaf node $N$ contains $m$ children nodes $N_i$ (for $1\le i\le m$), and each child node $N_i$ exclusively includes around $(1/m \times 100)\%$ similar vertices in $N$. Specifically, each leaf node in $ktT$ includes vertices $v\in G$ with the same or similar quasi-identifiers. Moreover, we use a hash function, $f(v.A_j)$, to hash value $v.A_j$ of a vertex $v\in G$ on non-sensitive attributes $A_j$ (for $1\le j\le d-1$) into a position in a bit vector with size $B$, where $B>1$ is an integer value. Each node $N\in ktT$ contains two information:
\begin{enumerate}
\item $(d-1)$ bit vectors, $A_j.vec$, obtained via \textit{bit OR} operators over all hash bit values of $v.A_j$ of vertex $v\in N$; and
\item a distance interval, $pvt.I$, of $n$-hop neighbor subgraphs between vertices in $N$ and tree node (pivot) $N$.
\end{enumerate}

\noindent {\bf Index Construction.} We build the $kt$-tree index, $ktT$, by clustering vertices $v$ in $G$ in a bottom-up strategy. First, we cluster all vertices $v\in G$ based on their quasi-identifier $v.QI$, and treat these clusters as the leaf nodes of $ktT$. For each leaf node including vertices $v$ with the same quasi-identifiers, we can get $(d-1)$ vectors $A_j.vec$ via hash function $f(v.A_j)$. Given $|PVTs|$ pivots $pvt$ obtained from Algorithm \ref{alg:pivot_set}, we can group leaf nodes (containing vertices with the same quasi-identifier) into $|PVTs|$ parent nodes via clustering, based on the graph edit distance between $n$-hop neighbor subgraphs of vertices $v$ and pivots $piv$. Furthermore, we select $\sqrt{|PVTs|}$ pivots from the $|PVTs|$ nodes, and cluster these nodes into $\sqrt{|PVTs|}$ supernodes with similar sizes. This process terminates until we reach the root of the $ktT$.

After we build the $ktT$, for each node in $ktT$, we calculate its stored information (i.e., $A_j.vec$ and $pvt.I$).

\noindent {\bf Vertex Pruning via $ktT$.} With the $ktT$, for each vertex $v\in G$, when we retrieve its candidate set $v.CS$, we can prune vertices $v^{'}$ in the leaf nodes $N\in ktT$ that cannot contain the same quasi-identifier $v^{'}.QI$ as $v.QI$, where \textit{bit and} operators over the $(d-1)$ vectors $A_j.vec$ of $v$ and $N$ return false. For vertices $v^{'}$ in the un-pruned leaf nodes $N\in ktT$, we calculate the graph edit distance between the $n$-hop neighbor subgraph of vertex $v$ and pivot $pvt$ (of node $N$), and then check whether or not $v'$ can be pruned via the pivot $pvt$ and the correspondingly stored distance interval $pvt.I$ of node $N$ (Lemma \ref{lem:lem1}).

\section{Experimental Evaluation}
\label{sec:experimentation_evaluation}

\subsection{Experimental Settings}
\label{subsec:exp_settings}

\noindent {\bf Real/Synthetic Data Sets.} We evaluate the performance of our $kt$-safety approach on 6 real and 3 synthetic data sets, as depicted in Table \ref{table:data_sets}. Please refer to Appendix \ref{sec:exp_data_sets} in our supplemental material for the detailed description of these tested data sets.

\begin{table}[t!]
\centering
\scriptsize
\caption{\small The tested data sets.} \label{table:data_sets}
\begin{tabular}{|c|c|c|}
\hline
\textbf{Data Sets} & \textbf{No. of Nodes} & \textbf{No. of Edges} \\
\hline
\hline
Cora~\cite{wang2019attributed} & 2,708 & 5,278 \\\hline
DBLP~\cite{zhou2009graph} & 79,593 & 201,334 \\\hline
Epinions~\cite{tang2012etrust} & 22,166 & 355,813 \\\hline
Facebook~\cite{leskovec2012learning} & 4,039 & 88,234 \\\hline
Wikipedia~\cite{leskovec2010predicting} & 7,115 & 103,689 \\\hline
Arxiv~\cite{leskovec2007graph} & 23,133 & 93,497 \\\hline
Uniform & 10,000 & 97,779  \\\hline
Gaussian & 10,000 & 163,973  \\\hline
Zipf & 10,000 & 56,067  \\\hline
\end{tabular}
\end{table}

\begin{table}[t!]
\scriptsize
\caption{\small The parameter settings.}\label{table:exp_parameter_setting}\hspace{4ex}
\begin{tabular}{|l|c|}
\hline
\qquad\qquad\qquad\qquad\textbf{Parameters} & \textbf{Values}\\
\hline
\hline
identity privacy threshold $k$ & 5, \textbf{10}, 15, 20  \\\hline
sensitive privacy threshold $\alpha$ & 0.1, \textbf{0.2}, 0.3 \\\hline
graph edit distance threshold $\epsilon$ &  3, 4, \textbf{5}, 6, 7 \\\hline
distance threshold, $t$, between  & \textbf{0.1}, 0.2, 0.3, 0.4  \\
two value distributions & \\\hline
the furthest distance, $n$, an attacker & \textbf{1}, 2 \\
can reach from a vertex & \\\hline
the maximum size, $\gamma$, of subgraphs & 500, \textbf{1000}, 1500, 2000 \\\hline
the number, $s$, of graph partitions &  2, 3, \textbf{4}, 5, 6  \\\hline
the size, $|QI|$, of vertices' quasi-identifiers & \textbf{10}, 20, 40, 80, 100  \\\hline
the graph size $|V(G)|$ &  \textbf{10K}, 20K, 40K, 80K, 100K \\\hline
\end{tabular}
\end{table}

\noindent {\bf Competitors.} We compare our $kt$-safe graph generation approach, denoted as $partition+pruning$, with four baseline approaches, namely $kl\text{-}graph$, $partition$, $pruning$, and $none$, where $partition+pruning$ is our approach that applies both graph partition and pruning strategies ($n = 1$ by default), $kl\text{-}graph$ uses an existing $k$-degree-$l$-diversity \cite{yuan2011protecting} graph mechanism ($l= 2$ by default), $partition$ considers our proposed graph partitioning approach only, $pruning$ uses our proposed pruning strategy only, and $none$ do not leverage any optimization method. 

\noindent {\bf Measures.} Following the literature \cite{zou2009k}, we report the \textit{distributions} of three graph utility measures: \textit{vertex degrees}, \textit{shortest path lengths}, \textit{cluster coefficient}. Specifically, the \textit{vertex degree distribution} is the distribution of all vertices in graph; the \textit{shortest path length distribution} is the distribution of the shortest path lengths between 1,000 randomly selected vertex pairs; and the \textit{cluster coefficient}, also known as \textit{transitivity}, is to measure the ratio of connected $1$-hop neighbor pairs of a vertex among all $1$-hop neighbor pairs of the vertex. Moreover, we report the mean and standard derivation of \textit{vertex degrees} and \textit{shortest path length} among 1,000 randomly selected vertices and vertex pairs, resp. Furthermore, we evaluate and report the sizes of largest components in the anonymized graphs. In addition, We evaluate the percentage of $kt$-safety anonymized nodes after applying the $k$-degree-$l$-diversity \cite{yuan2011protecting}. Finally, we report the \textit{wall clock time} (i.e., CPU time) and anonymization cost to anonymize a graph as $kt$-safety. 

\noindent {\bf Parameter Settings.} Table \ref{table:exp_parameter_setting} depicts experimental settings, where default values are in bold. We run our experiments on a PC with Intel(R) Core(TM) i7-6600U CPU 2.70 GHz and 32 GB memory. All algorithms were implemented by C++. The code is available at \textit{\url{http://www.cs.kent.edu/~wren/kt-safety/}}.

\begin{figure}[t!]
\centering
\subfigure[][{\small distributions of degrees}]{
\scalebox{0.17}[0.17]{\includegraphics{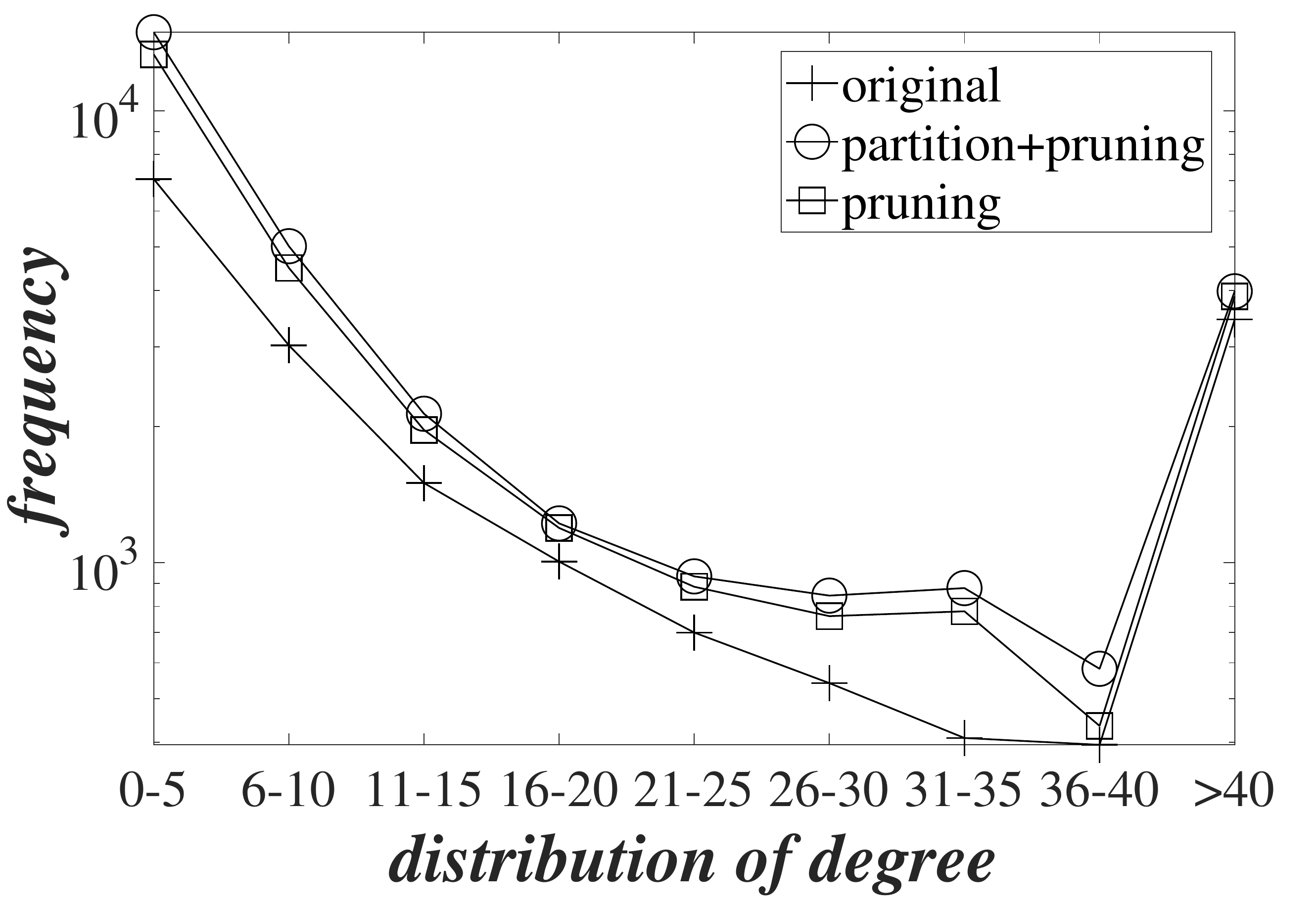}}
\label{subfig:Epinions_vs_degree}
}
\subfigure[][{\small distributions of the shortest path lengths}]{\hspace{-2ex}                  
\scalebox{0.2}[0.2]{\includegraphics{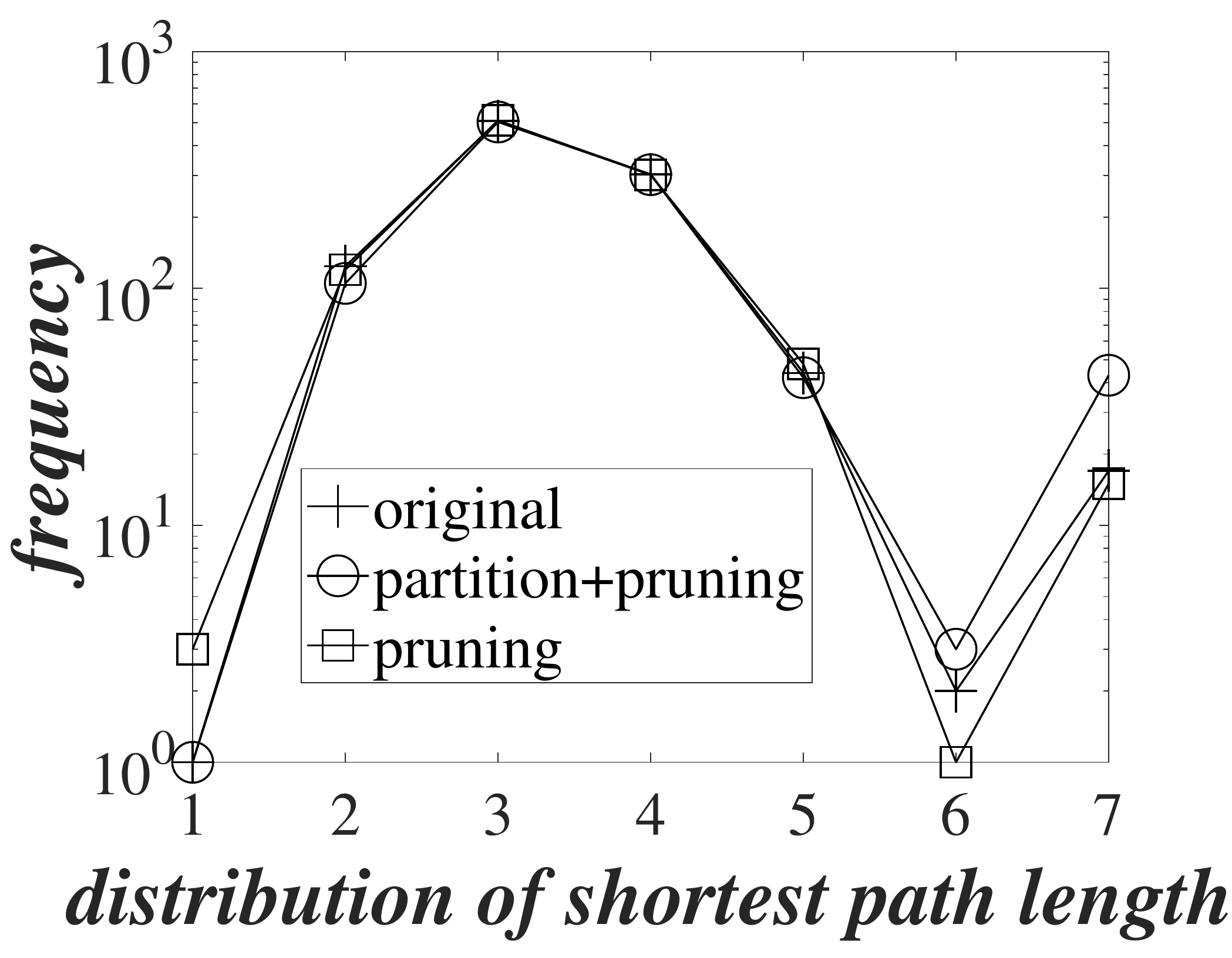}}
\label{subfig:Epinions_vs_spl}
}
\caption{\small Utility evaluation on $Epinions$.}
\label{fig:utility_Epinions}
\end{figure}

\subsection{Utility Evaluation}
\label{subsec:utility_eva}
Due to space limitations, we report the utility result of our $kt$-safe graph generation approach on $Epinions$ in this section and put the results of other 8 data sets to Appendix \ref{subsec:more_utility_eva} in our supplemental material. In order to improve readability, we only report the utility evaluation results of the approaches $partition+pruning$ and $pruning$, since the anonymized graphs via $partition+pruning$ and $pruning$ have the same utility performance as that of $partition$ and $none$, resp.

\noindent {\bf Utility Evaluation vs. Vertex Degree Distribution.} Figure \ref{subfig:Epinions_vs_degree} evaluates the degree distributions of our proposed $kt$-safe graph generation approach over $Epinions$. Specifically, we illustrates the degree distribution of the original graph, which can be regarded as a standard. In figures, we can see that the generated $kt$-safe graphs by our methods (including the 3 variants) vividly reveal the degree distribution trend of original graph. Moreover, from figures, $partition+pruning$ and $partition$ add more fake nodes in the anonymization process than $pruning$ and $none$, since they need more fake nodes to make each partitioned subgraph as $kt$-safe.

\noindent {\bf Utility Evaluation vs. Shortest Path Length Distribution.} Figure \ref{subfig:Epinions_vs_spl} demonstrates the distributions of shortest path lengths among 1,000 randomly selected vertex pairs over $Epinions$. From figures, all the four approaches perform well for revealing the shortest path length distribution of the unanonymized graph. This is reasonable, since we only consider addition operations in graphs. 

Please refer to Appendix \ref{subsec:more_utility_eva} in our supplemental material for the evaluation of distributions of cluster coefficient, the mean and standard derivation of degrees and shortest path length among 1,000 randomly selected vertices and vertex pairs resp., and the largest component size over real/synthetic data. 

\begin{figure}[t!]
\centering
\subfigure[][{\small wall clock time}]{\hspace{-2ex}                  
\scalebox{0.18}[0.18]{\includegraphics{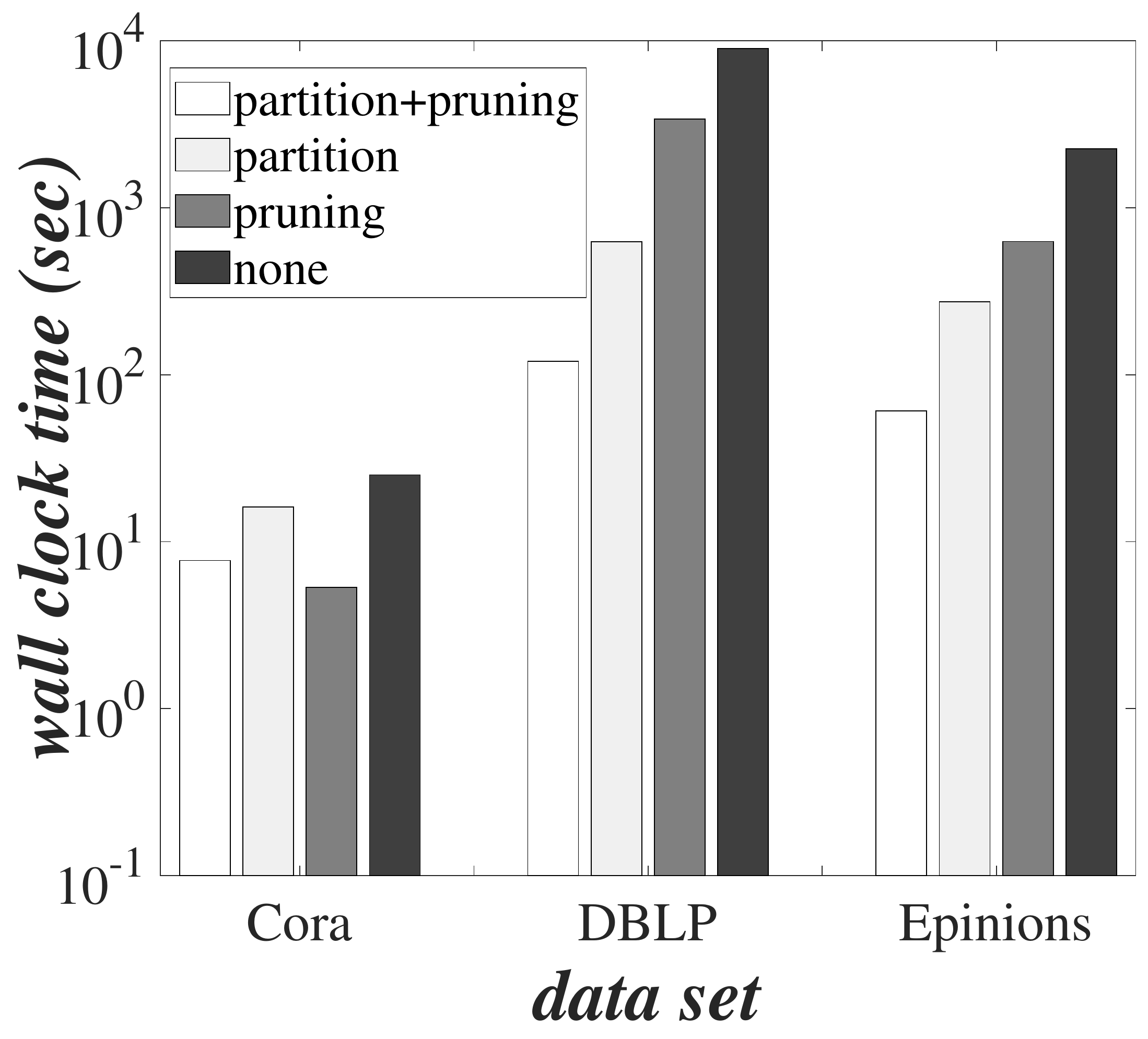}}
\label{fig:exper:time_cost_vs_dataset}
}\quad
\subfigure[][{\small anonymization cost}]{\hspace{-2ex}                  
\scalebox{0.18}[0.18]{\includegraphics{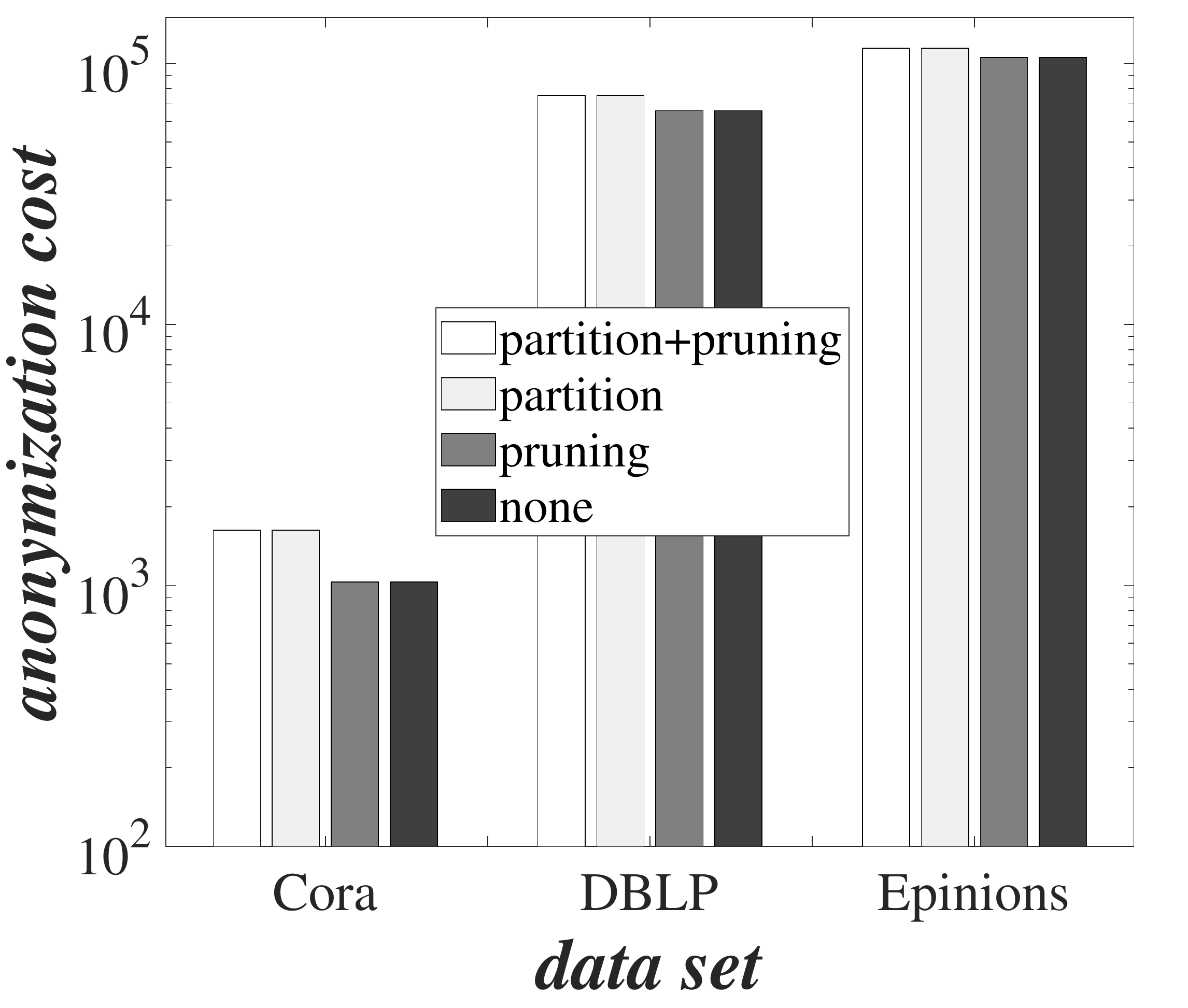}}
\label{fig:exper:anonymization_cost_vs_dataset}
}
\caption{\small Time and anonymization cost vs. real/synthetic data sets.}
\label{fig:performance_vs_datasets}
\end{figure}

\begin{figure}[t!]
\centering
\subfigure[][{\small wall clock time}]{\hspace{-2ex}                  
\scalebox{0.25}[0.25]{\includegraphics{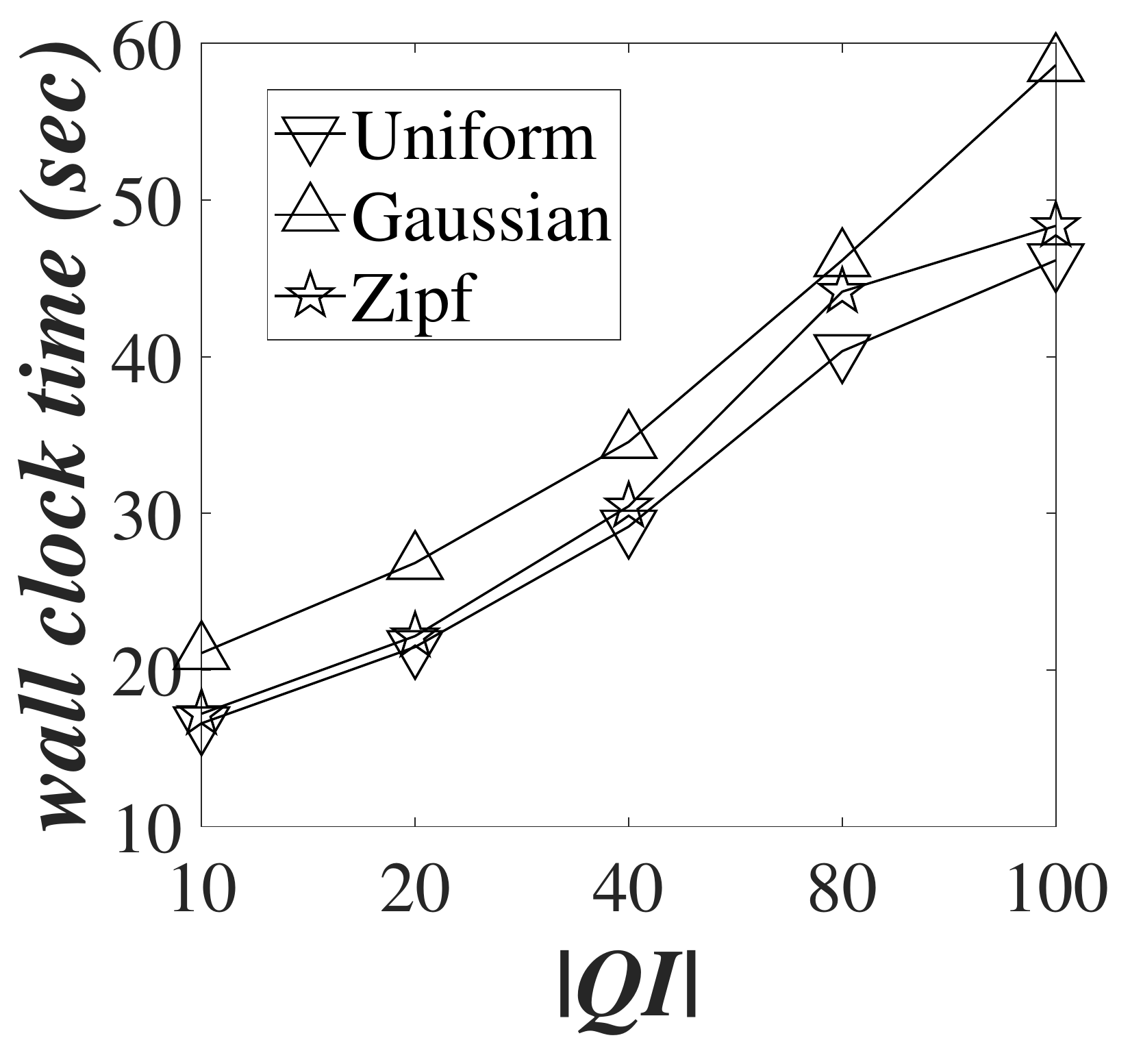}}
\label{fig:time_cost_QI}
}\quad
\subfigure[][{\small anonymization cost}]{\hspace{-2ex}                  
\scalebox{0.25}[0.25]{\includegraphics{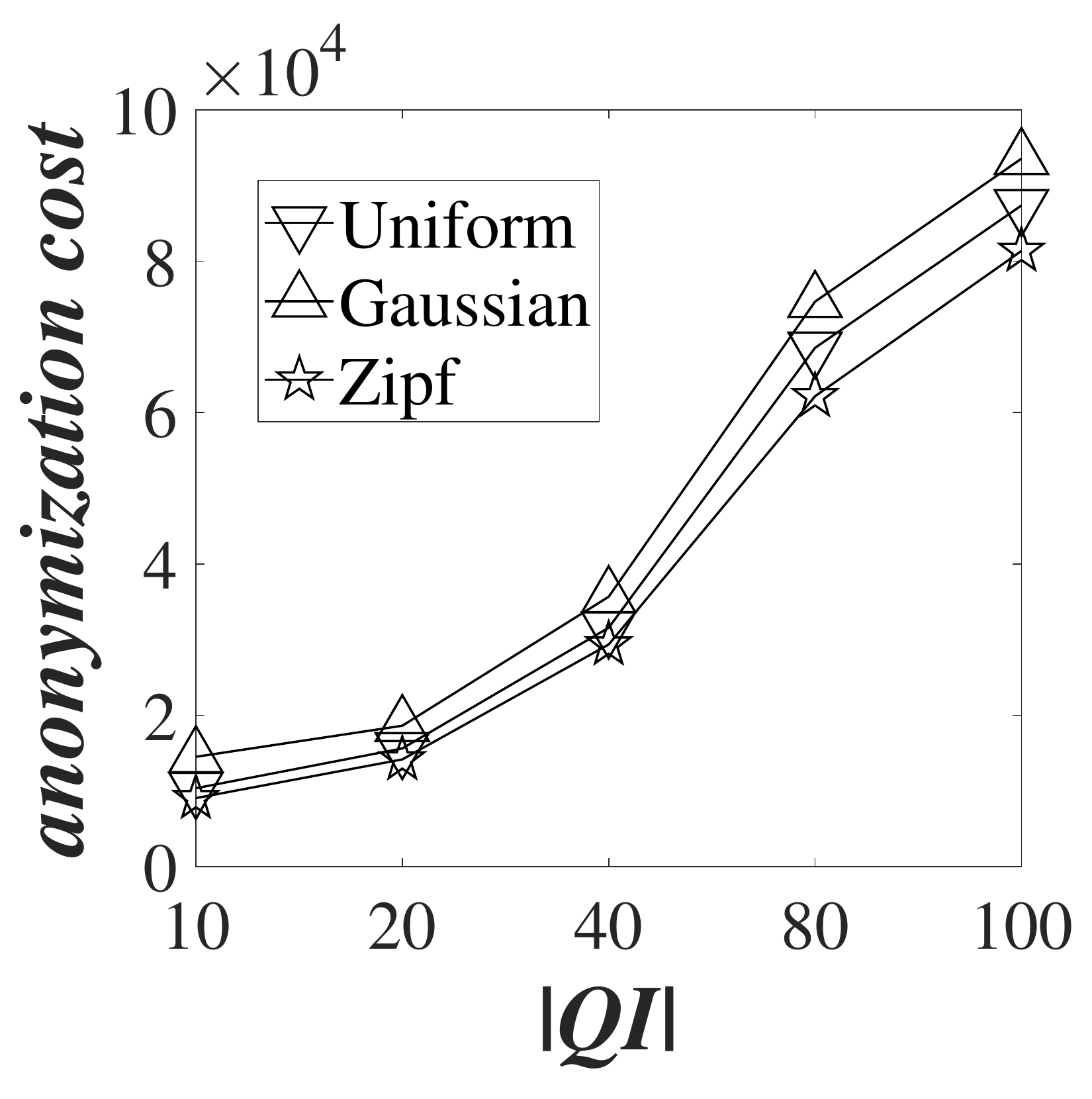}}
\label{fig:anonymization_cost_QI}
}
\caption{\small The time and anonymization cost of $partition$$+$$pruning$ vs. the size, $|QI|$, of vertices' quasi-identifiers.}
\label{fig:time_vs_QI}
\end{figure}

\begin{figure}[t!]
\centering
\subfigure[][{\small wall clock time}]{\hspace{-2ex}                  
\scalebox{0.18}[0.18]{\includegraphics{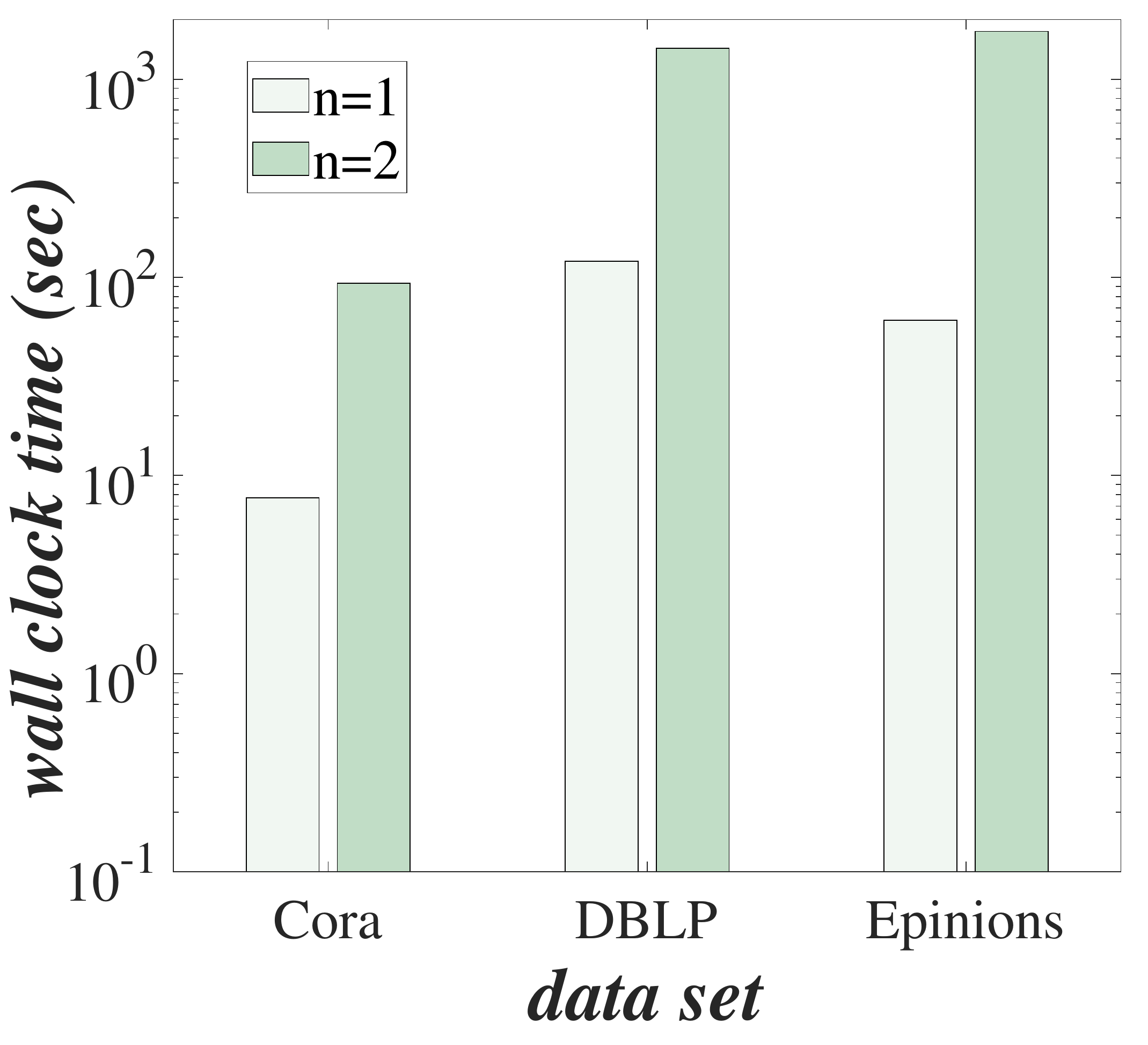}}
\label{fig:time_cost_n}
}\quad
\subfigure[][{\small anonymization cost}]{\hspace{-2ex}                  
\scalebox{0.18}[0.18]{\includegraphics{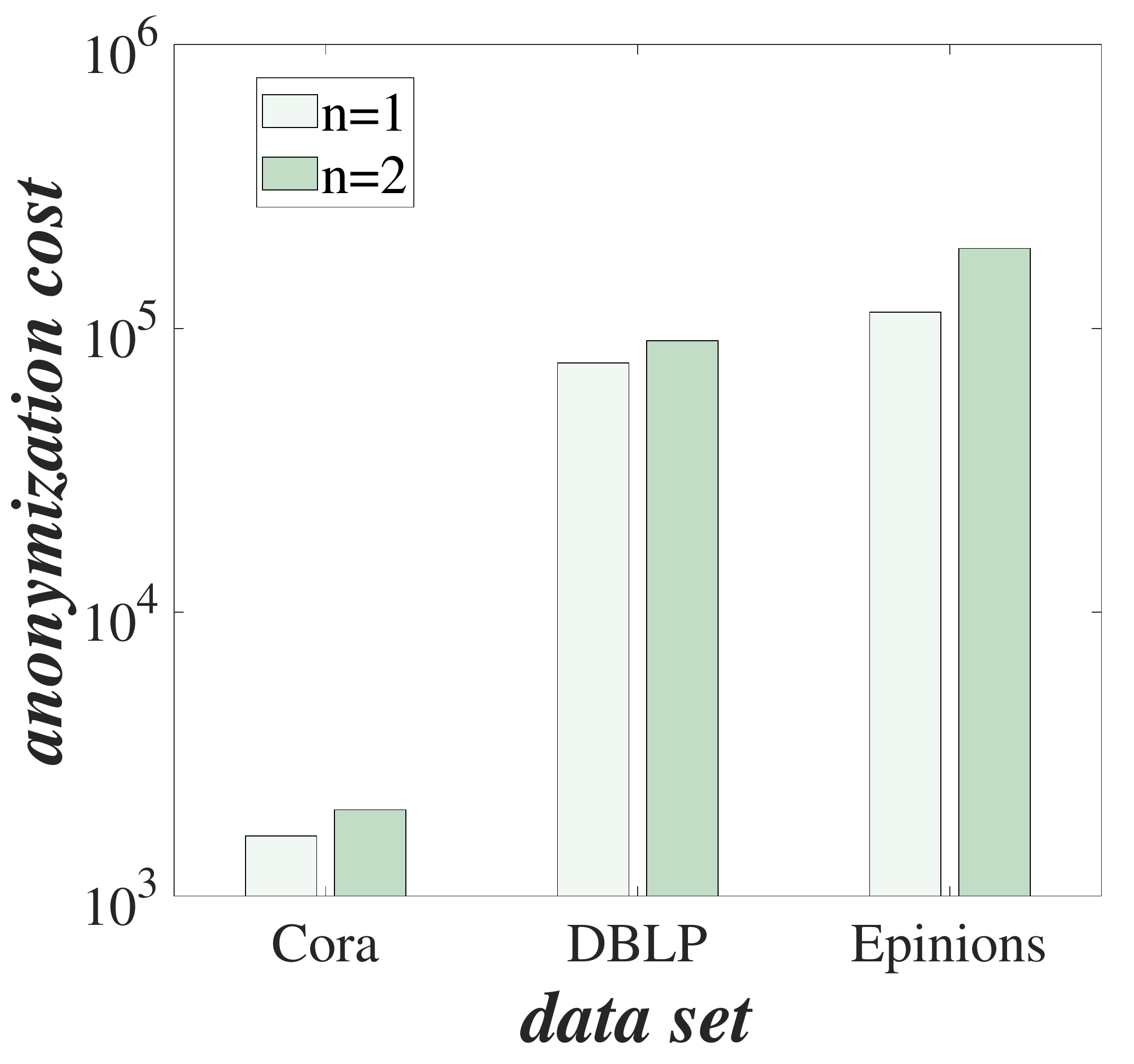}}
\label{fig:anonymization_cost_n}
}
\caption{\small Time and anonymization cost of $partition$$+$$pruning$ vs. the furthest distance, $n$, of an attacker can reach.}
\label{fig:time_vs_n}
\end{figure}

\begin{table*}[h]
\centering
\caption{\small The percentage of $kt$-safety anonymized nodes after applying the $kl$-graph.} \label{table:privacy_power}
\begin{tabular}{|c|c|c|c|c|c|c|c|c|c|}
\hline
\textbf{Data Sets} & \textbf{Cora} & \textbf{DBLP} & \textbf{Epinions} & \textbf{Facebook} & \textbf{Wikipedia} & \textbf{Arxiv} & \textbf{Uniform} & \textbf{Gaussian} & \textbf{Zipf} \\
\hline
\hline
Ratio (\%) & 56.53 & 45.36 & 38.32 & 38.1 & 24.53 & 51.42 & 32.1 & 34.33 & 29.51 \\\hline
\end{tabular}
\end{table*}

\subsection{Comparisons of Privacy Preservation Power}
\label{subsec:privacy_evaluation}

Table \ref{table:privacy_power} depicts the percentage of $kt$-safety anonymized nodes after applying $kl$-graph technique \cite{yuan2011protecting}. From the results, we can see that the anonymized graphs via $kl\text{-}graph$ cannot fully satisfy our $kt\text{-}safety$ requirement, since $kl\text{-}graph$ does not consider the intersection attack between quasi-identifiers (from both nodes and their neighbors) and graph structure. Note that, for an anonymized node via our $kt\text{-}safety$ approach, we cannot guarantee that it has at least $k$ candidates with the same degrees. However, since we do not lease the statistics of graph edit operators, the attackers cannot uniquely identify any target vertex and its sensitive attributes. Thus, $kt\text{-}safety$ is a better, feasible graph privacy mechanism (with stronger privacy preservation power) than $kl\text{-}graph$ for attributed graphs.

\subsection{Efficiency Evaluation}
\label{subsec:effi_scal_eva}
Similar to utility evaluation, we only report the time and anonymization cost on $Cora$, $DBLP$ and $Epinions$, and put the result on the other 6 graphs to Appendix \ref{subsec:more_efficiency_eva} in our supplemental material.

\noindent {\bf Efficiency Evaluation vs. Data Sets.} Figure \ref{fig:exper:time_cost_vs_dataset} illustrates the wall clock time of anonymizing the graph for $Cora$, $DBLP$, and $Epinions$. From the figure, we can see that our approach, $partition+pruning$, can achieve the lowest wall clock time on $DBLP$ and $Epinions$, which indicates the efficiency of our graph anonymization approach. Note that, $partition+pruning$ needs more time cost than $pruning$ on $Cora$, since it needs to partition graph into subgraphs prior to the $kt$-safe anonymization process. This indicates that we may not need to partition graphs in the preprocessing phase for small graphs, by setting the subgraph size threshold $\gamma$ as a large value. For the three variants of our methods, $partition$ (via divide-and-conquer) has the smallest wall clock time, which indicates the necessity of graph partitioning, especially for large graphs. Moreover, $none$ is expected to perform the worst. Please refer to Appendix \ref{subsec:more_efficiency_eva} in our supplemental material for the break-up cost analysis of $partition+pruning$ on $Cora$, $DBLP$, and $Epinions$.

\noindent {\bf Anonymization Cost Evaluation.} Figure \ref{fig:exper:anonymization_cost_vs_dataset} illustrates the anonymization cost of all the approaches on $Cora$, $DBLP$ and $Epinions$. From the figure, $partition+pruning$ and $partition$ have slightly higher cost than $pruning$ and $none$, since they need to add more fake nodes to the partitioned subgraphs than that of the original single graph. However, our approach requires lower time cost for anonymizing large graphs (as shown in Figure \ref{fig:exper:time_cost_vs_dataset}). Therefore, $partition+pruning$ can be used for anonymizing large graphs (e.g., $DBLP$), whereas $pruning$ can be adopted over small graphs (e.g., $Cora$).

\noindent {\bf Time and Anonymization Cost Evaluation vs. the size, $|QI|$, of vertices' quasi-identifiers.} Figure \ref{fig:time_vs_QI} reports the performance of our $partition+pruning$ approach on three synthetic data sets, by varying $|QI|$ from 10 to 100, where other parameters are set to their default values in Table \ref{table:exp_parameter_setting}. From the figures, we can see that larger $|QI|$ leads to both higher time and anonymization costs. This is reasonable, since larger $|QI|$ will generate a more complex (stricter) neighborhood environment for each vertex to be anonymized. Therefore, for each vertex, we need more time to check its $n$-hop neighborhood and add more fake nodes to make each vertex $kt$-safe.

\noindent {\bf Time and Anonymization Cost Evaluation vs. $n$.} Figure \ref{fig:time_vs_n} demonstrates the effect of the parameter $n$ on our $Partition+pruning$ approach on $Cora$, $DBLP$, and $Epinions$, where $n = $ 1 and 2, and other parameters are by default. From the figures, when $n$ increases, both time and anonymization costs increase, since we need to consider a larger range of neighborhood information for each vertex. Specifically, as shown in Figure \ref{fig:time_cost_n}, when $n$ changes from 1 to 2, the wall clock time of our approach increases by 1 order of magnitude (due to the NP-hardness of computing the GED between subgraphs). In Figure \ref{fig:anonymization_cost_n}, the anonymization cost via our approach slightly increases for larger $n$.

We also evaluated our $partition+pruning$ approach by systematically varying other parameters in Table \ref{table:exp_parameter_setting}. Due to space limitations, we will not report similar experimental results here. For interested readers, please refer to Appendix \ref{sec:more_results} in our supplemental material for more efficiency and anonymization cost evaluations. 

\section{Related Work}
\label{sec:related_work}

\noindent {\bf Privacy Preservation for Tabular Data.} The $k$-anonymity \cite{sweeney2002k} mechanism was first proposed to study the privacy issue of the released tabular table. In particular, it requires that any tuple $t$ in the released tabular table have a candidate set $t.CS$ containing at least $k$ indistinguishable candidate records. After that, many works with various privacy preservation constraints have been proposed, including $l$-diversity \cite{machanavajjhala2006diversity}, $t$-closeness \cite{li2007t}, $\delta$-presence \cite{nergiz2007hiding}, $m$-invariance \cite{xiao2007m}, $\beta$-likeness \cite{cao2012publishing}, differential privacy \cite{dwork2011differential}, and so on.

Differential privacy (DP) \cite{dwork2011differential} is a more rigorous privacy preservation mechanism, requiring the inference probability of the presence of any element in a data set should be smaller than $exp(\epsilon)$, where $\epsilon \in [0, 1]$ is a privacy indicator. There are many existing works of the DP variant (e.g., \cite{kotsogiannis2020one}) and applications (e.g., \cite{li2020estimating,cormode2018marginal,zhang2016privtree,yu2019differentially}).

Our work focuses on the privacy preservation over graph data (instead of tabular tables), which contain both vertex attributes and graph structures. We cannot directly borrow previous techniques for tabular data, due to the privacy preservation with graph structures. 

\noindent {\bf Privacy Preservation for Graph Data.} Existing works on the graph privacy can be classified into two categories: the ones applying the differential privacy (e.g., \cite{chen2014correlated,day2016publishing,jorgensen2016publishing,hay2009accurate,kifer2011no,yang2012differential}) and those adopting other privacy mechanisms like $k$-anonymity (e.g., \cite{zou2009k,cheng2010k, boldi2012injecting,liu2008towards,yuan2010personalized,yuan2011protecting,zhou2008preserving}). Moreover, there are some works (e.g., \cite{backstrom2007wherefore,hay2008resisting}) studying possible attacker models for anonymized graphs.


\underline{Graph privacy via differential privacy.}
Hay et al. \cite{hay2009accurate} first applied the differential privacy on graph data, and proposed two privacy-preservation mechanisms, edge-based and node-based differential privacies, which require the presence of any edge (e.g., relationship) and node (e.g., person) not be disclosed with high inferred confidences, respectively. Then, related works on  differential privacy over graphs have two directions: publishing graph statistics (e.g., degree distribution \cite{day2016publishing}) and publishing the entire graph \cite{hay2009accurate,kifer2011no,yang2012differential,chen2014correlated,jorgensen2016publishing}. 

Although differential privacy is a very robust privacy mechanism (e.g., future proofed), it may not be a good choice for an attributed graph under our attacker model (Definition \ref{def:ak}), which is the intersection of quasi-identifiers and $n$-hop neighbors (including graph structure and their quasi-identifiers). Therefore, we do not consider differential privacy as a competitor/baseline in our experiments.

\underline{Graph privacy via other privacy mechanisms.}
Other existing works (e.g., \cite{zou2009k,cheng2010k, boldi2012injecting,liu2008towards,yuan2010personalized,yuan2011protecting,zhou2008preserving,song2012sensitive}) mainly leveraged the idea of $k$-anonymity \cite{sweeney2002k}. Specifically, Song et al. \cite{song2012sensitive} proposed $l$-sensitive-label-diversity mechanism, for protecting from 1-hop neighborhood attack, but cannot resist from an attacker with $n$-hop ($n>1$) neighbor information, which is the focus of our work.

Prior works mainly focused on the graph structure attack only (i.e., $n$-hop neighbor attack) \cite{zou2009k,cheng2010k}. Although some existing works \cite{yuan2010personalized,yuan2011protecting,song2012sensitive} studied both 1-hop neighbors and attribute attacks (i.e., quasi-identifier of a vertex), they cannot protect the graph privacy if an attacker knows the vertex's $n$-hop ($n>1$) neighbor information (containing both structural and quasi-identifier information). Thus, their proposed techniques cannot be directly applied to our $kt$-safe graph anonymization problem.

\section{Conclusions}
\label{sec:conclusions}
In this paper, we formalize an important problem, \textit{kt-safety graph anonymization}, which is useful for protecting the privacy of the released graph data, in many real-world applications such as social networks, communication networks, citation networks, and so on. Due to the NP-hardness of the \textit{kt-safety graph anonymization} problem, we propose a $kt$-safe graph generation framework, which adopts the idea of divide-and-conquer to anonymize the graph data. In order to optimize the generation process of the $kt$-safe graph, we design a cost-model-based graph partitioning strategy, and derive effective pruning and indexing mechanisms to improve the anonymization efficiency. Through extensive experiments on real/synthetic data, we confirm the effectiveness and efficiency of our proposed $kt$-safe graph generation approach.
\ifCLASSOPTIONcompsoc
  \section*{Acknowledgments}
\else
  \section*{Acknowledgment}
\fi

Xiang Lian was supported by NSF OAC (No. 1739491) and Lian Startup (No. 220981) from Kent State University. 

{\small
\bibliographystyle{abbrv}
\bibliography{kt-safety}  
}

\begin{IEEEbiography}[{\includegraphics[width=1in,height=1.25in,clip,keepaspectratio]{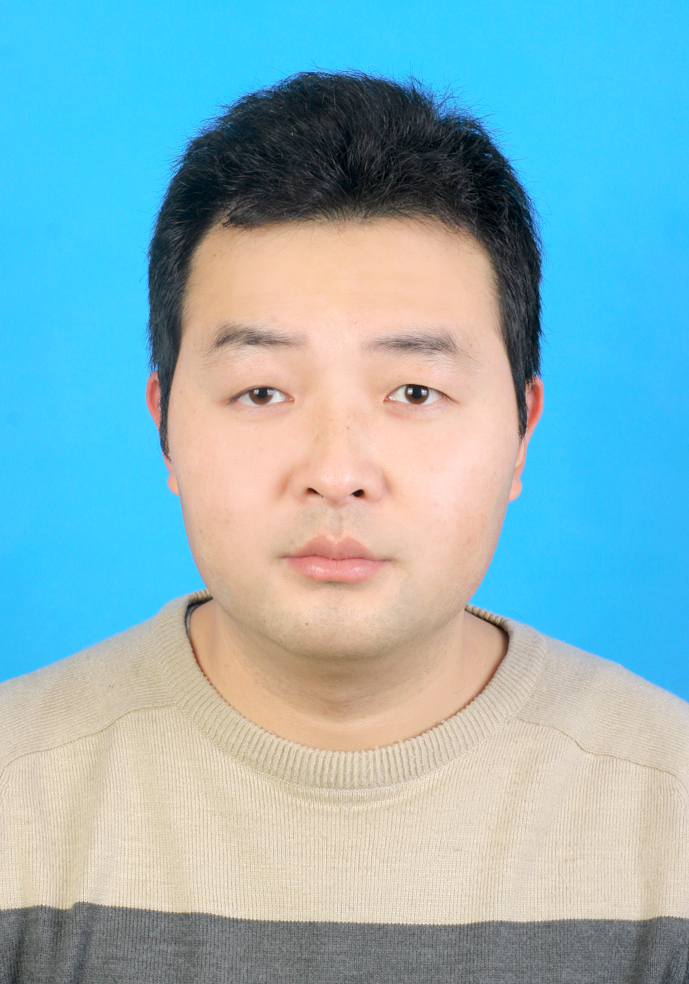}}]{Weilong Ren}
received the Ph.D. degree in Computer Science at Kent State University in 2021. He is now a research scientist in Shenzhen Institute of Computing Sciences, Shenzhen, China. His current research interests include incomplete data management, data integration, and data privacy. 
\end{IEEEbiography}

\begin{IEEEbiography}[{\includegraphics[width=1in,height=1.25in,clip,keepaspectratio]{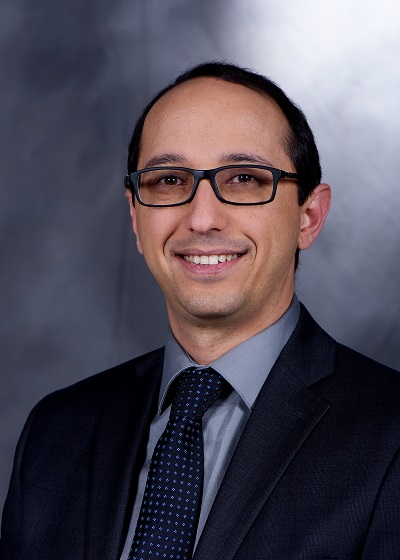}}]{Kambiz Ghazinour}
is an Associate Professor at The State University of New York, Canton. He is directing the Advanced Information Security and Privacy Research Lab and has advised more than 40 research projects related to data security and privacy. Dr. Ghazinour has more than 60 peer reviewed publications and is actively working on privacy-awareness and educating people with usable security and privacy techniques and detect privacy breaches and recommend better privacy practices for the users whose information is widely (and mostly unintentionally) available throughout the Web.
\end{IEEEbiography}

\begin{IEEEbiography}[{\includegraphics[width=1in,height=1.25in,clip,keepaspectratio]{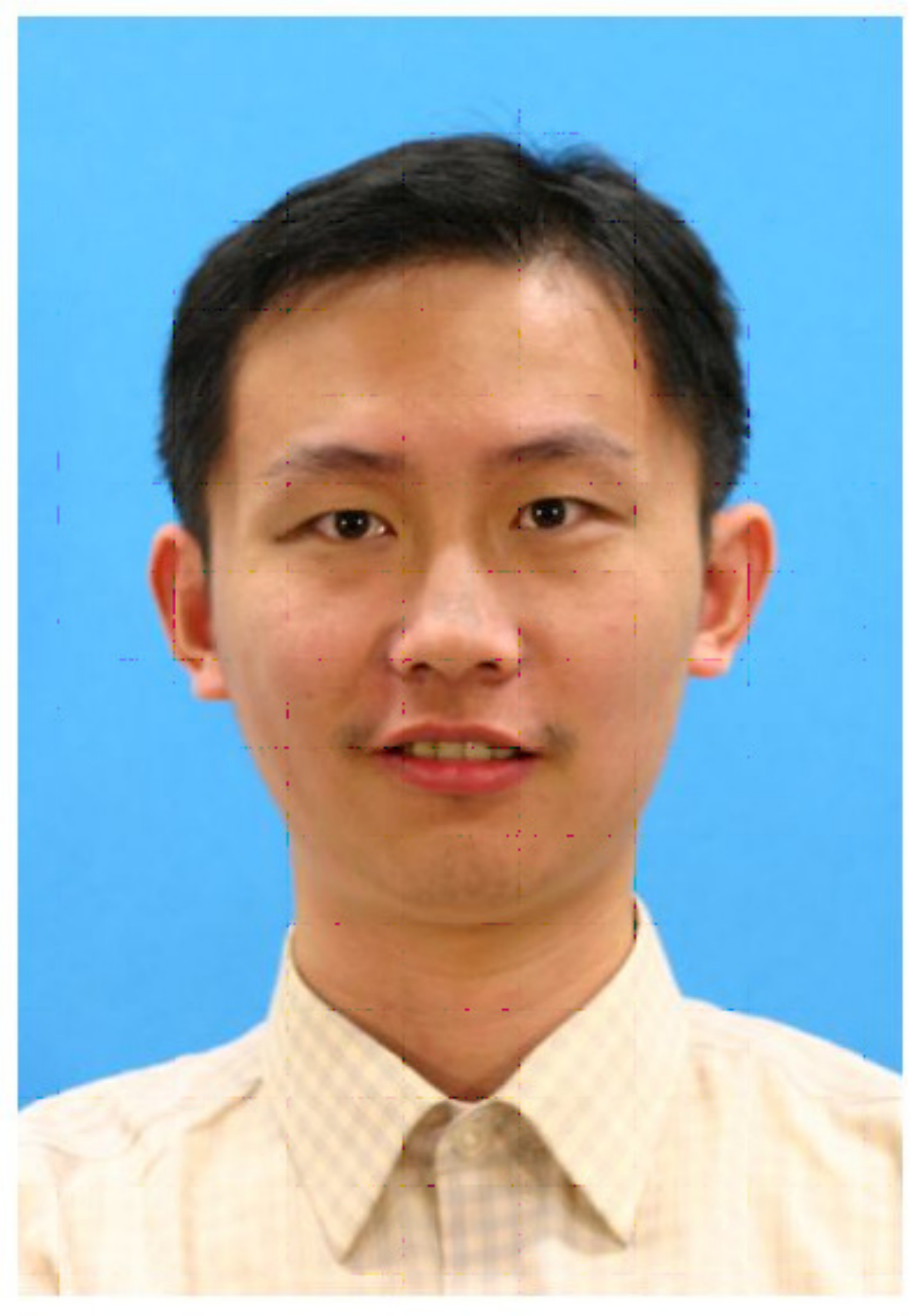}}]{Xiang Lian}
received the BS degree from the Department
of Computer Science and Technology,
Nanjing University, China, and the PhD degree in computer
science from the Hong Kong University of
Science and Technology, Hong Kong. He is now an associate
professor in the Computer Science Department
at the Kent State University, Kent, Ohio, USA. His research interests
include probabilistic/uncertain/inconsistent databases and uncertain/certain graph databases.
\end{IEEEbiography}

\newpage

\noindent {\bf \LARGE Supplemental Materials}

\small

\appendices

\section{More Technical Details}
\label{sec:more_algs}

\subsection{Proof of Theorem~\ref{theo:NP_hardness}}
\label{subsec:proof_of_NPhard}
\noindent {\it Proof:} We introduce a special instance of the $kt$-safe graph anonymization problem in Definition~\ref{def:kt_graph_problem} that resists from any structural attacks but ignores attribute-related attacks. We illustrate that this special instance itself is NP-hard, which can in turn prove the NP-hardness of the $kt$-safe graph anonymization problem in Definition \ref{def:kt_graph_problem}. Specifically, we prove the NP-hardness of the structural-attack-resistance-only instance by reducing from the $k$-automorphism problem \cite{zou2009k} which is an NP-hard problem. We provide a 4-step detailed proof as follows. 

\noindent {\bf The $k$-automorphism problem (Step 1).} We give an instance of the $k$-automorphism problem which takes a non-attributed graph $G$ and a parameter $k$ as input and aims to construct an anonymized graph $G'$ via graph edits with the minimal cost $Cost(G, G')$. As a result, for each vertex $v_i\in G'$, there exist $(k-1)$ automorphic functions $F_a(\cdot)$ ($1\leq a\leq k-1$), where $F_a(v_i)$ returns a different vertex $v_i'$ with exactly the same $n$-hop neighborhood structure as that of the vertex $v_i$ and $F_{a1}(v_i)\neq F_{a2}(v_i)$ ($a1\neq a2$). That is, the anonymized graph $G'$ can protect its vertices from any ($n$-hop) structural attack.

\noindent {\bf The structural-attack-resistance-only instance of the $kt$-safe graph anonymization problem (Step 2).} Given a non-attributed graph $G=(V(G), E(G))$, and thresholds $k$, $t=1$, $\epsilon=0$, and $\alpha=1$, the problem of the structural-attack-resistance-only instance of the $kt$-safe graph anonymization problem is to produce an anonymized $kt$-safe graph $G'$ from $G$, with minimal anonymization cost $Cost(G, G')$ (given by Eq.~(\ref{eq:eq1})).

\noindent {\bf The setup of graph model and thresholds (Step 3).} There are two features for the instance of anonymized $kt$-safe graph above. First, we simplify our graph model in Definition~\ref{def:lg} from attributed graphs $G=(V(G),E(G),A(G))$ to non-attributed ones $G=(V(G),E(G))$ by removing the set of vertices' attributes (or equivalently setting them to the same value), and set attribute-related thresholds $t=1$ and $\alpha=1$. As a result, the conditions $v_i.QI = v_m.QI$ and $att\_dist(HN(v_i,L), HN(v_m,L), A_j) \leq t$ in Definition~\ref{def:anonymization_set} and $P_{sens}(v_i.PS, A_d) \le \alpha$ in Definition~\ref{def:KT_Safe_Vertex} will always hold. Moreover, we strengthen the resistance of the anonymized $kt$-safe graph from any ($n$-hop) structural attacks, by setting $\epsilon=0$. 

\noindent {\bf Reduction from the instance of the $k$-automorphism problem to the structural-attack-resistance-only instance of the $kt$-safe anonymization problem (Step 4).} Given a non-attribute graph $G$ and a parameter $k$ as input, we can reduce the instance of the $k$-automorphism problem to the structural-attack-resistance-only instance of the $kt$-safe anonymization problem, by replacing the requirement of $(k-1)$ automorphic functions $F_a$ for each vertex $v_i$ with the functionally equivalent limits of a $kt$-safe vertex ($|v_i.PS|\geq k$ and $ged(HN(v_i,n), HN(v_m,n))=0$), and setting three parameters $t=1$, $\epsilon=0$, and $\alpha=1$ (as required by the structural-attack-resistance-only instance of the $kt$-safe anonymization problem). This reduction transformation takes polynomial time. We can verify that the anonymized graph $G'$ is a $k$-automorphism network with the minimal anonymization cost $Cost(G, G')$, if and only if $G'$ is a produced $kt$-safe graph with the minimal anonymization cost to the structural-attack-resistance-only instance of the $kt$-safe anonymization problem. Thus, the structural-attack-resistance-only instance of the $kt$-safe anonymization problem is NP-hard, and in turn the $kt$-safety anonymization problem is also NP-hard.  \qquad $\square$

\subsection{Proof of Lemma~\ref{lem:lem1}}
\label{subsec:proof_lem1}
\noindent {\it Proof:} Let $ged(HN(v_i,n), HN(v_p,n)) = dist_1$ and $ged(HN(v_m,n),$ $HN(v_p,n))$ $= dist_2$, based on triangle inequality, we have $|dist_1 - dist_2| \leq ged(HN(v_i,n), HN(v_m,n))$. From the lemma assumption, $|dist_1 - dist_2| > \epsilon$ holds. By using the inequality transition, we can deduce that $ged(HN(v_i,n), HN(v_m,n)) > \epsilon$, which violates the candidate set requirement (i.e., $ged(HN(v_i,n), HN(v_m,n))$ $\le \epsilon$). Thus, vertex $v_m$ cannot be in the candidate set $v_i.CS$.
\qquad $\square$

\subsection{Attack Resistance Discussions on the Anonymized Graph via Algorithm \ref{alg:kt_safe_frame}}
\label{subsec:attack_resistance_discussion}
\noindent {\bf Attack Resistance Discussions on the Anonymized Graph via Algorithm \ref{alg:kt_safe_frame}.} Note that, the anonymized $kt$-safe graph via Algorithm \ref{alg:kt_safe_frame} can resist from existing attacker models such as minimality attack \cite{wong2007minimality}. Given a graph $G$ and its external knowledge defined in Definition \ref{def:ak}, the anonymized $kt$-safe graph $G^{'}$ of $G$ via Algorithm \ref{alg:kt_safe_frame} will not suffer from the minimality attack, i.e., an attacker cannot locate any node and identify its sensitive attribute values with probabilities larger than $1/k$ and $\alpha$ (as defined in Definition \ref{def:KT_Safe_Vertex}), respectively. The reasons for this are as follows. First, the one-to-one mapping of nodes between $G^{'}$ and $G$ does not hold here, which is the basis of minimality attack. This is because, instead of generalizing attribute values, we produce $G^{'}$ by inserting into $G$ duplicate/fake attributed nodes. Moreover, when releasing the resulting anonymized graph $G'$, we will {\it not} release statistics of graph edit operators, such as the number of the inserted nodes/edges/attributes. As a result, the attacker cannot identify whether a node in $kt$-safe graph $G^{'}$ is real or fake, and cannot uniquely identify any target vertex in $G'$ and its sensitive attribute values with high confidences.

\subsection{Graph Partitioning}
\label{subsec:graph_partition}

\begin{algorithm}[t!]\small
\KwIn{a graph $G$, the maximum size, $\gamma$, of subgraphs, and the number, $s$, of partitions/subgraphs}
\KwOut{a set, $\mathcal{G}$, of partitioned subgraphs}
$\mathcal{G} \leftarrow \emptyset$

\If{$|V(G)| \leq \gamma$}{
    \If{$G$ is the original graph to be anonymized}{
        $\mathcal{G} \leftarrow G$
    }\Else{
        add the expanded graph of $G$ to $\mathcal{G}$
    }
}
\Else{
    partition $G$ into $s$ subgraphs, $g_1$, $g_2$, ..., and $g_s$, with similar sizes
    
    \For{$i=1$ to $s$}{
        $\mathcal{G} \leftarrow \mathcal{G} \cup partition\_graph(g_i, \gamma, s)$
    }
}

return $\mathcal{G}$ containing partitioned subgraphs
\caption{$partition\_graph(G, \gamma, s)$}
\label{alg:graph_partition}
\end{algorithm}

Given a graph $G$, the maximum size, $\gamma$, of subgraphs, and the number, $s$, of partitions, Algorithm \ref{alg:graph_partition} produces a set, $\mathcal{G}$, of partitioned subgraphs. Specifically, we will first check whether or not the number, $|V(G)|$, of vertices in a graph $G$ is larger than the maximum size $\gamma$. If $|V(G)|\le \gamma$ holds for the original graph $G$, we do not need to partition $G$. If $|V(g)|\le \gamma$ holds for a subgraph $g$, we do not need to further partition $g$. Since our anonymization process involves $n$-hop neighbors of each vertex in $G$, we will thus expand subgraph $G$ by including those vertices that are within $n$ hops away from border vertices in $G$, and add the resulting expanded graph of $G$ to $\mathcal{G}$ (lines 2-6). When $|V(G)| > \gamma$ holds, we partition graph $G$ based on the size and divide $G$ into $s$ subgraphs $g_i$ ($1\le i\le s$) with similar sizes (i.e., $|V(g_i)|$), and partition each subgraph $g_i$ by recursively invoking function $partition\_graph(\cdot)$ (lines 7-10). Finally, we return the final partitioning set $\mathcal{G}$ (line 11). 

\subsection{Discussions on Parameters $\gamma$ and $s$}
\label{subsec:gamma_s_appendix}
\noindent \textbf{Discussions on Parameters $\gamma$ and $s$.} Parameters $\gamma$ and $s$ are mainly used for the trade-off between the anonymization cost and efficiency. Intuitively, for larger $\gamma$ (e.g., close to or even greater than $|V(G)|$) and/or smaller $s$ (e.g., close to 2), the partitioned subgraphs $g_i\subseteq G$ contain more vertices and each vertex in $g_i$ has more candidates (line 6 in Algorithm \ref{alg:kt_safe_frame}), which leads to lower anonymization cost and higher time cost to evaluate. On the other hand, with smaller $\gamma$ (e.g., $\leq$ 1,000) and/or larger $s$ (e.g., $\geq$ 10), we need less time but higher anonymization cost to anonymize $G$, since each vertex in $g_i$ has fewer candidates.

Moreover, the setup of $\gamma$ and $s$ may also affect the usability of the anonymized graph. Given larger $\gamma$ and/or smaller $s$, the anonymized graph will have better usability, since there are fewer partitioned subgraphs with lower anonymization cost. Instead, smaller $\gamma$ and/or larger $s$ will lead to worse usability of the anonymized graph. Therefore, for small graphs with fewer vertices, we can use larger $\gamma$ (e.g., close to or even greater than $|V(G)|$) and/or smaller $s$ (e.g., close to 2) for a better utility with lower anonymization cost; for large graphs with unbearable anonymization time cost, we can use smaller $\gamma$ (e.g., $\leq$ 1,000) and/or larger $s$ (e.g., $\geq$ 10) for a better efficiency.

\subsection{Candidate Set Detection}
\label{subsec:cand_set_detect}

\begin{algorithm}[t!]\small
\KwIn{a graph $G$, a vertex $v_i$, and thresholds $k$, $t$, $\epsilon$, and $\alpha$}
\KwOut{a candidate set, $v_i.CS$, of vertex $v_i$}
$v_i.CS \leftarrow null$

\For{each $v_m\in V(G)$}{
    \If{$v_m.QI = v_i.QI$}{
        \If{$ged(HN(v_i,n), HN(v_m,n)) \le \epsilon$}{
            $v_i.CS \leftarrow v_i.CS \cup \{v_m\}$
        }
    }
}

return $v_i.CS$

\caption{$initial\_candidate(v_i, G)$}
\label{alg:init_cand}
\end{algorithm}

As illustrated in Algorithm \ref{alg:init_cand}, for each vertex $v_m\in V(G)$, we will first check whether or not $v_m$ has the same quasi-identifier with $v_i$, and the graph structure of the $n$-hop neighbors, $HN(v_m,n)$, similar to that of $v_i$ (lines 3-4). If the answer is yes, we will add $v_m$ to $v_i.CS$ as a potential candidate vertex in the protection set $v_i.PS$ (line 5). Finally, we return the candidate set $v_i.CS$ (line 6).

\subsection{Discussions on Parameters $k$, $t$, $\epsilon$ and $\alpha$}
\label{subsec:k_t_epsilon_alpha_appendix}
\noindent {\bf Discussions on Parameters $k$, $t$, $\epsilon$ and $\alpha$.} These 4 parameters are mainly related to the trade-off between privacy and usability of the anonymized graph. Specifically, in order to achieve a stronger privacy protection, we can set $k$ to a large value (e.g., $\geq 10$) and the other three to smaller values (e.g., close to 0), which in return may lead to worse usability. This is because, under this parameter setting, for each $kt$-safe vertex $v_i$ in the anonymized graph, it requires that 1) its protection set $v_i.PS$ contains more vertices $v_i^{'}$ (restricted by $k$); 2) the value distributions between $HN(v_i,n)$ and $HN(v_i^{'},n)$ on non-sensitive attributes are similar for each $v_i^{'}\in v_i.PS$ (restricted by $t$); 3) the graph edit distance between $HN(v_i,n)$ and $HN(v_i^{'},n)$ is small for each $v_i^{'}\in v_i.PS$ (restricted by $\epsilon$); and 4) a small fraction of vertices $v_i^{'}\in v_i.PS$ have sensitive values (restricted by $\alpha$). As a result, we need to add more fake vertices and edges for a safer $kt$-safe graph. The case of smaller $k$ and larger $t$, $\epsilon$ and $\alpha$ is opposite and omitted here.

\begin{algorithm}[t!]\small
\KwIn{a graph $G$, a sample graph size $|V(g_s)|$, an iteration threshold $iter$, and a pivot set size $|PVTs|$}
\KwOut{a pivot set $PVTs$}
$g_s \leftarrow$ a sample subgraph with $V(g_s)$ vertices from $G$

$cnt.max \leftarrow null$

$pvts \leftarrow null$

\For{$i=1$ to $iter$}{
    $cnt.now \leftarrow 0$

    \If{$i=1$}{
        $PVTs \leftarrow$ $|PVTs|$ random vertices $v_p \in G$
        
        \For{each vertex $v\in g_s$}{
            \For{each vertex $v'\in g_s$ ($v'\ne v$)}{
                \If{$v'$ can be pruned via Lemma \ref{lem:lem1}}{
                    $cnt.now \leftarrow cnt.now + 1$
                }
            }
        }
    }
    \Else{
        $pvts \leftarrow$ pivot set $PVTs$ after randomly replacing a pivot in $PVTs$ with a different vertex in $G$
        
        $cnt.now \leftarrow$  the count of pruned vertices in $g_s$ via lines 8-11
    }
    
    \If{$cnt.now > cnt.max$}{
        $cnt.max \leftarrow cnt.now$
        
        $PVTs \leftarrow pvts$
    }
    
}

return $PVTs$
\caption{Pivot Set Selection}
\label{alg:pivot_set}
\end{algorithm}

\subsection{The Selection of a Pivot Set}
\label{subsec:pivot_set_appendix}
\noindent {\bf The Selection of a Pivot Set.}  The basic idea of $PVTs$ selection is as follows. We sample a subgraph $g_s$ from $G$. Then, we select a pivot set $PVTs$ from $G$ that can prune a large number of vertices $v_m^{'}\in g_s$ that cannot be the candidate of $v_i^{'}.CS$ for each $v_i^{'}\in g_s$. Finally, for each vertex $v_i$ in graph $G$, we leverage the selected $PVTs$ to prune the vertex $v_m\in G$ that are not in its candidate set $v_i.CS$, which accelerate the execution of candidate set detection of Algorithm \ref{alg:init_cand}.

Specifically, given a graph $G$, a sample graph size $|V(g_s)|$, an iteration threshold $iter$, and a pivot set size $|PVTs|$, Algorithm \ref{alg:pivot_set} illustrates the details of retrieving such a good pivot (vertex) set $PVTs$. In particular, we have two variables, $cnt.max$ and $cnt.now$, to record the maximum and current numbers of pruned vertices for different pivot sets in iterations, respectively, where vertices are pruned via \textit{Lemma} \ref{lem:lem1}. Algorithm \ref{alg:pivot_set} first samples a subgraph $g_s$ from $G$ (line 1). In the first iteration, we randomly select a pivot set $PVTs$ from $G$ and record the number of pruned vertices via Lemma \ref{lem:lem1} (lines 6-11). In each of following iterations, we obtain a new pivot set $pvts$ by randomly replacing an old pivot $v_p\in PVTs$ and recount the number of pruned vertices (lines 12-14). Finally, Algorithm \ref{alg:pivot_set} obtains and returns the pivot set which prunes the maximum number of vertices.

\begin{figure*}[ht]
\centering
\subfigure[][{\small Graph partitioning}]{\hspace{0.5ex}  
\scalebox{0.2}[0.2]{\includegraphics{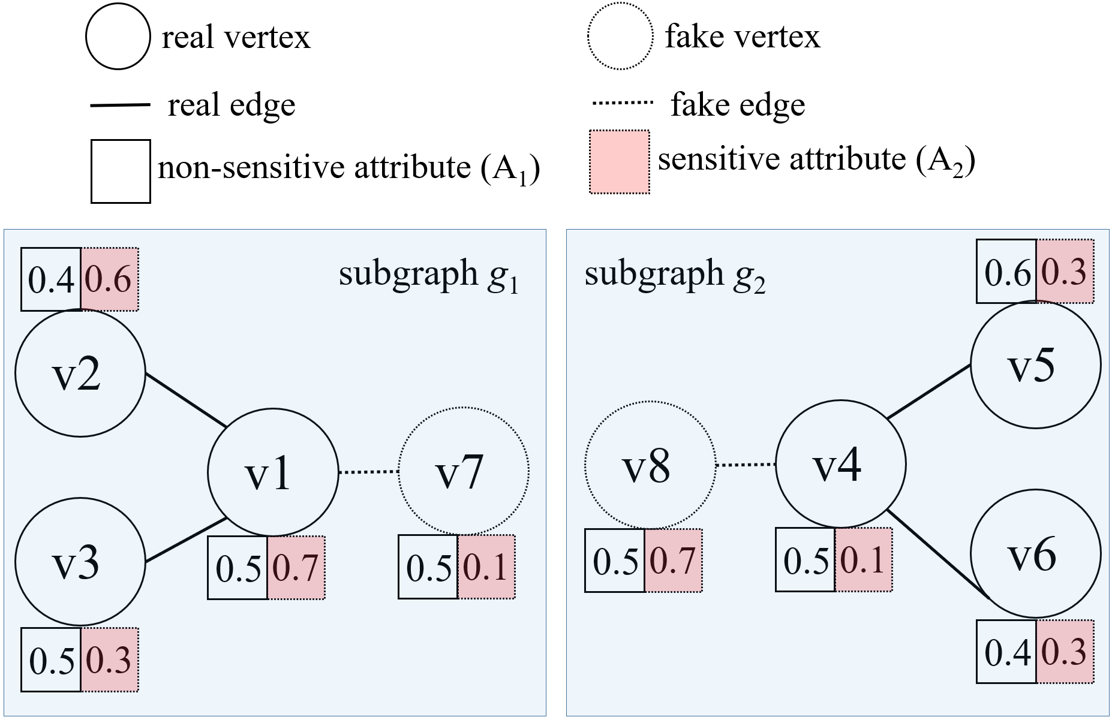}}
\label{subfig:partition}
}
\subfigure[][{\small $kt$-safe subgraph generation}]{\hspace{0.5ex}                
\scalebox{0.2}[0.2]{\includegraphics{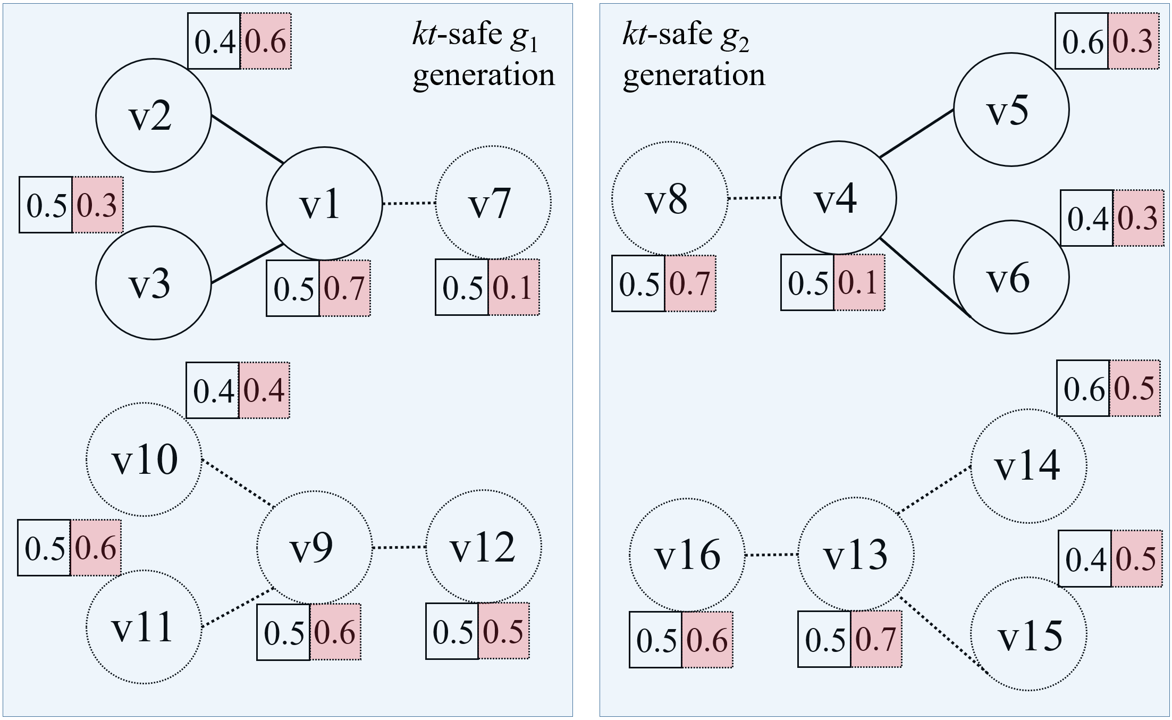}}
\label{subfig:kt_safe_subgraph}
}
\subfigure[][{\small Subgraph merging}]{\hspace{0.5ex}                   
\scalebox{0.2}[0.2]{\includegraphics{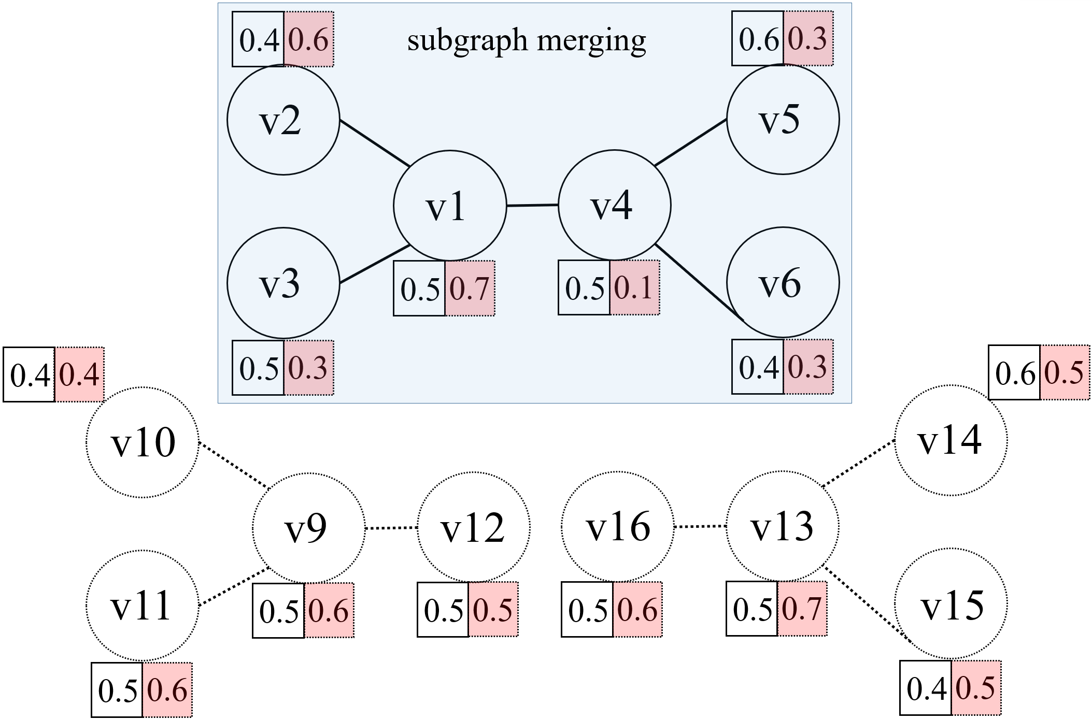}}
\label{subfig:kt_safe_G}
}
\caption{\small A $kt$-safe graph generation demo of graph $G$ in Fig. \ref{fig:motivation1} via Algorithm \ref{alg:kt_safe_frame} (with $k=2$, $t=0$, $\epsilon=0$, $n=1$, $\alpha=0.5$, $\gamma=3$, and $s=2$).} 
\label{fig:approach_demo}
\end{figure*}

\subsection{A $kt$-safe graph generation demo of graph $G$ in Fig. \ref{fig:motivation1} via Algorithm \ref{alg:kt_safe_frame}}
\label{subsec:demo}
\begin{example} {\bf (Graph Partition)}
Continue with Example \ref{example_2}. Assuming $k=2$, $t=0$, $\epsilon=0$, $n=1$, $\alpha=0.5$, $\gamma=3$, and $s=2$, we can divide the graph $G$ in Figure \ref{fig:motivation1} into two subgraphs, $g_1$ and $g_2$, by cutting the edge connecting vertices $v_1$ and $v_4$ via Algorithm \ref{alg:graph_partition}, as shown in Figure \ref{subfig:partition}. Specifically, for $g_1$, we add a duplicate vertex $v_7$ of $v_4$ with the same attribute values, since the 1-hop neighbor $v_4$ of $v_1$ is excluded from $g_1$ due to the graph partitioning. For subgraph $g_2$, the operations are similar and thus omitted. 
\label{example_4}
\end{example}

\begin{example} {\bf ($kt$-Safe Subgraph Generation)}
Continue with Example \ref{example_4}. Given the partitioned subgraphs in Figure \ref{subfig:partition}, Figure \ref{subfig:kt_safe_subgraph} shows a possible case of the generated $kt$-safe subgraphs via Algorithm \ref{alg:kt_safe_vertex}. For subgraph $g_2$, in order to make it $kt$-safe, we add 4 fake vertices, $\{v_4,v_5,v_6,v_8\}$, which contain the same quasi-identifiers but different sensitive attribute values as vertices $\{v_{13},v_{14},v_{15},v_{16}\}$, respectively. The case for $g_1$ is similar and discussion thus omitted. 
\label{example_5}
\end{example}

\begin{example} {\bf (Subgraph Merging)}
Continue with Example \ref{example_5}. Given the partitioned $kt$-safe subgraphs in Figure \ref{subfig:kt_safe_subgraph}, as shown in Figure \ref{subfig:kt_safe_G}, we can merge subgraphs $g_1$ and $g_2$, since the 1-hop neighbors of border vertices $v_1$ and $v_4$ are unchanged in Figure \ref{subfig:kt_safe_subgraph} w.r.t. that of $v_1$ and $v_4$ in the original graph (Figure \ref{fig:motivation1}).
\label{example_6}
\end{example}

\begin{figure}[H]
\centering
\scalebox{0.22}[0.22]{\includegraphics{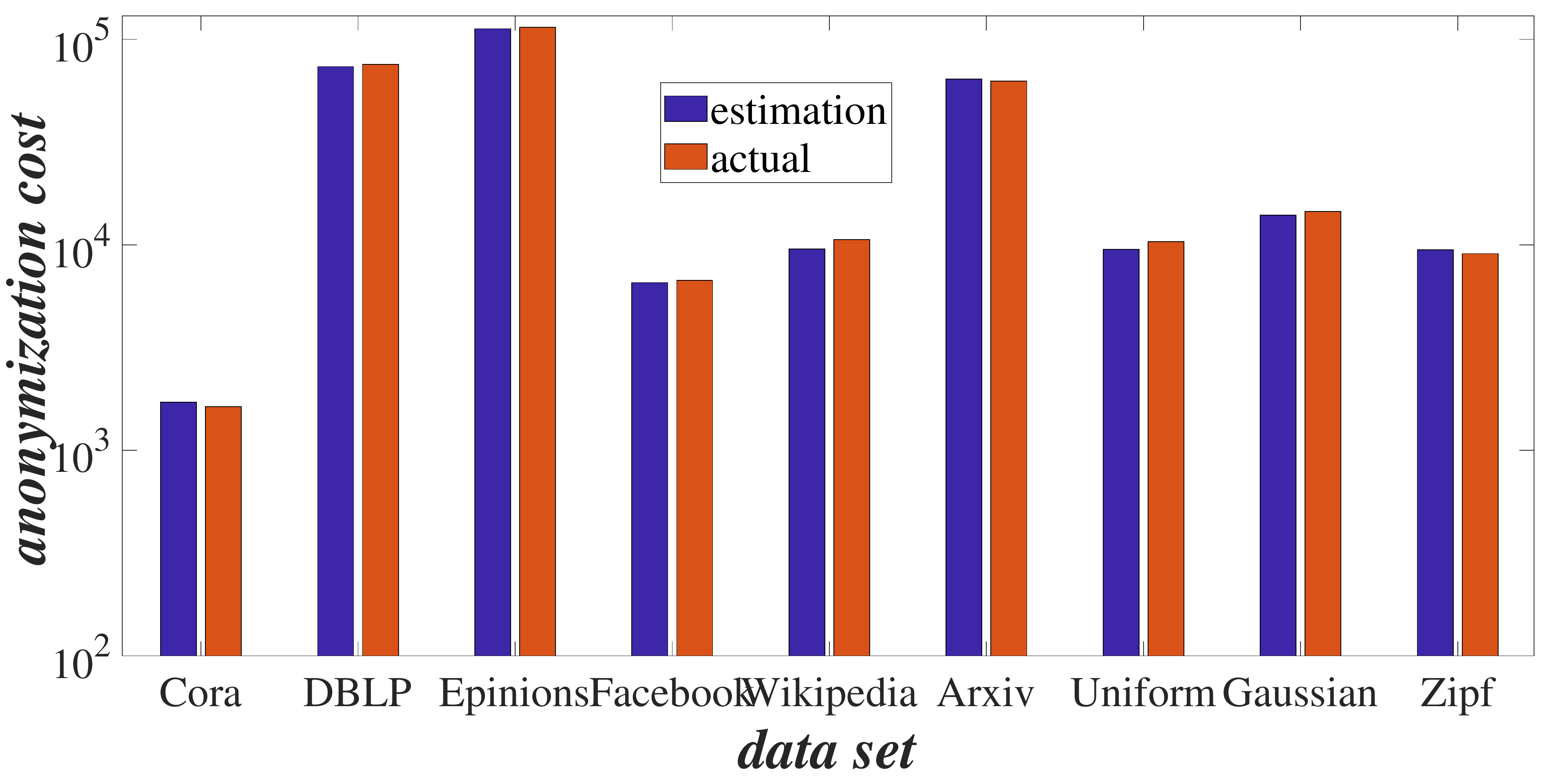}}
\caption{\small Cost model verification (Eq. \ref{eq:cost_model_for_gp}) for the anonymization cost.}
\label{fig:cost_model_eva}
\end{figure}

\section{Description of Experimental Data Sets}
\label{sec:exp_data_sets}
\underline{Real data sets.} 
We used six real-world graphs, $Cora$, $DBLP$, $Epinions$, $Facebook$, $Wikipedia$, and $Arxiv$. $Cora$ \cite{wang2019attributed} contains 2,708 nodes (scientific publications from seven classes) and 5,278 edges (citations among these publications); $DBLP$ \cite{zhou2009graph} is a co-author network from four research areas with 79,593 nodes (authors) and 201,334 edges (co-author relationships); $Epinions$\footnote{\scriptsize\url{https://www.cse.msu.edu/~tangjili/datasetcode/truststudy.htm}} \cite{tang2012etrust} contains 22,166 nodes (users) and 355,813 edges (trust relationships), crawled from viewer site Epinions in May, 2011; $Facebook$\footnote{\scriptsize\url{http://snap.stanford.edu/data/ego-Facebook.html}} \cite{leskovec2012learning} contains 4,039 nodes (users addresses) and 88,234 edges (friendships), collected from survey participants; $Wikipedia$\footnote{\scriptsize\url{http://snap.stanford.edu/data/wiki-Vote.html}} \cite{leskovec2010predicting} includes 7,115 nodes (users) and 103,689 edges (votes), from Wikipedia till January 2008; $Arxiv$\footnote{\scriptsize\url{https://snap.stanford.edu/data/ca-CondMat.html}} \cite{leskovec2007graph} has 23,133 nodes (authors) and 93,497 edges (co-authorships), collected from Condense Matter category in the period from January 1993 to April 2003. Note that, for all graphs, we treat them as undirected graphs.

For each publication (node) in $Cora$, we consider the number of unique vocabulary words it contains and the class it belongs to as its attributes. For each author (node) in $DBLP$, we associate it with two attributes, prolific and research topics, where authors with $\geq$ 20, $\geq 10$ and $<10$ papers are labelled as highly prolific, prolific and low prolific, resp. For each user (node) in $Epinions$, we randomly retrieve one rating item he/she gave, and set the attributes of this rating as the attributes of the node. The remaining three real data sets do not contain attribute values. Therefore, we fill the attribute values of vertices on $Facebook$, $Wikipedia$, and $Arxiv$, with real values from Intel lab data\footnote{\scriptsize\url{http://db.csail.mit.edu/labdata/labdata.html}}, US \& Canadian city weather data\footnote{\scriptsize\url{https://www.kaggle.com/selfishgene/historical-hourly-weather-data}}, and S\&P 500 stock data\footnote{\scriptsize\url{https://www.kaggle.com/camnugent/sandp500}}, respectively. Specifically, $Intel$ contains 2.3 million data, collected from 54 sensors deployed in Intel Berkeley Research lab on Feb. 28-Apr. 5, 2014; $Weather$ contains $45.3K$ historical weather (temperature) data for 30 US and Canadian Cities during 2012-2017; $Stock$ has $619K$ historical stock data for all companies found on the S\&P 500 index till Feb. 2018. For each data set, we retrieved 4 attributes, where we consider 3 non-sensitive attributes and one sensitive attribute. Attribute values are normalized to integers from 0 to 5, which leads to at most 125 combinations of different \textit{quasi\text{-}identifiers}.

\underline{Synthetic data sets.} We produce 3 types of synthetic graph data containing 10,000 nodes, whose degree distributions follow uniform, Gaussian, or Zipf (a special case of heavy-tail) distribution, denoted as $Uniform$, $Gaussian$, and $Zipf$, respectively, as shown in Table \ref{table:data_sets}. Specifically, $Uniform$ holds nodes with degrees falling in $[0, 40]$, $Gaussian$ generates nodes via a model taking 20 and 8 as \textit{mean} and \textit{sigma}, respectively, and $Zipf$ produces nodes with scale invariance 0.8. Moreover, for attributes, $Uniform$, $Gaussian$, and $Zipf$ data contain $d$-dimensional $(=4)$ attribute values, following uniform, Gaussian, and Zipf distributions, respectively, where each dimension takes 5 different values (0$\sim$5). Our codes for generating synthetic graphs are available at \textit{\url{http://www.cs.kent.edu/~wren/kt-safety/}}. 

\section{More Experimental Results}
\label{sec:more_results}

\begin{figure}[H]
\centering
\scalebox{0.3}[0.3]{\includegraphics{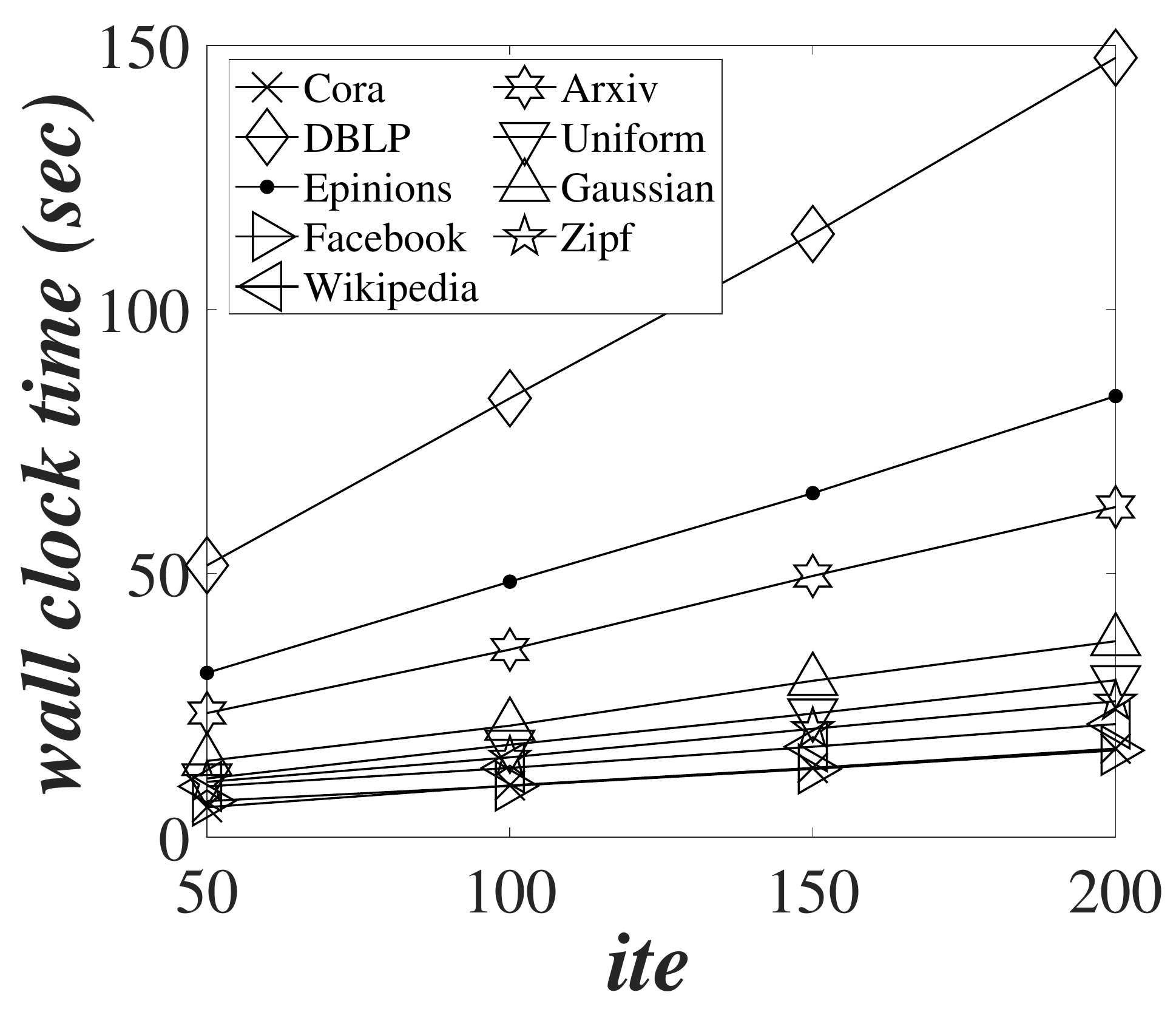}}
\caption{\small The cost evaluation of selecting a good partitioning via Algorithm \ref{alg:partition_set}.}
\label{fig:time_cost_partition_ite}
\end{figure}

\begin{figure}[H]
\centering
\scalebox{0.3}[0.3]{\includegraphics{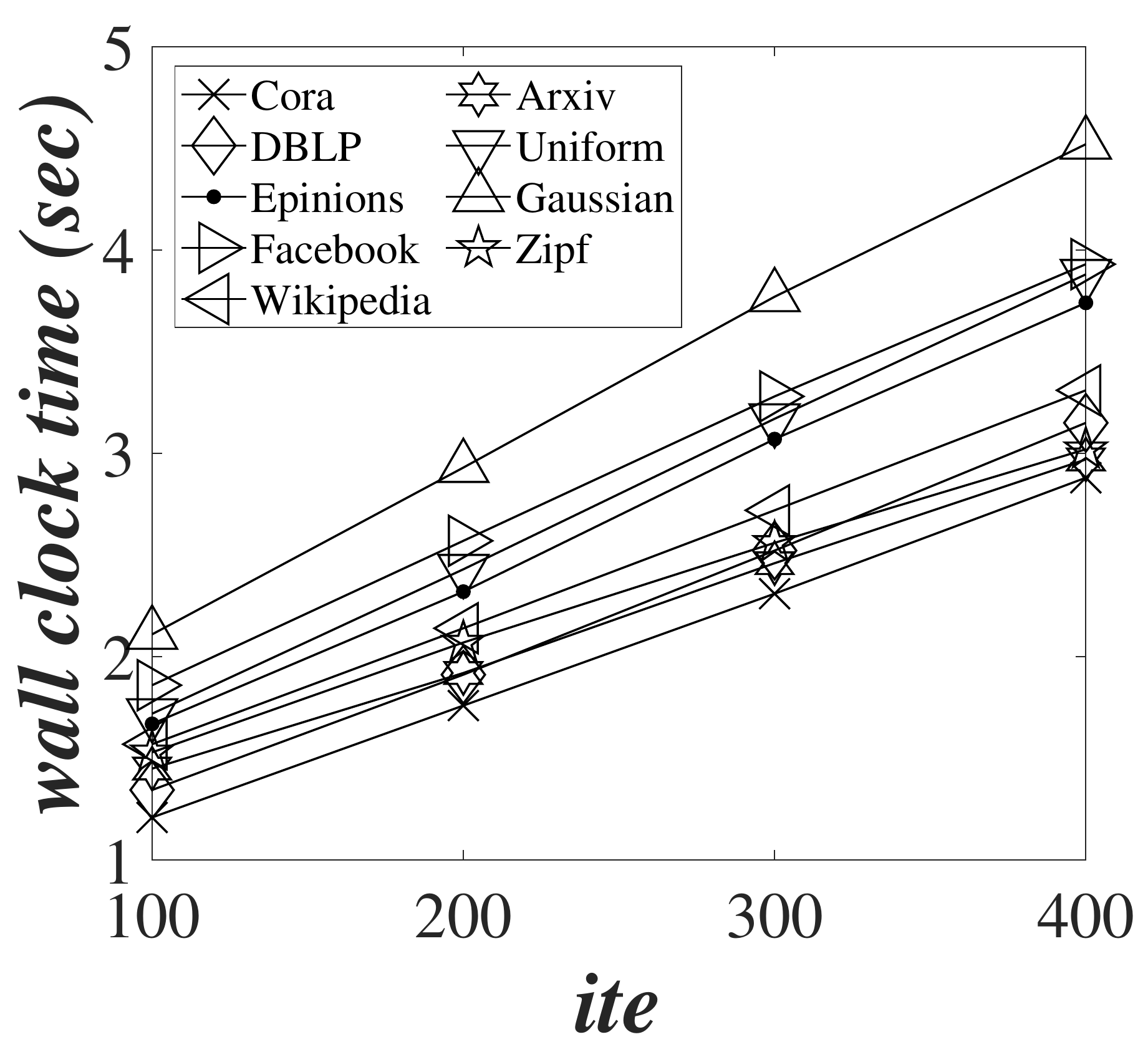}}
\caption{\small The cost evaluation of selecting a good pivot set via Algorithm \ref{alg:pivot_set}.}
\label{fig:time_cost_pivot_ite}
\end{figure}

\begin{table*}[t!]
\centering
\scriptsize
\caption{\small The additional cost of a generated $kt$-safe graph w.r.t. a near optimal $kt$-safe graph for $Cora$ ($\gamma=3000$).} \label{table:additional_cost}
\begin{tabular}{|c|c|c|}
\hline
\textbf{Approach} & \textbf{anonymization cost} & \textbf{wall clock time (sec)} \\
\hline
parition+pruning & 1031 & 5.314 \\\hline
near optimal & 993 & 17,511.2\\\hline
\end{tabular}
\end{table*}

\subsection{Verification of the Cost Model and algorithms}
\label{subsec:cost_model_eva}

\noindent {\bf Cost Model Verification for the Anonymization Cost Estimation via Eq. (\ref{eq:cost_model_for_gp}).} We verify our cost model in Eq. (\ref{eq:cost_model_for_gp}) for estimating the cost of anonymizing a graph to be $kt$-safe (Eq. (\ref{eq:eq1}) in Definition \ref{def:anony_cost}). Figure \ref{fig:cost_model_eva} illustrates the estimated anonymization cost of $partition+pruning$ on both real and synthetic data, where the retrieved sample graphs with 1,000 vertices and all parameters follow their default values in Table \ref{table:exp_parameter_setting}. We also compare our estimated anonymization cost with the actual one (via $partition+pruning$) on all data sets. From the figure, we can see that our proposed cost model can well estimate the anonymization cost of our proposed approach, which demonstrates the effectiveness of our cost model in Eq. (\ref{eq:cost_model_for_gp}).

\noindent {\bf Cost Evaluation of Selecting a Good Partitioning via Algorithm \ref{alg:partition_set}.} Figure \ref{fig:time_cost_partition_ite} shows the time cost of Algorithm \ref{alg:partition_set} to select a good partitioning set of subgraphs on all tested data sets, where we vary the iteration number $ite$ from 50 to 200. From the figure, as $ite$ increases, the CPU time increases linearly. The partitioning strategy via Algorithm \ref{alg:partition_set} is still efficient even when $ite=100$ (e.g., $<147.37$ sec for $DBLP$). Intuitively, more iterations will lead to a partitioning set with a low estimated anonymization cost. For all our experiments, we set $ite$ to 50 by default, which can achieve a good partitioning set as evaluated in our cost model verification (Figure \ref{fig:cost_model_eva}).

\noindent {\bf Cost Evaluation of Selecting a Good Pivot Set via Algorithm \ref{alg:pivot_set}.} Figure \ref{fig:time_cost_pivot_ite} reports the CPU time of selecting a set of 10 pivots on real/synthetic data via Algorithm \ref{alg:pivot_set}, where we vary the iteration number from 100 to 400 and set the size of sample graphs to 1,000. From the figure, we can see that the time cost linearly increases as $ite$ increases. When $ite=400$, Algorithm \ref{alg:pivot_set} is still very efficient on all data sets (i.e., $2.88$ sec $\sim 4.52$ sec). In this paper, we set $ite$ to 100 by default. As shown in Figure \ref{fig:exper:time_cost_vs_dataset}, $pruning$ outperforms $none$ on all data sets, which demonstrates the effectiveness of the pivot set selection strategy via Algorithm \ref{alg:pivot_set}.

\noindent {\bf Additional Cost Evaluation of the Generated $kt$-Safe Graph.} Table \ref{table:additional_cost} demonstrates the anonymization and time costs of generating a $kt$-safe graph on $Cora$ via our $partition+pruning$ approach and that of producing a near optimal $kt$-safe graph. Note that, there are $|V(G)|!$ execution orders for producing a $kt$-safe graph, which leads to an unbearable time cost to obtain the optimal anonymized graph. Instead, we randomly sample 3,000 (i.e., $\gamma$) different execution orders for producing a $kt$-safe graph for $Cora$, and select the one with the lowest anonymization cost as the near optimal $kt$-safe graph. From the table, we can see that the anonymized graph via our approach has a cost (i.e., 1031) similar to the near optimal one (i.e., 993). Moreover, our $partition+pruning$ approach needs less time cost than the one for generating the near optimal anonymized graph, where we estimate the near optimal time by summing up the total time of the 3,000 orders.

\begin{figure*}[ht]
\centering
\subfigure[][{\small Cora}]{
\scalebox{0.15}[0.15]{\includegraphics{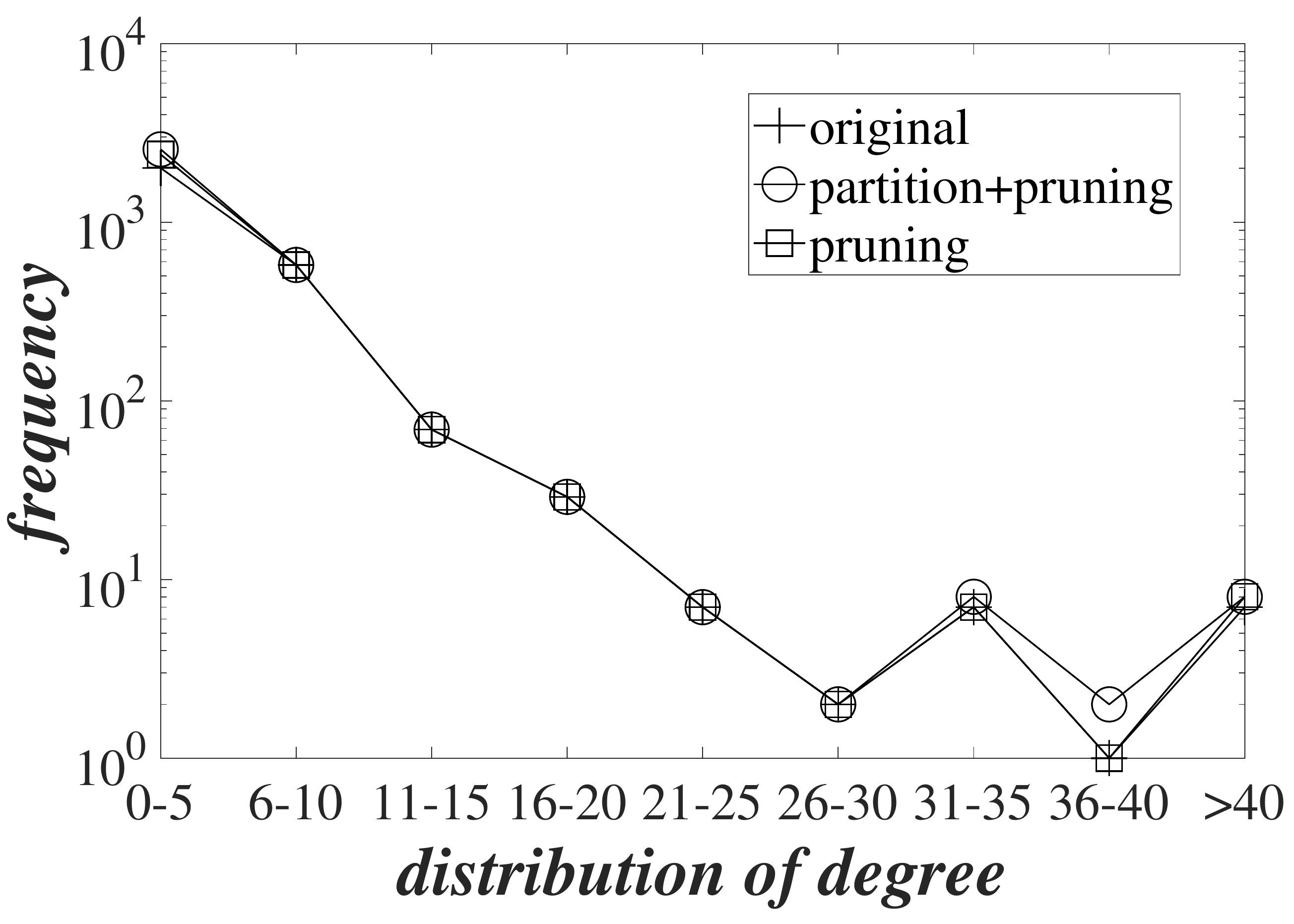}}
\label{subfig:Cora_vs_degree}
}
\subfigure[][{\small DBLP}]{
\scalebox{0.15}[0.15]{\includegraphics{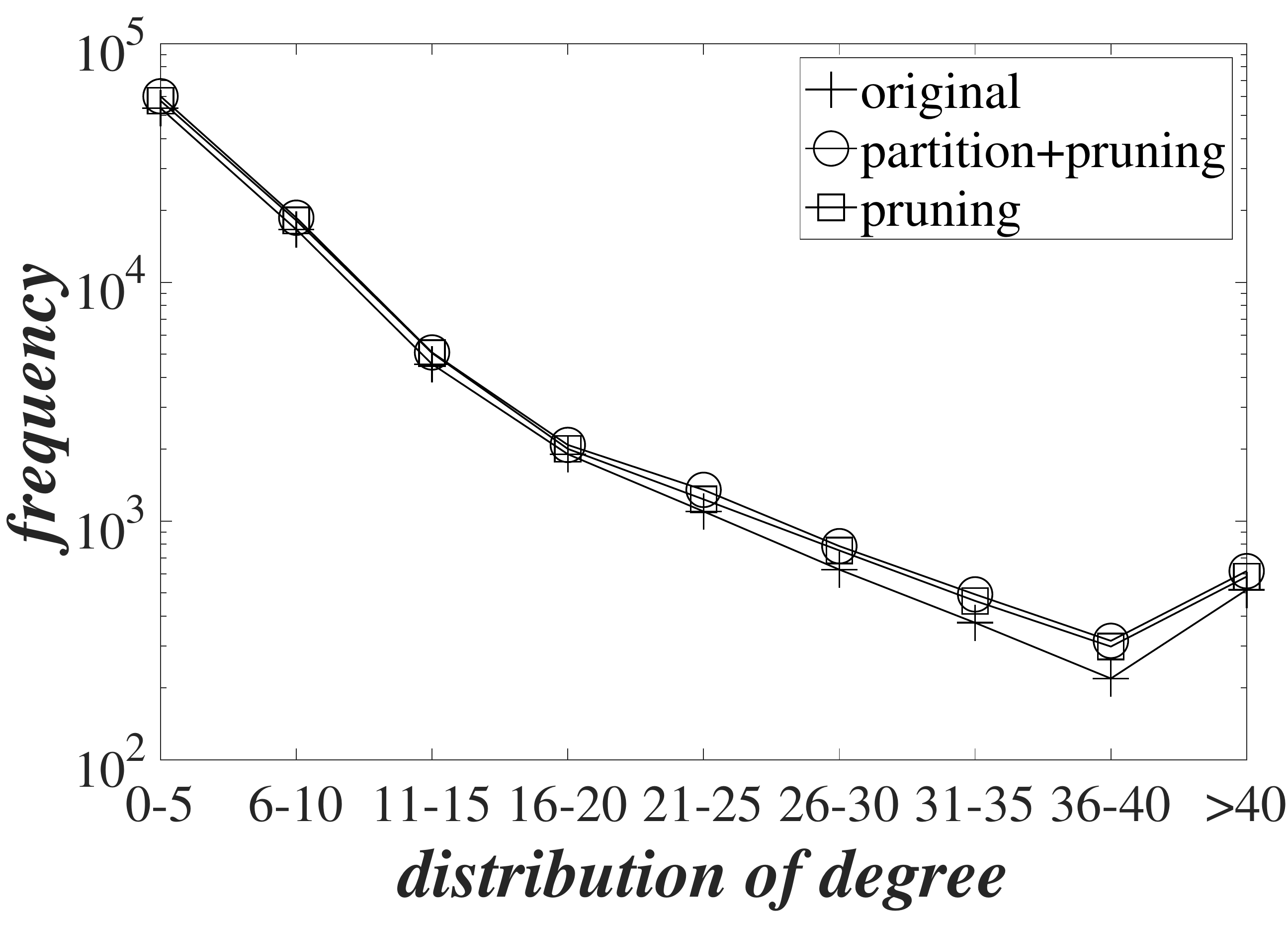}}
\label{subfig:DBLP_vs_degree}
}
\subfigure[][{\small Facebook}]{
\scalebox{0.15}[0.15]{\includegraphics{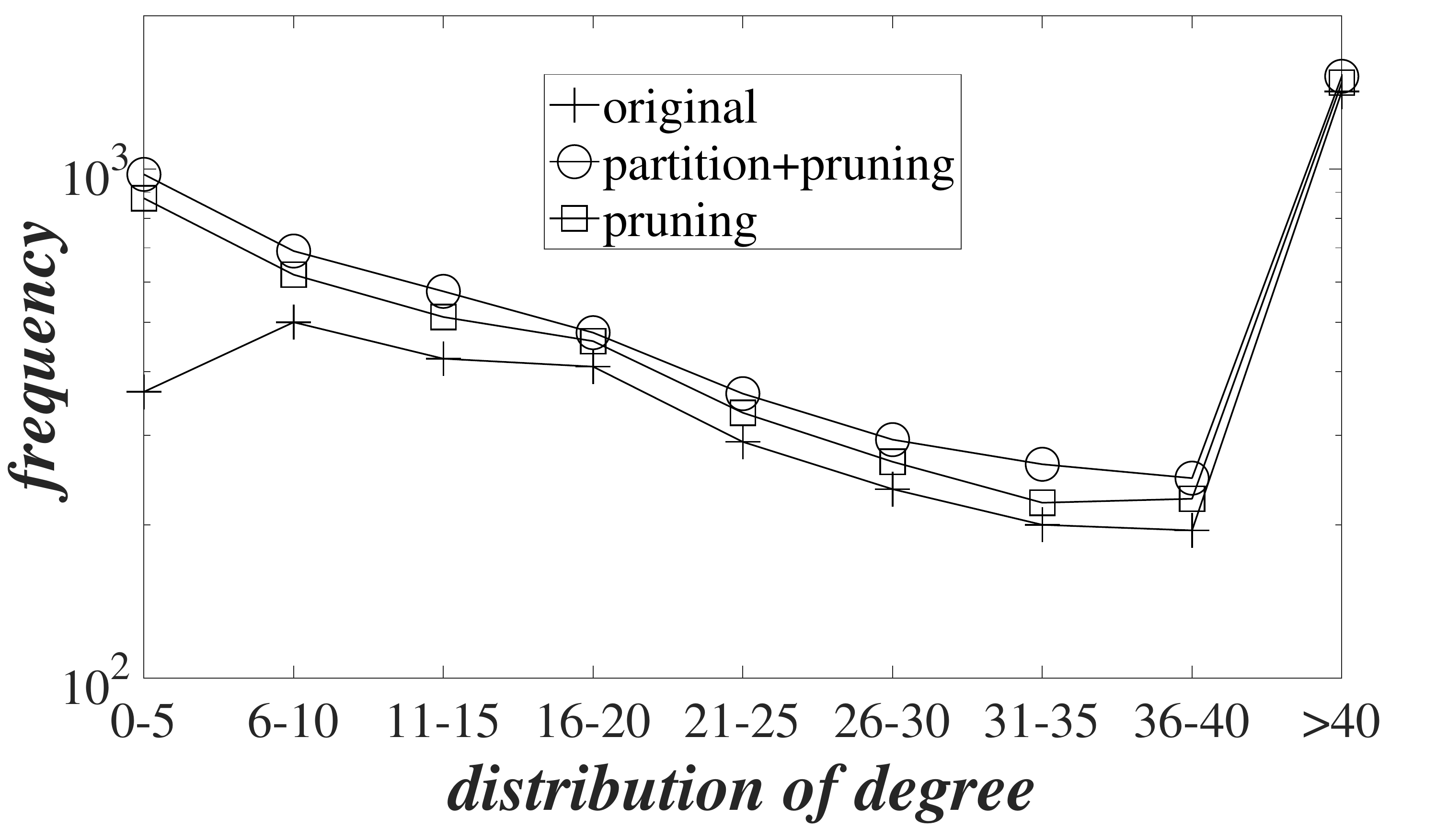}}
\label{subfig:Facebook_vs_degree}
}
\subfigure[][{\small Wikipedia}]{\hspace{-2ex}                
\scalebox{0.15}[0.15]{\includegraphics{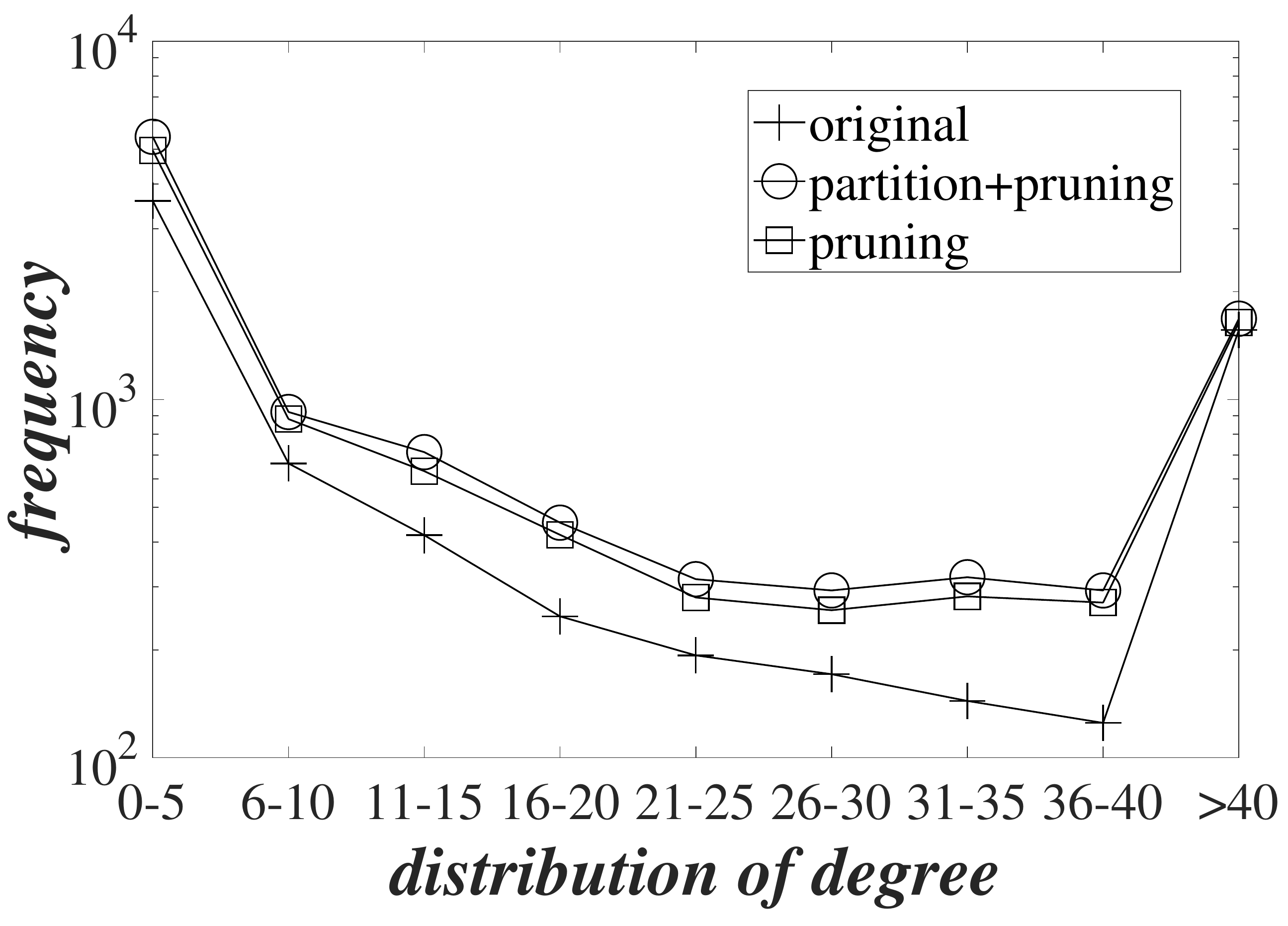}}
\label{subfig:Wikipedia_vs_degree}
}\\
\subfigure[][{\small Arxiv}]{
\scalebox{0.15}[0.15]{\includegraphics{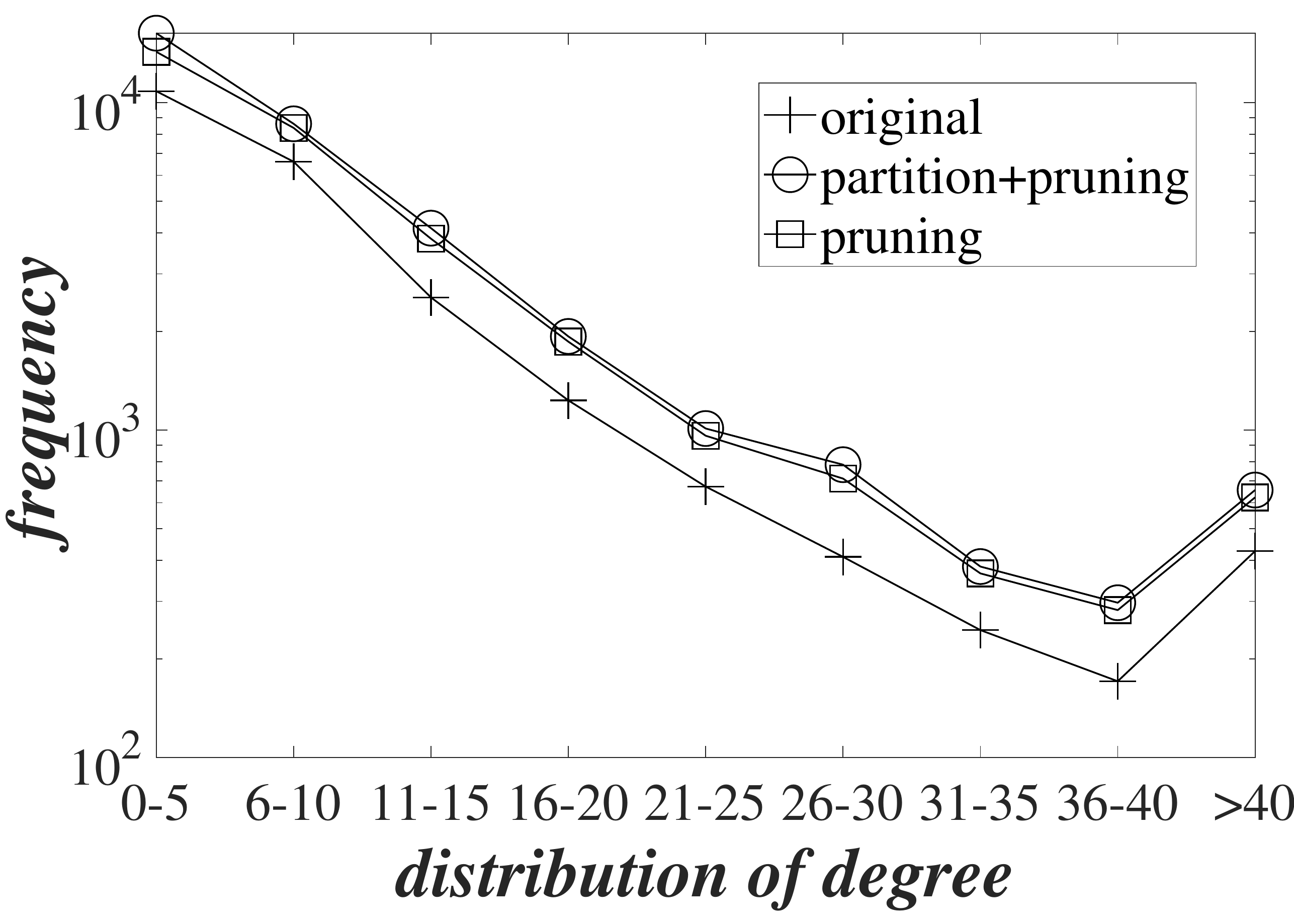}}
\label{subfig:Arxiv_vs_degree}
}
\subfigure[][{\small Uniform}]{
\scalebox{0.15}[0.15]{\includegraphics{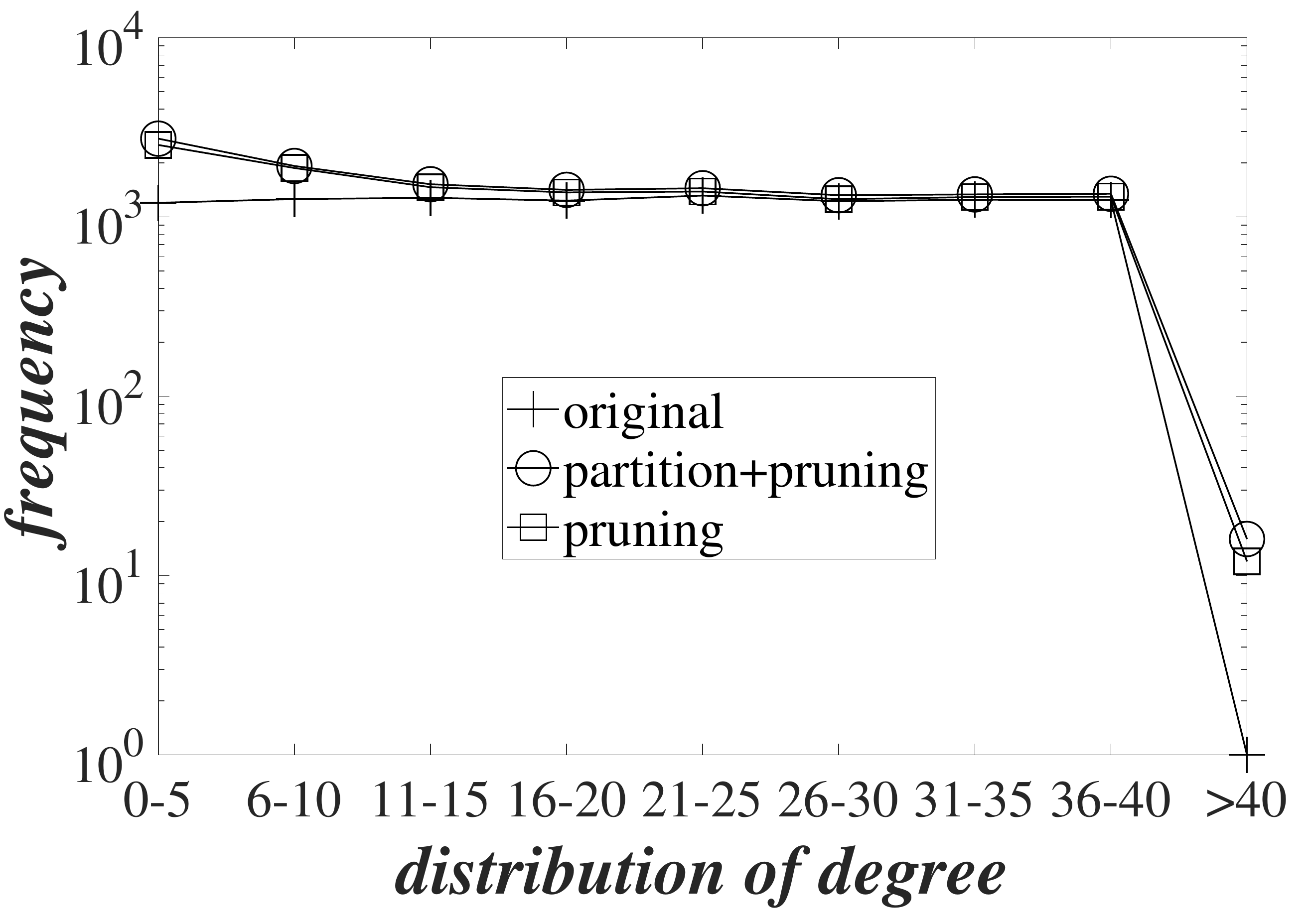}}
\label{subfig:uniform_vs_degree}
}
\subfigure[][{\small Gaussian}]{
\scalebox{0.15}[0.15]{\includegraphics{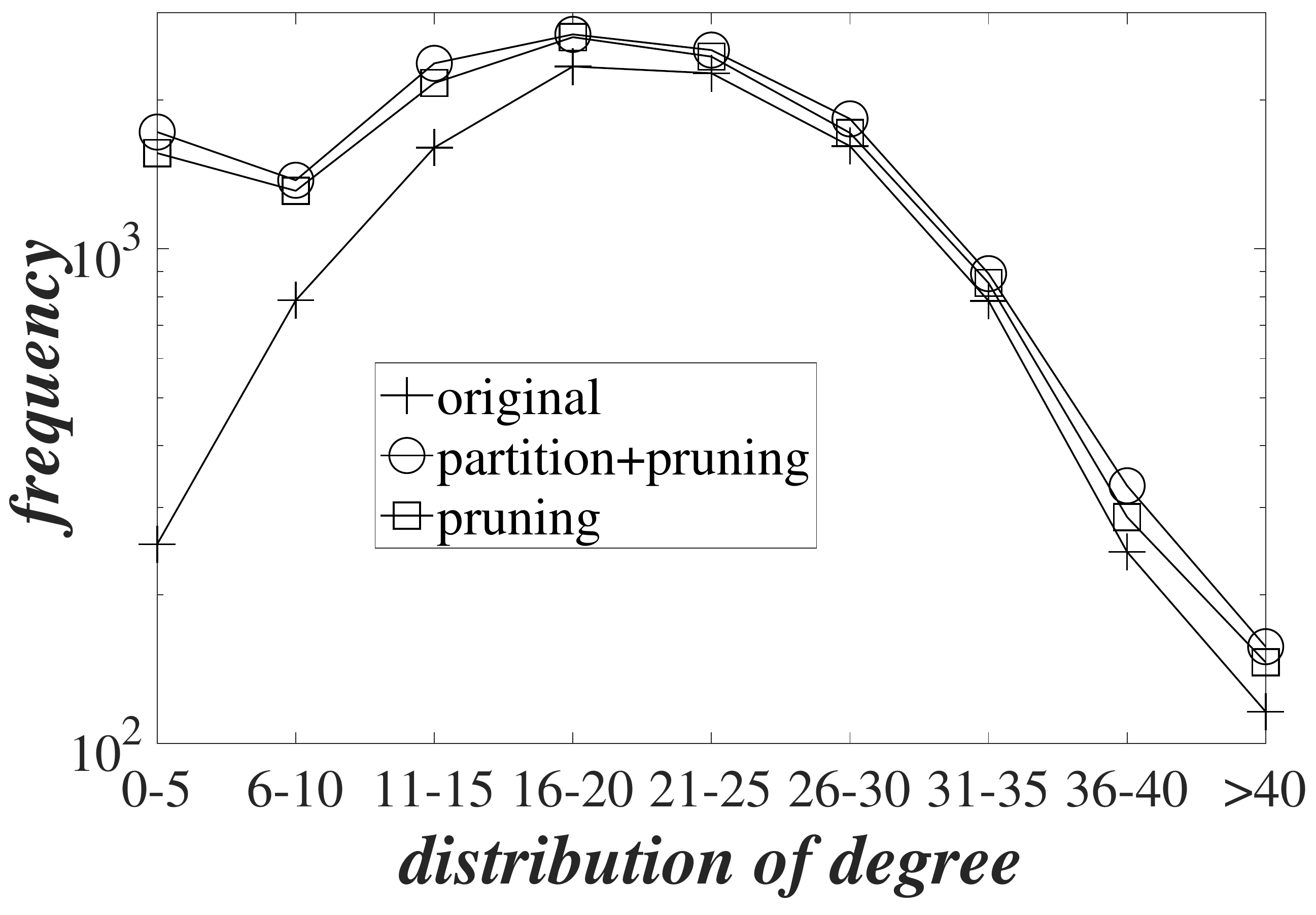}}
\label{subfig:gaussian_vs_degree}
}
\subfigure[][{\small Zipf}]{
\scalebox{0.15}[0.15]{\includegraphics{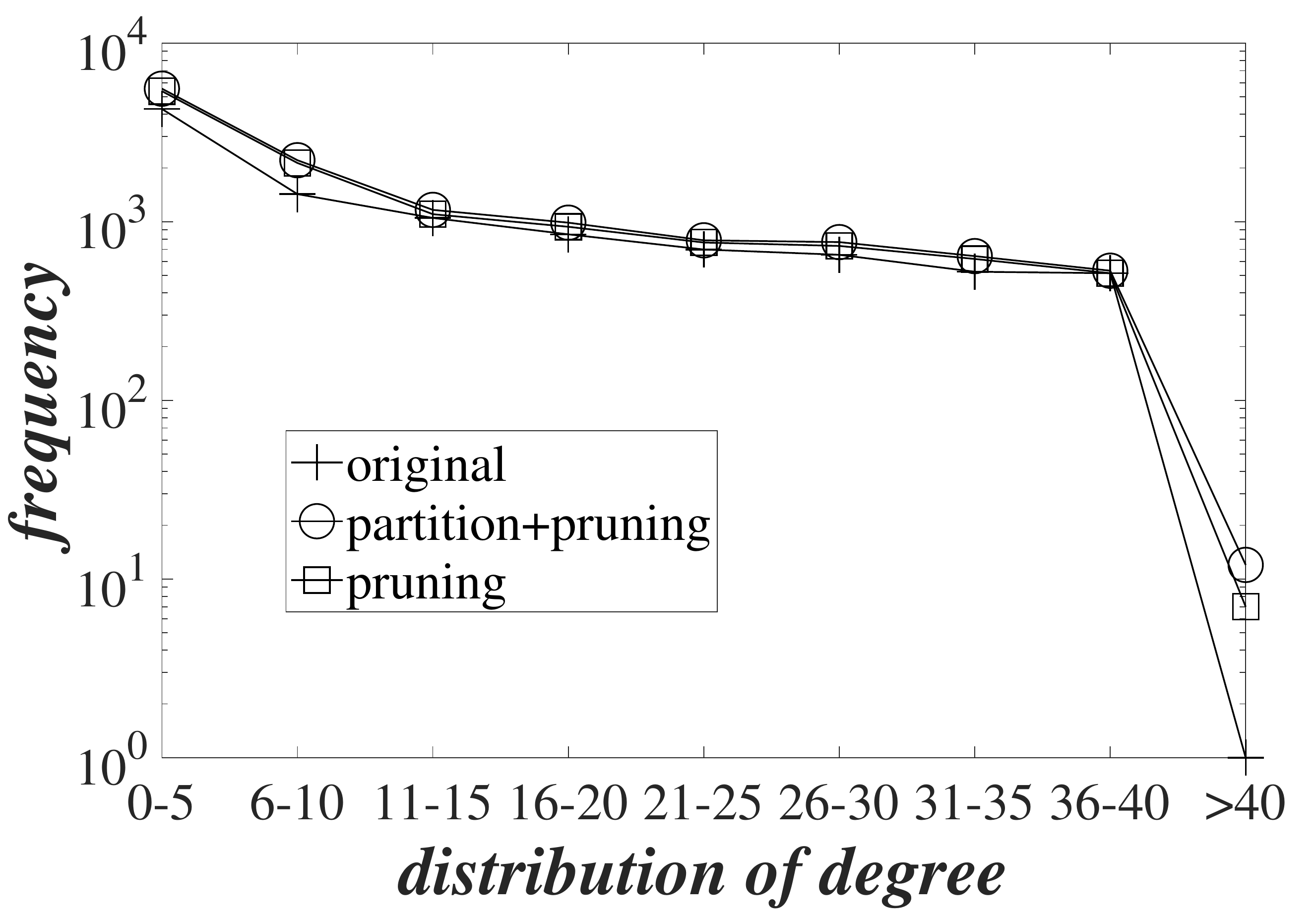}}
\label{subfig:zipf_vs_degree}
}
\caption{\small Distributions of Degrees.} 
\label{exper:more_distribution_vs_degree}
\end{figure*}

\begin{figure*}[ht]
\centering
\subfigure[][{\small Cora}]{
\scalebox{0.18}[0.18]{\includegraphics{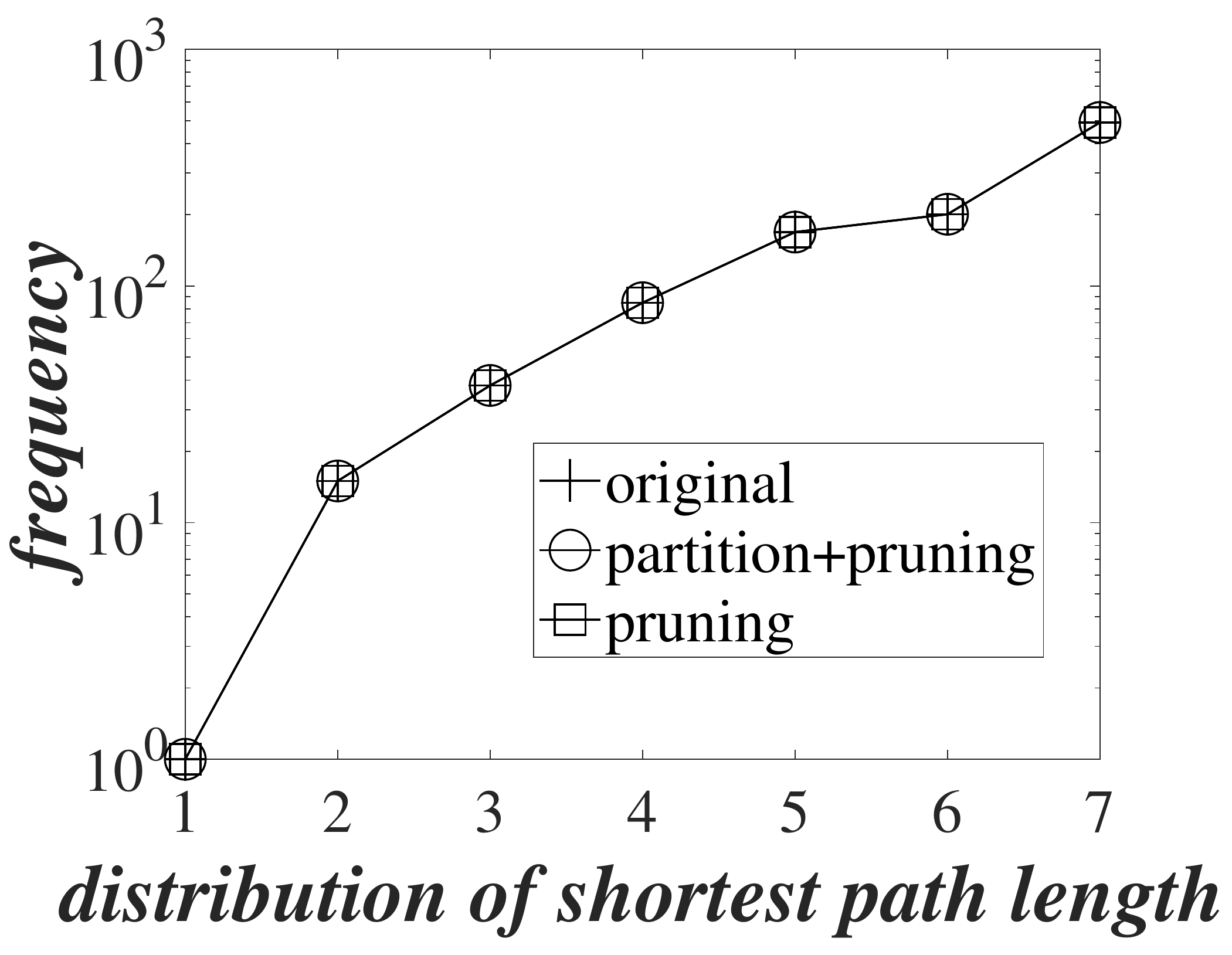}}
\label{subfig:Cora_vs_spl}
}
\subfigure[][{\small DBLP}]{
\scalebox{0.18}[0.18]{\includegraphics{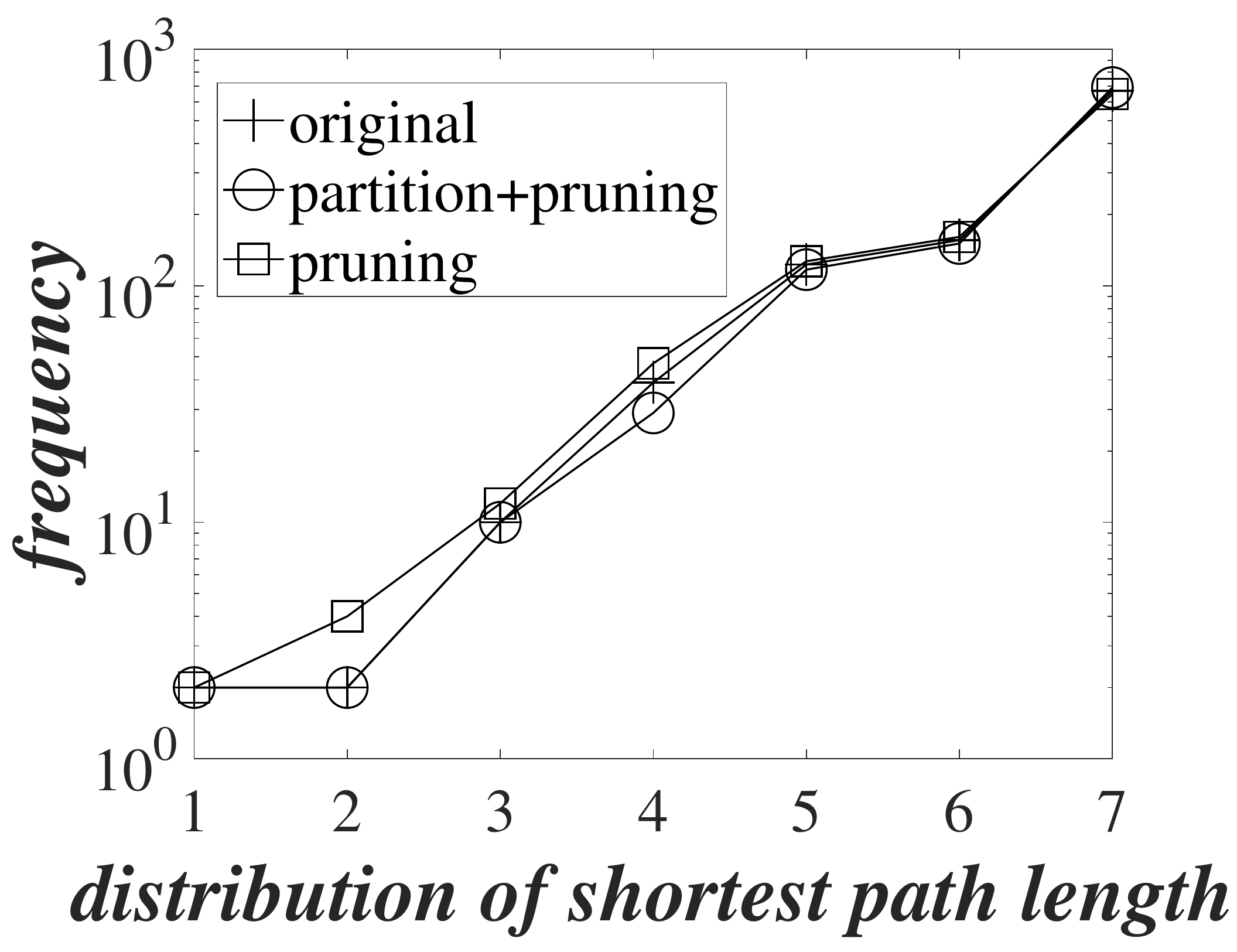}}
\label{subfig:DBLP_vs_spl}
}
\subfigure[][{\small Facebook}]{
\scalebox{0.18}[0.18]{\includegraphics{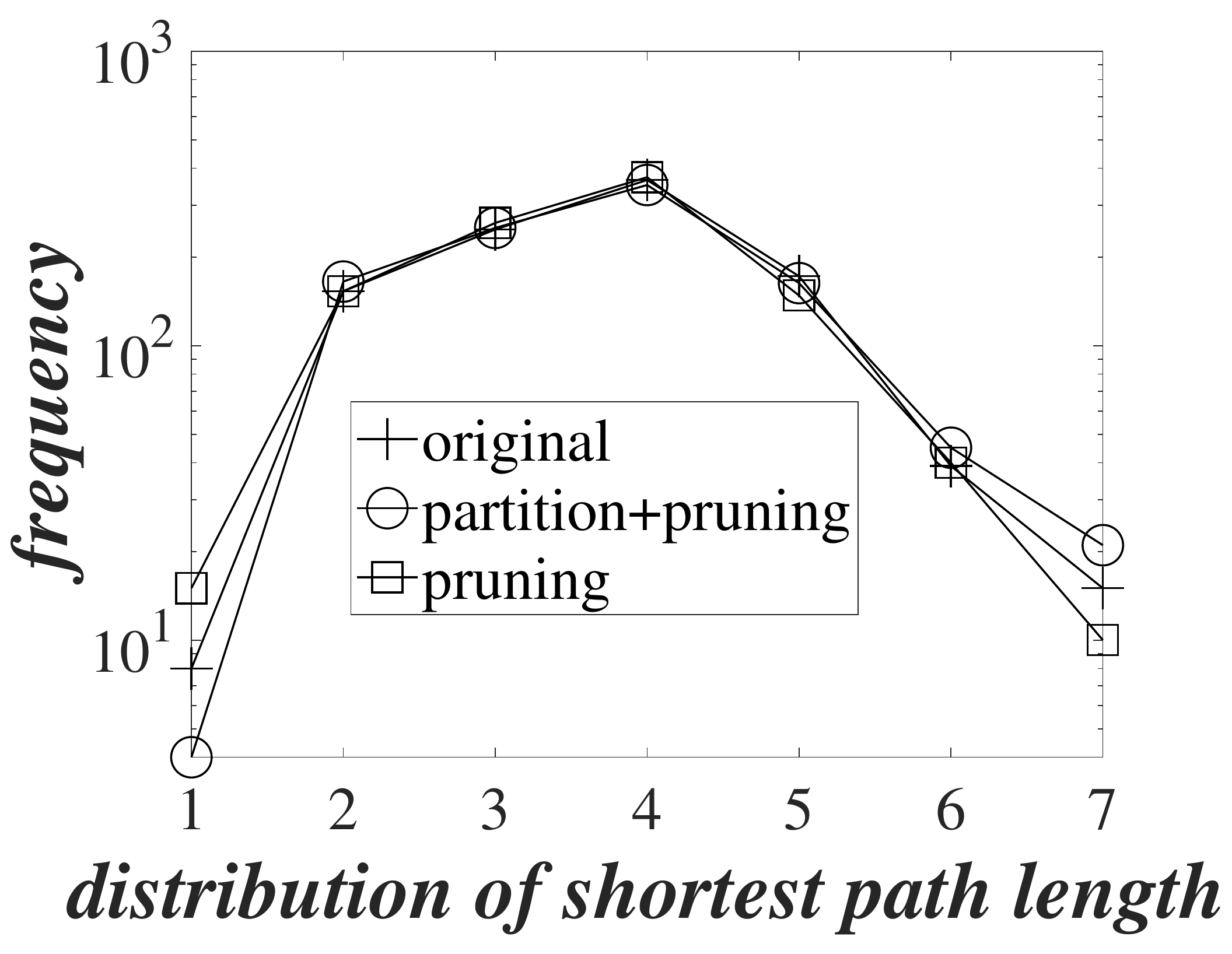}}
\label{subfig:Facebook_vs_spl}
}
\subfigure[][{\small Wikipedia}]{
\scalebox{0.18}[0.18]{\includegraphics{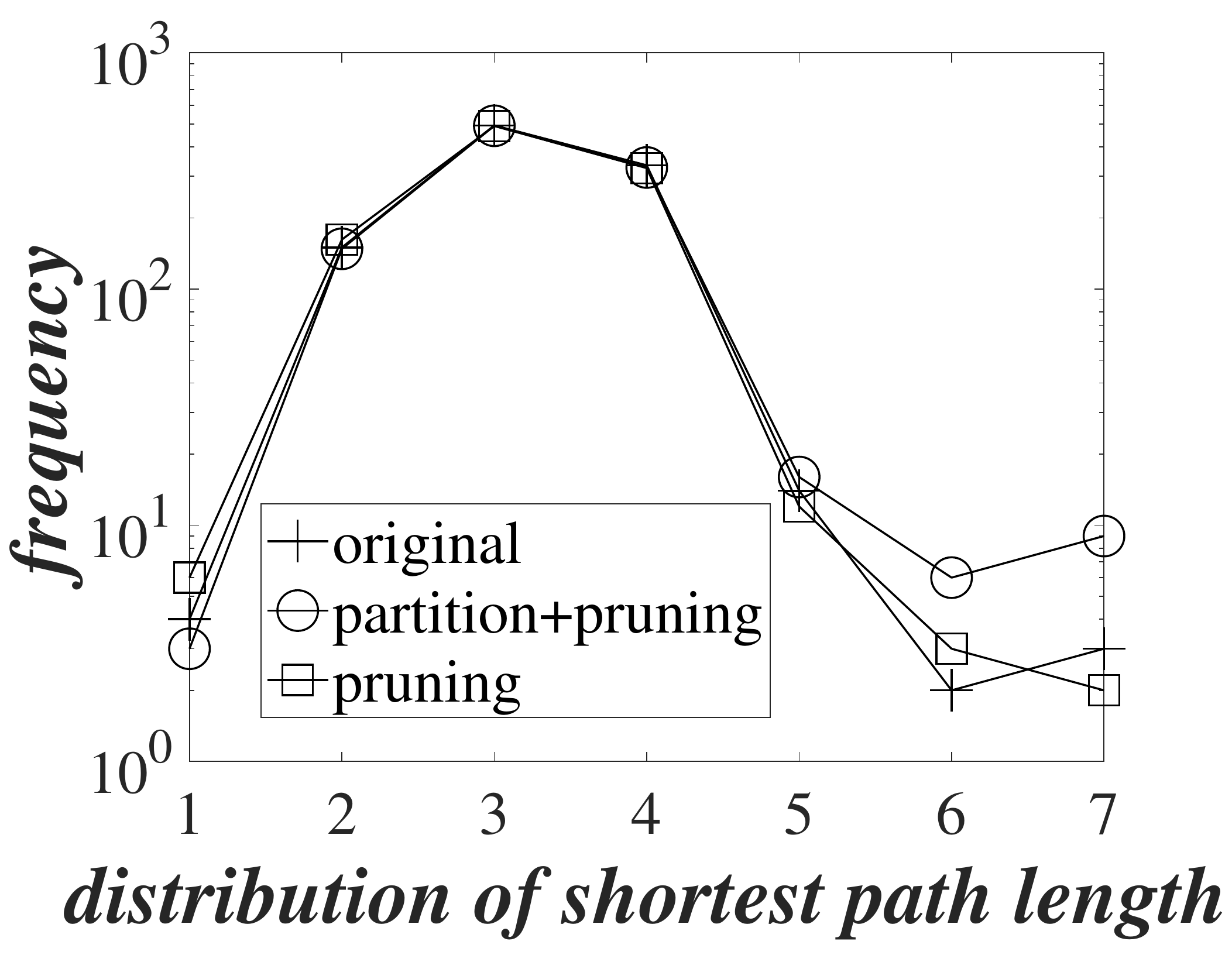}}
\label{subfig:Wikipedia_vs_spl}
}\\
\subfigure[][{\small Arxiv}]{
\scalebox{0.18}[0.18]{\includegraphics{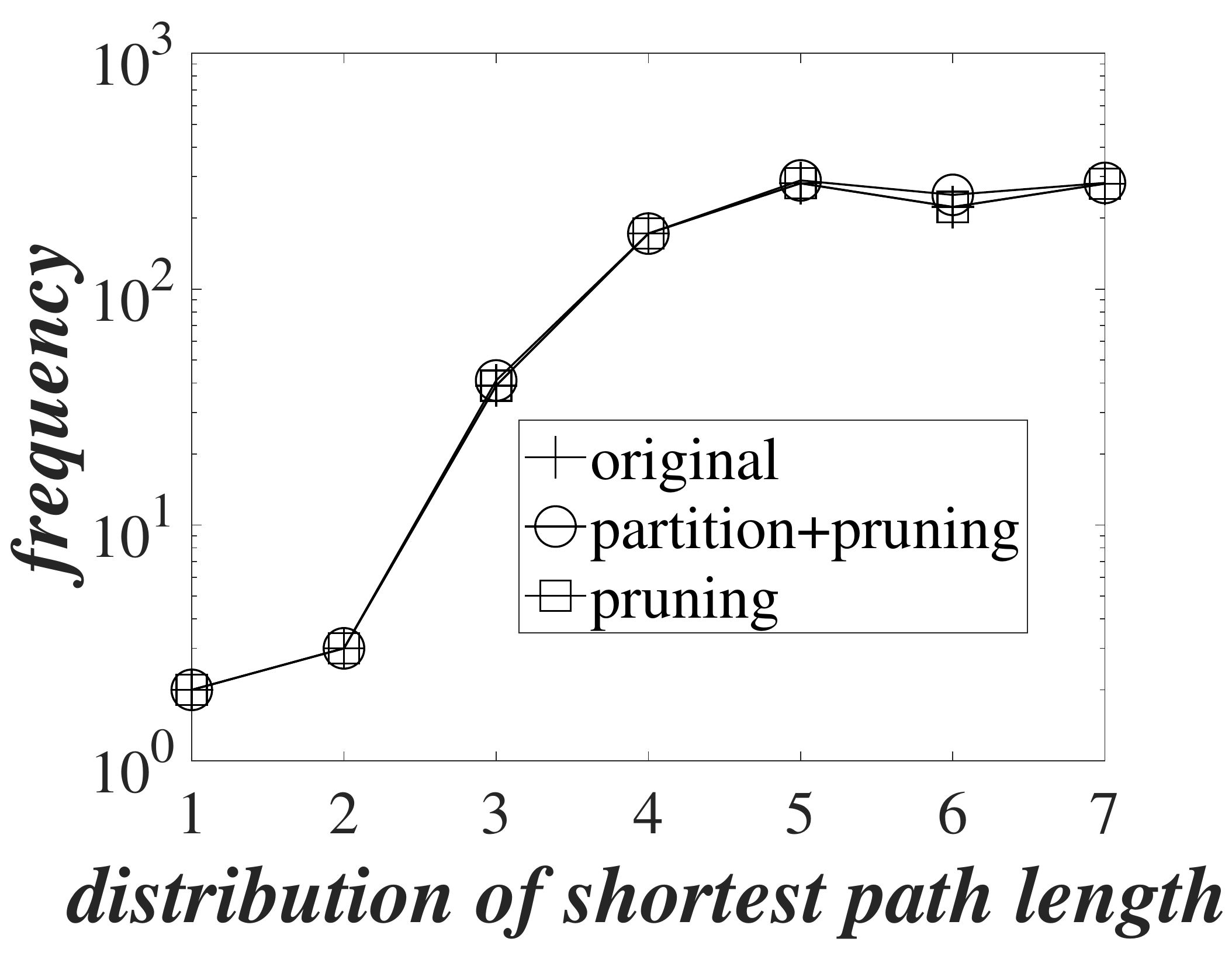}}
\label{subfig:Arxiv_vs_spl}
}
\subfigure[][{\small Uniform}]{
\scalebox{0.18}[0.18]{\includegraphics{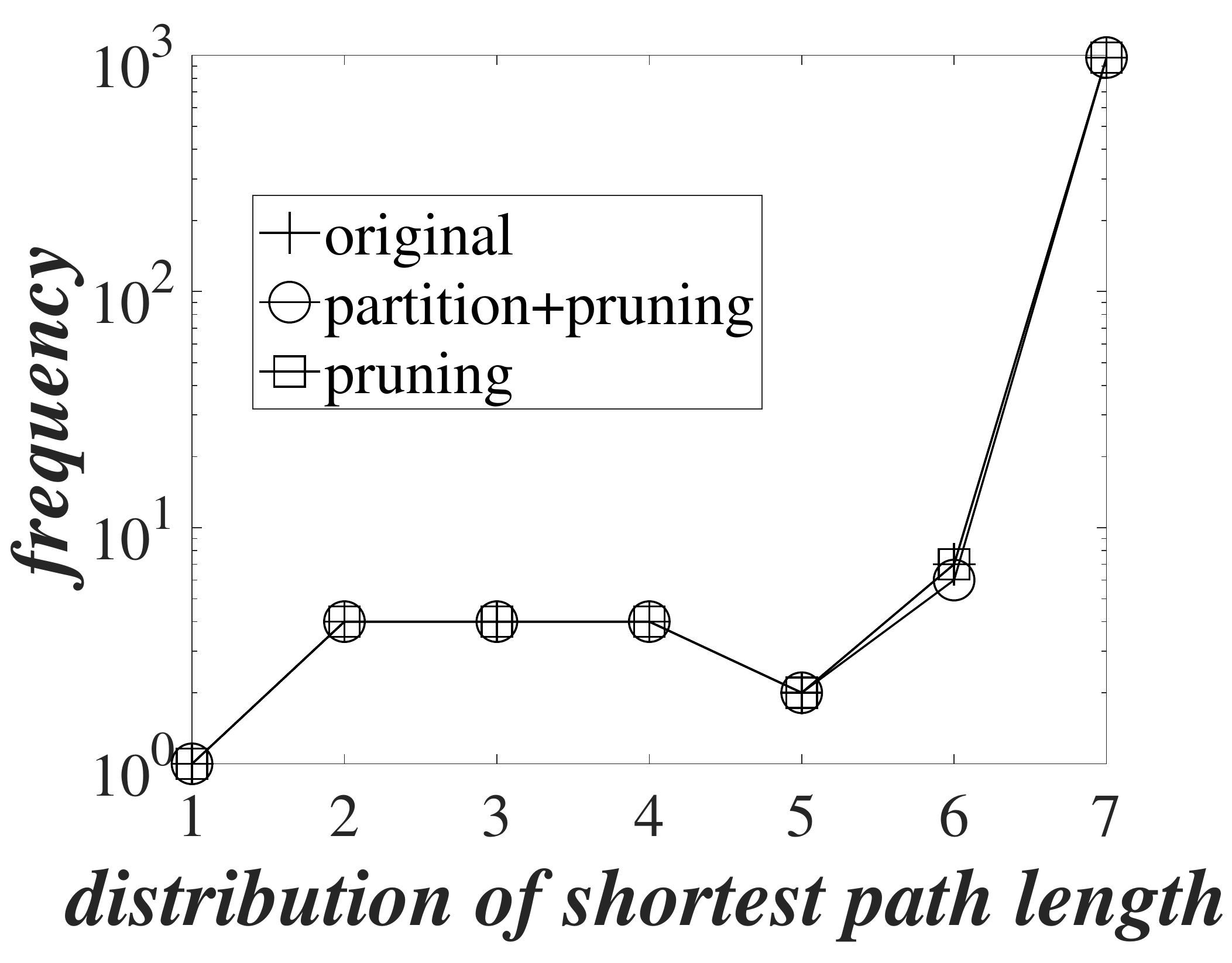}}
\label{subfig:uniform_vs_spl}
}
\subfigure[][{\small Gaussian}]{
\scalebox{0.18}[0.18]{\includegraphics{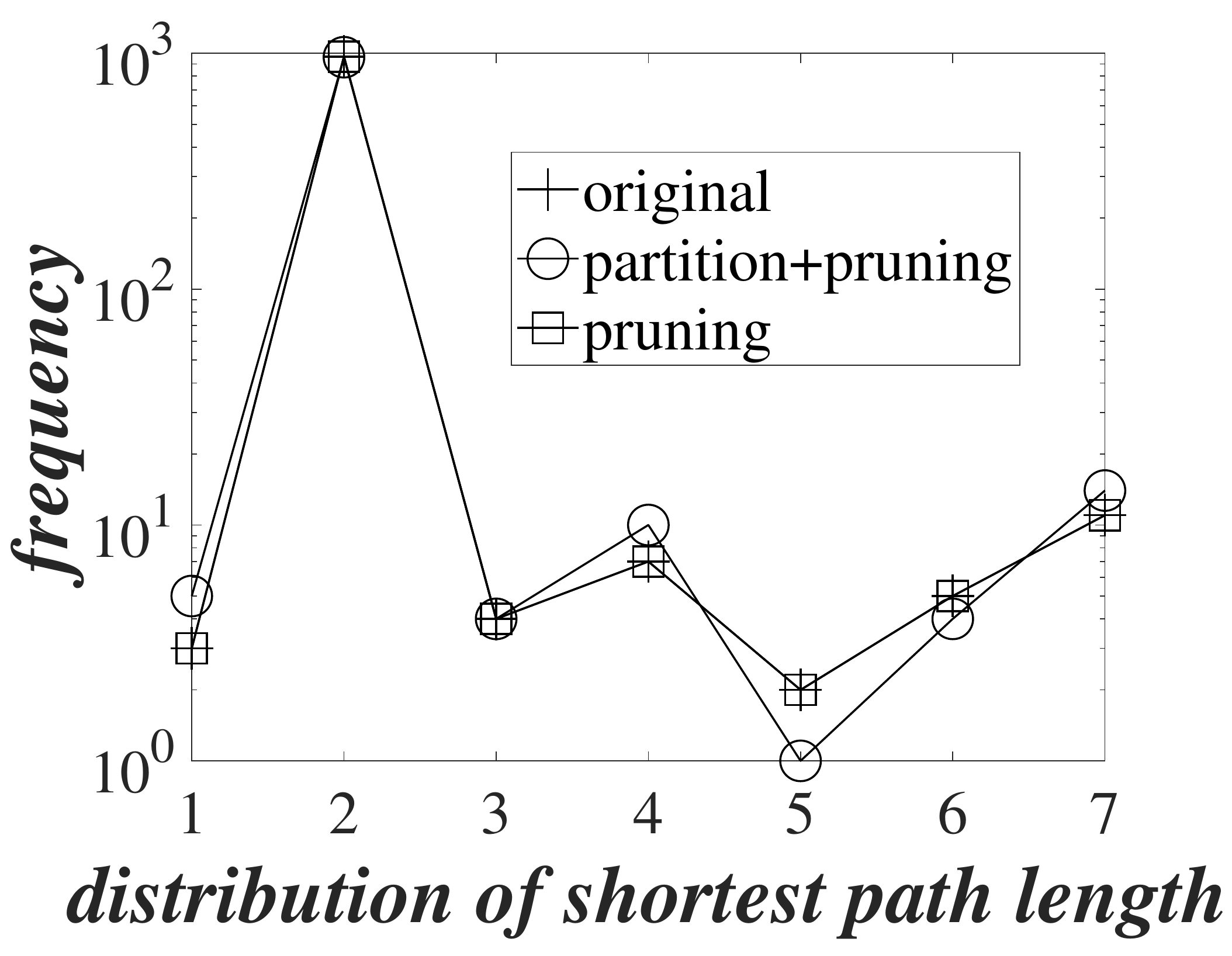}}
\label{subfig:gaussian_vs_spl}
}
\subfigure[][{\small Zipf}]{
\scalebox{0.18}[0.18]{\includegraphics{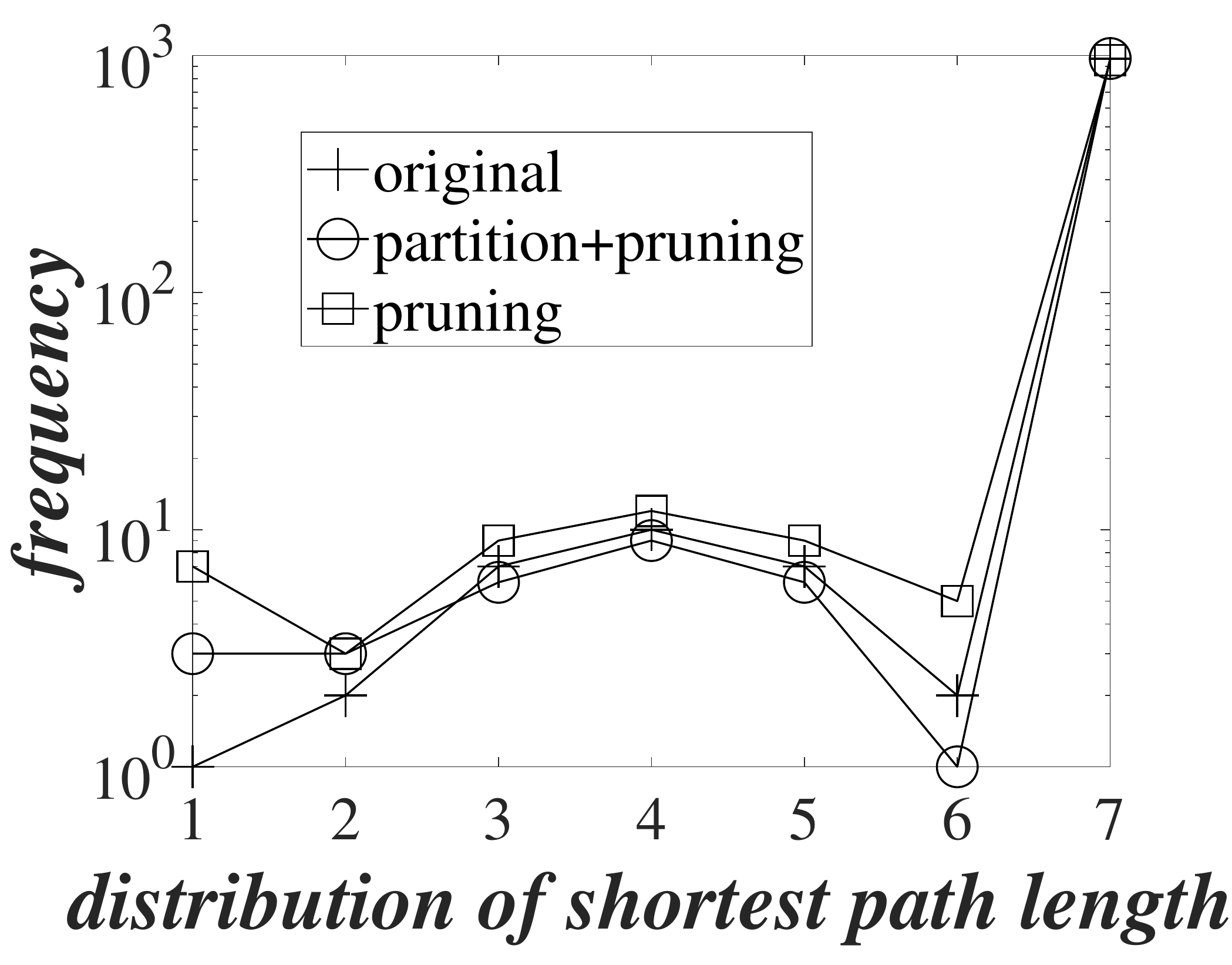}}
\label{subfig:zipf_vs_spl}
}
\caption{\small Distributions of Shortest Path Lengths.} 
\label{exper:more_distribution_vs_shortest_path} 
\end{figure*} 

\begin{figure*}[ht]
\centering
\subfigure[][{\small Cora}]{\hspace{-1.5ex}                   
\scalebox{0.2}[0.2]{\includegraphics{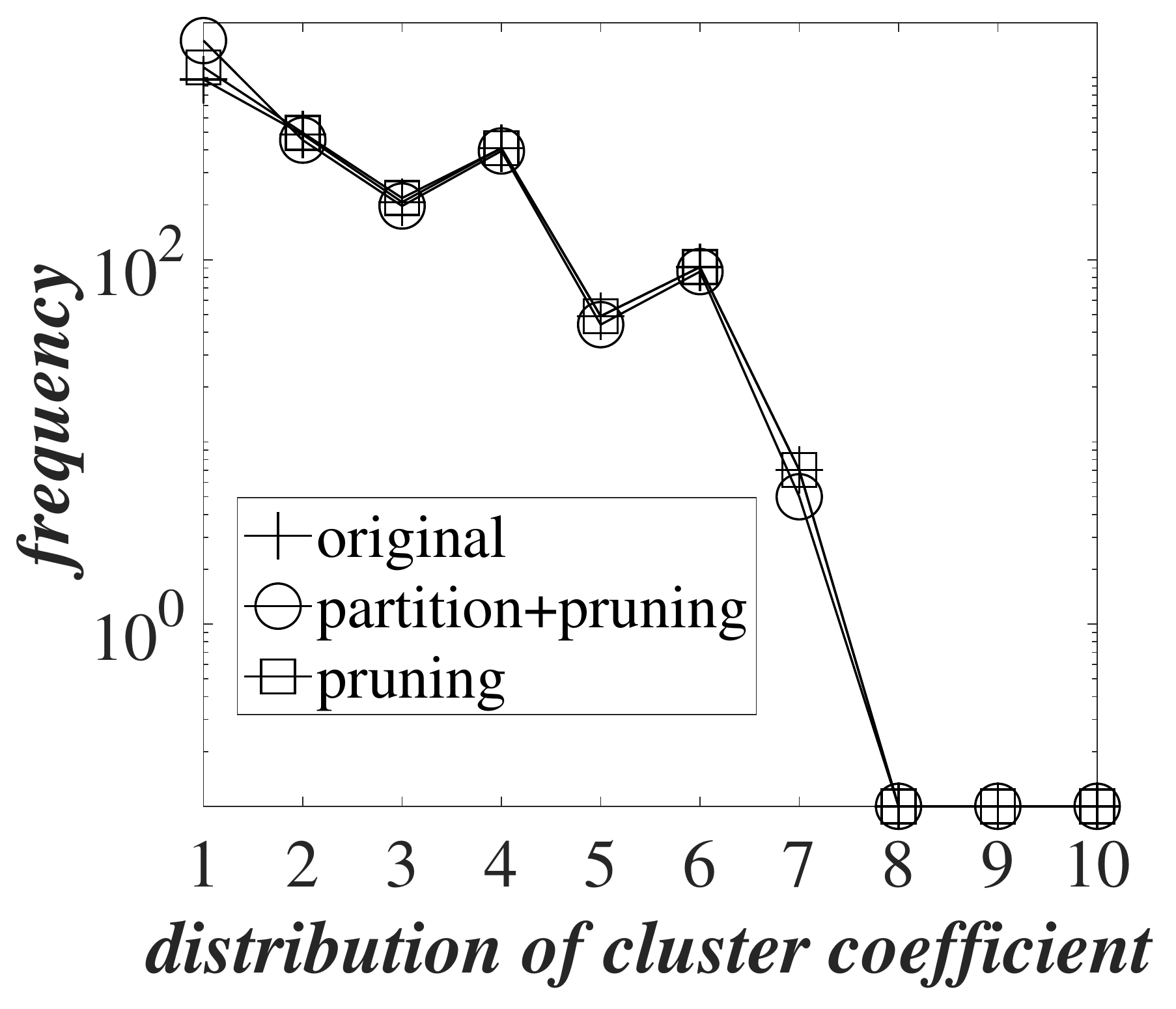}}
\label{subfig:Cora_vs_clc}
}
\subfigure[][{\small DBLP}]{\hspace{-1.5ex}                   
\scalebox{0.2}[0.2]{\includegraphics{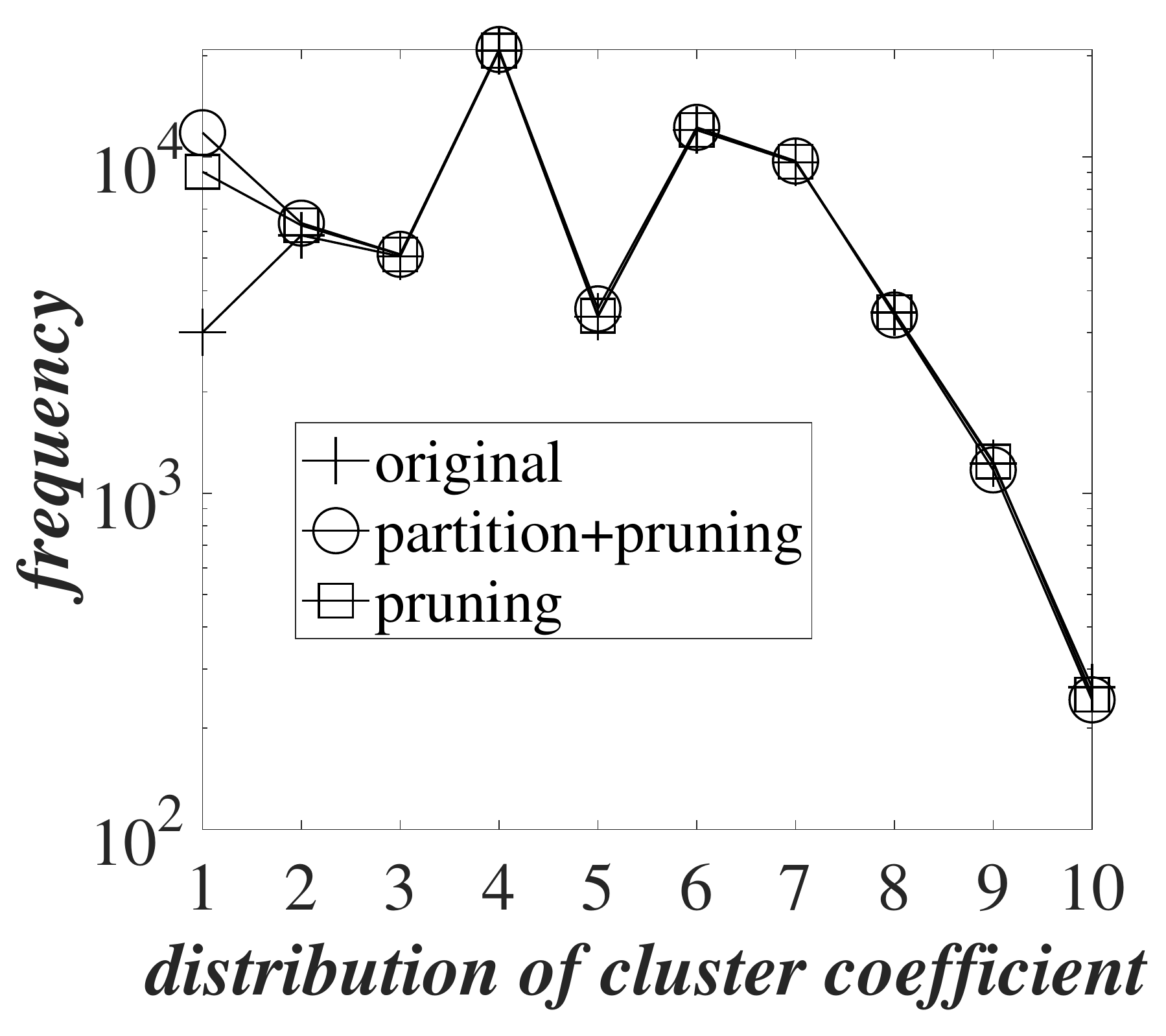}}
\label{subfig:DBLP_vs_clc}
}
\subfigure[][{\small Epinions}]{\hspace{-1.5ex}                   
\scalebox{0.2}[0.2]{\includegraphics{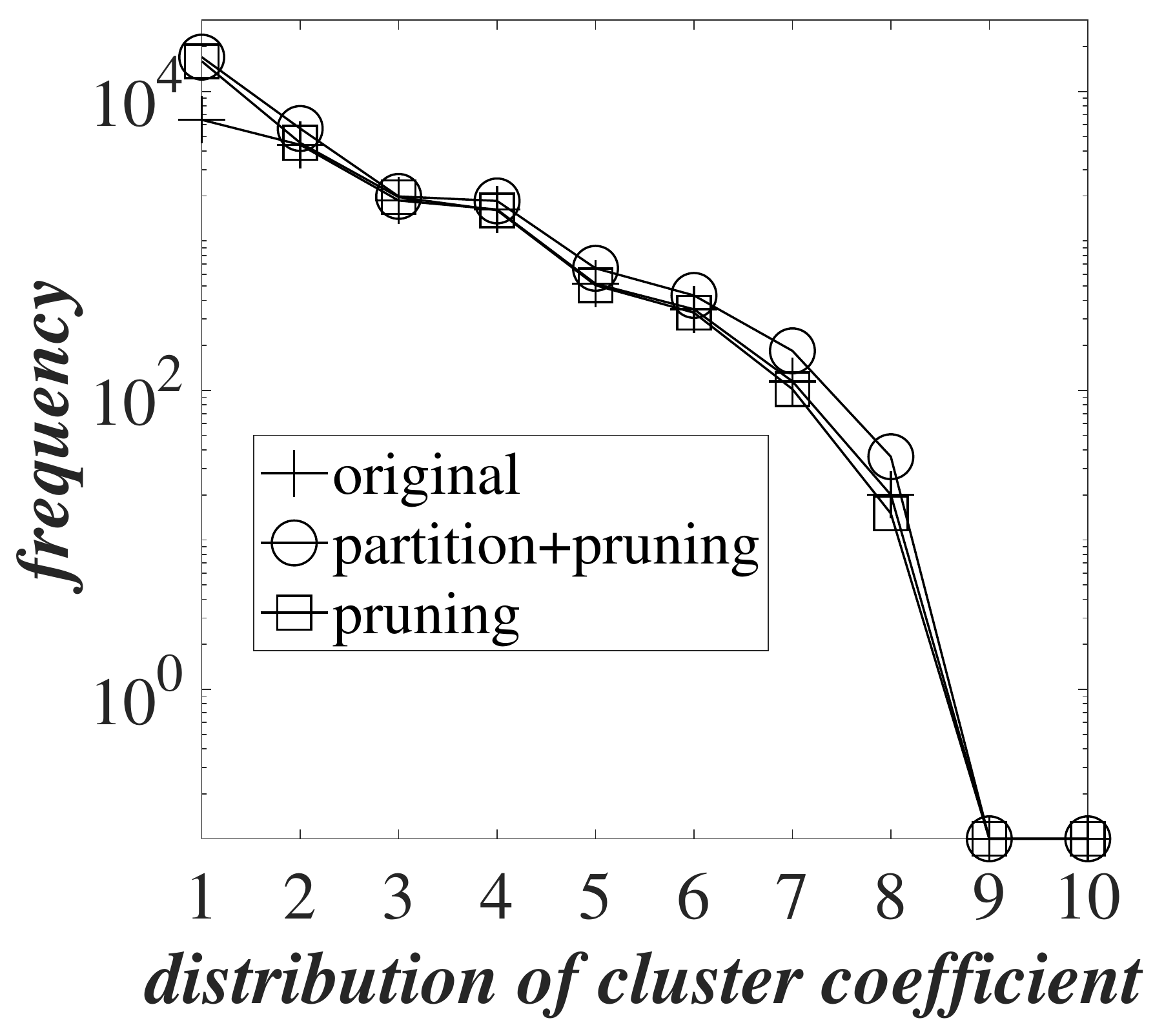}}
\label{subfig:Epinions_vs_clc}
}
\subfigure[][{\small Facebook}]{\hspace{-1.5ex}  
\scalebox{0.2}[0.2]{\includegraphics{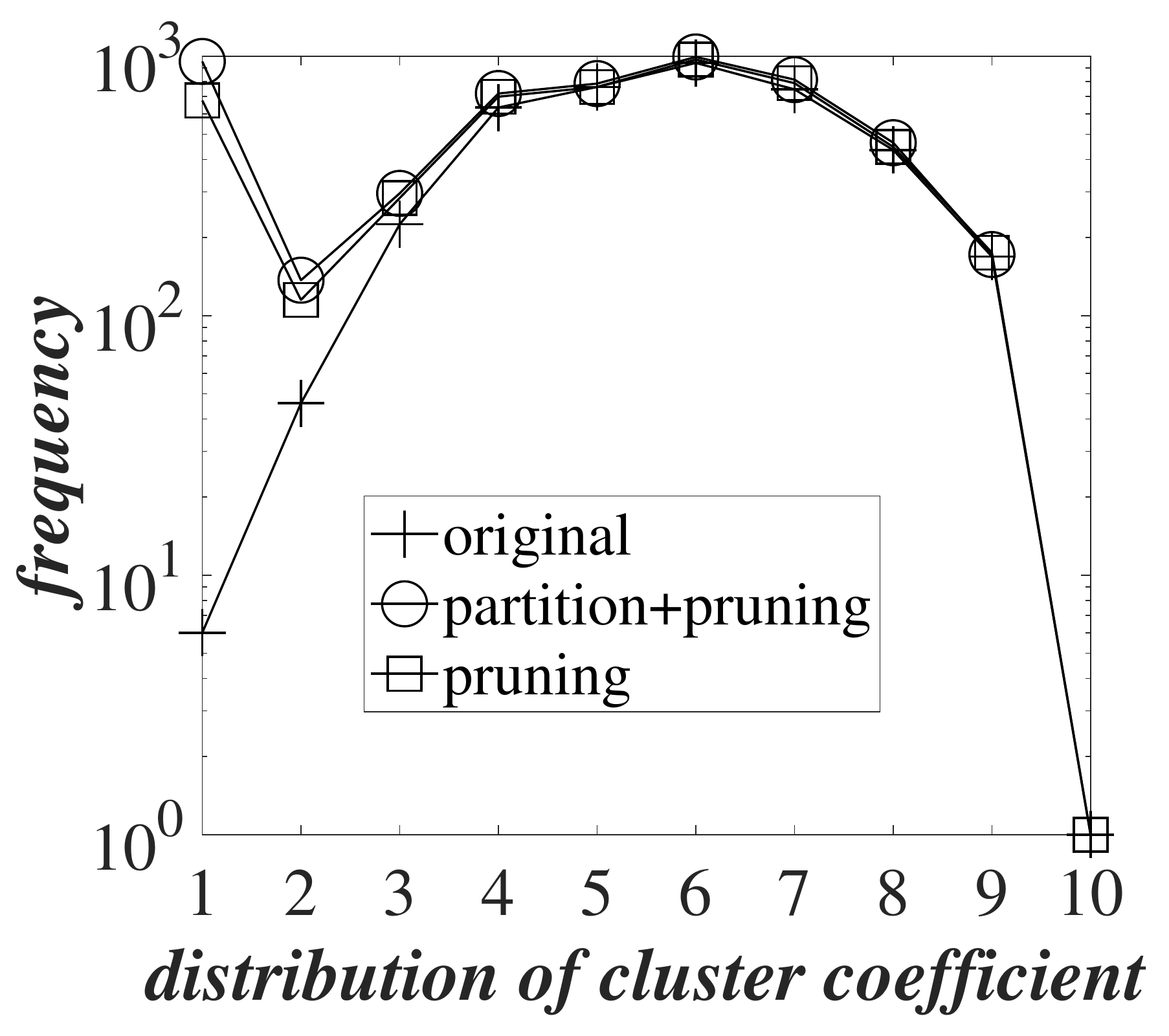}}
\label{subfig:Facebook_vs_clc}
}
\subfigure[][{\small Wikipedia}]{\hspace{-2ex}                
\scalebox{0.2}[0.2]{\includegraphics{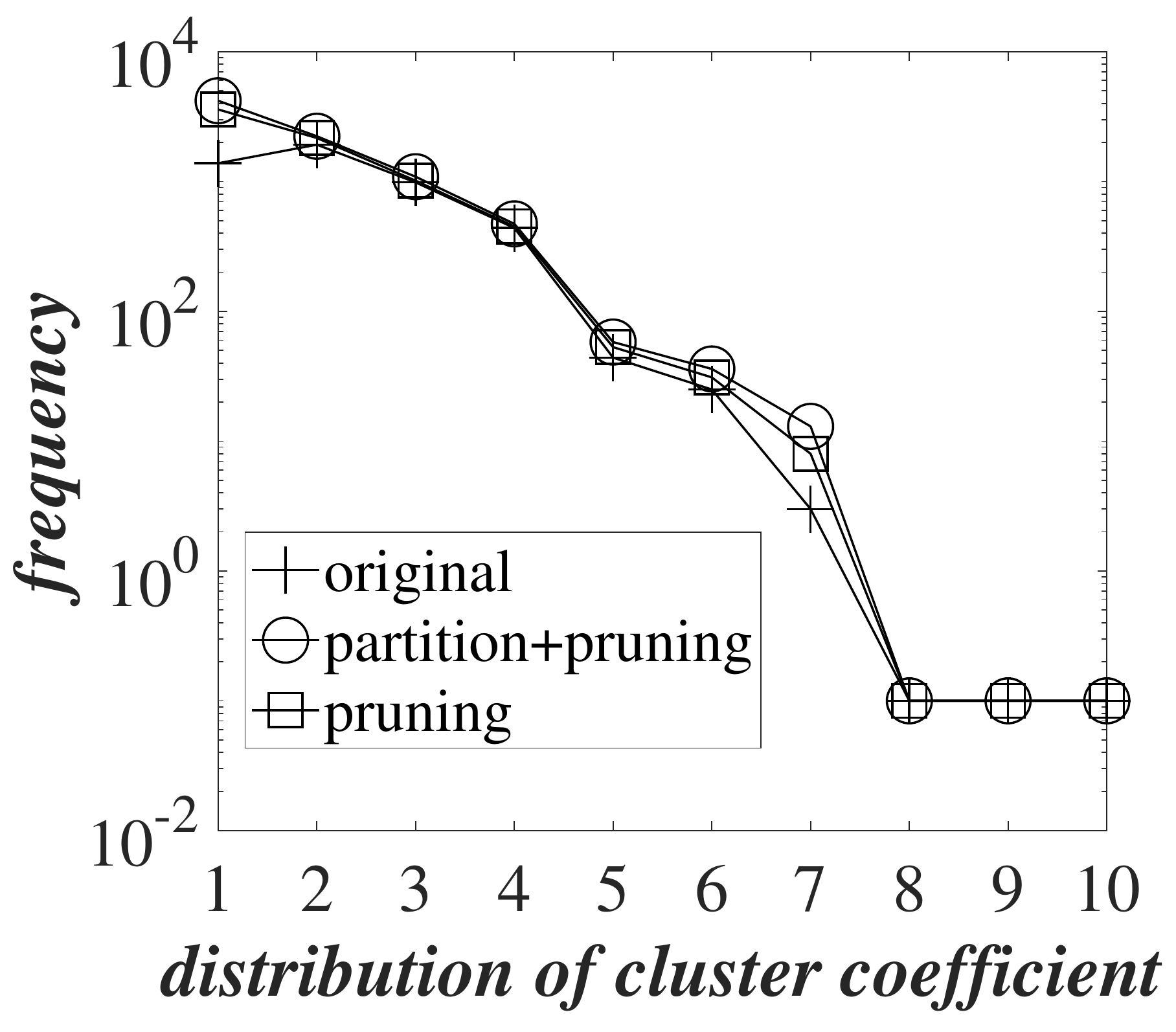}}
\label{subfig:Wikipedia_vs_clc}
}\\
\subfigure[][{\small Arxiv}]{\hspace{-2ex}                   
\scalebox{0.2}[0.2]{\includegraphics{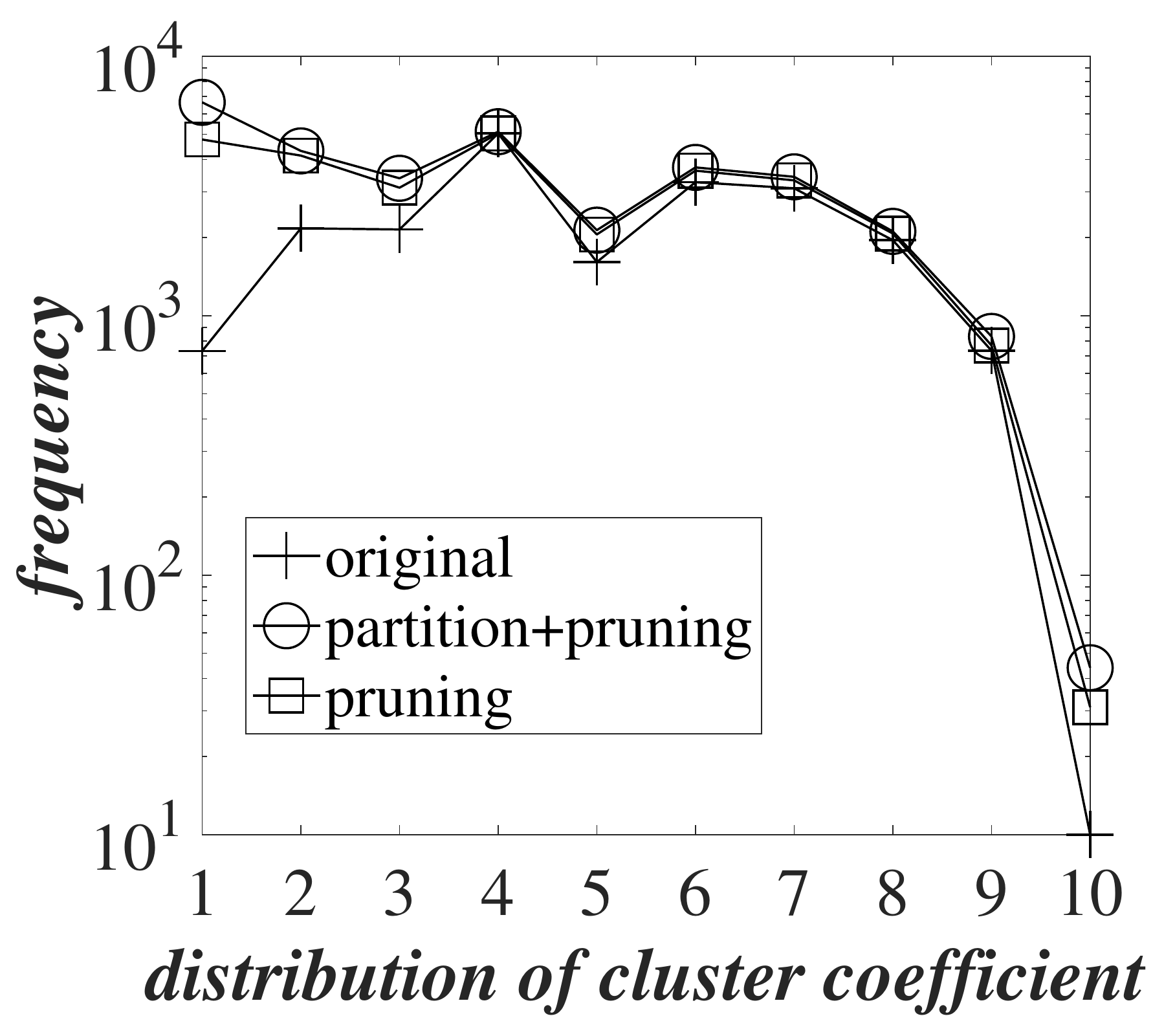}}
\label{subfig:Arxiv_vs_clc}
}
\subfigure[][{\small Uniform}]{\hspace{-1.5ex}                   
\scalebox{0.2}[0.2]{\includegraphics{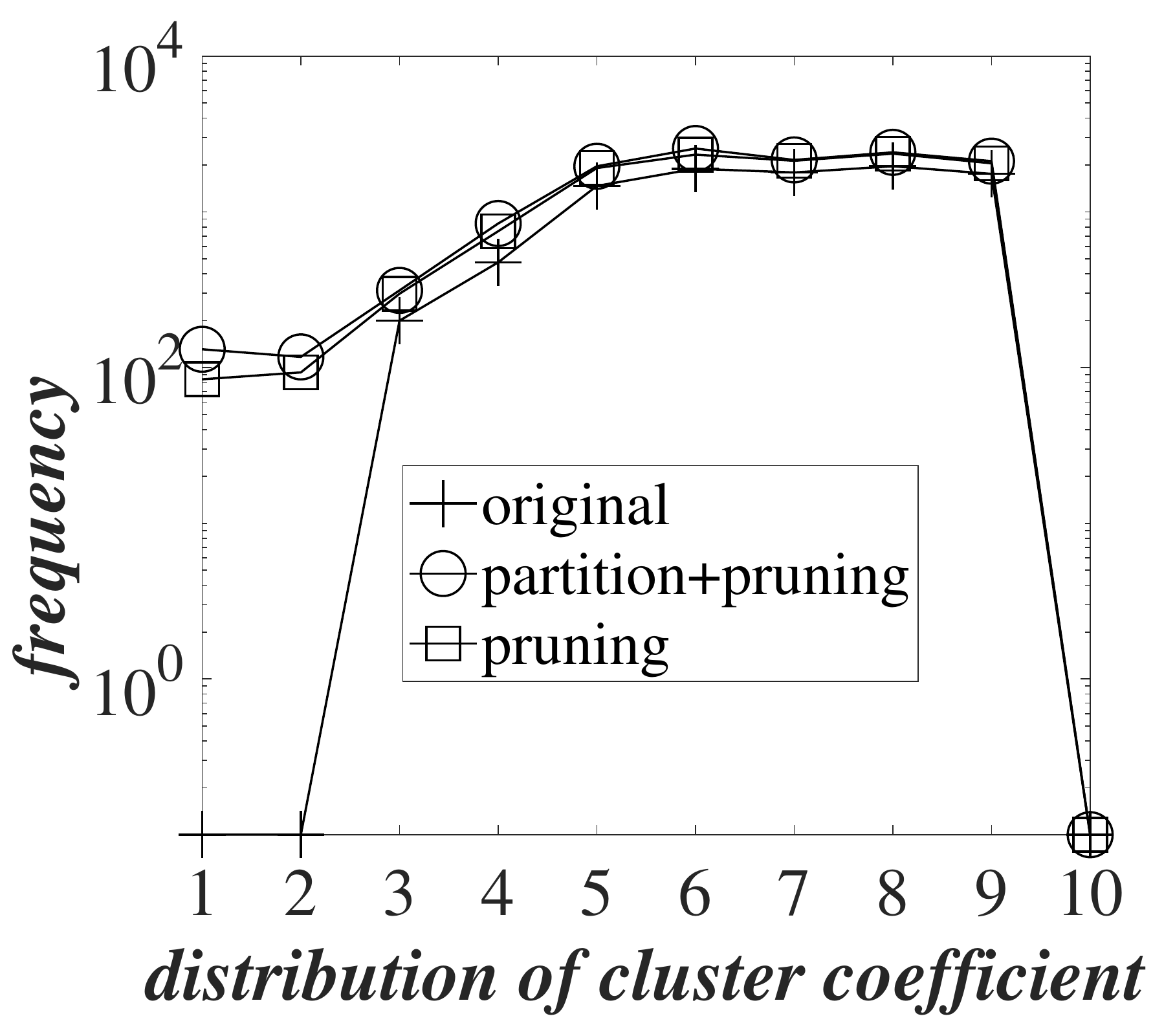}}
\label{subfig:uniform_vs_clc}
}
\subfigure[][{\small Gaussian}]{\hspace{-1.5ex}                  
\scalebox{0.2}[0.2]{\includegraphics{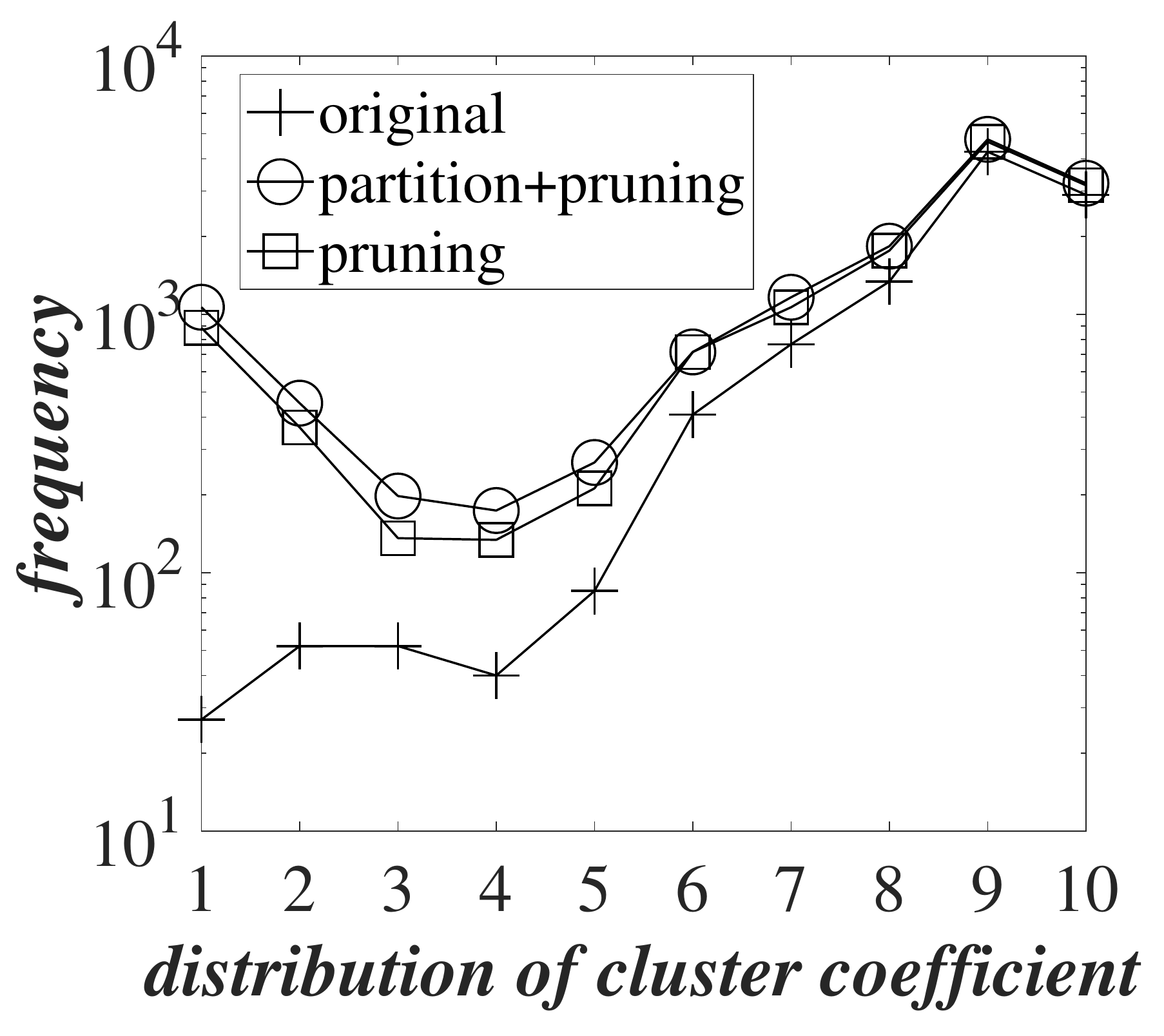}}
\label{subfig:gaussian_vs_clc}
}
\subfigure[][{\small Zipf}]{\hspace{-2ex}                   
\scalebox{0.2}[0.2]{\includegraphics{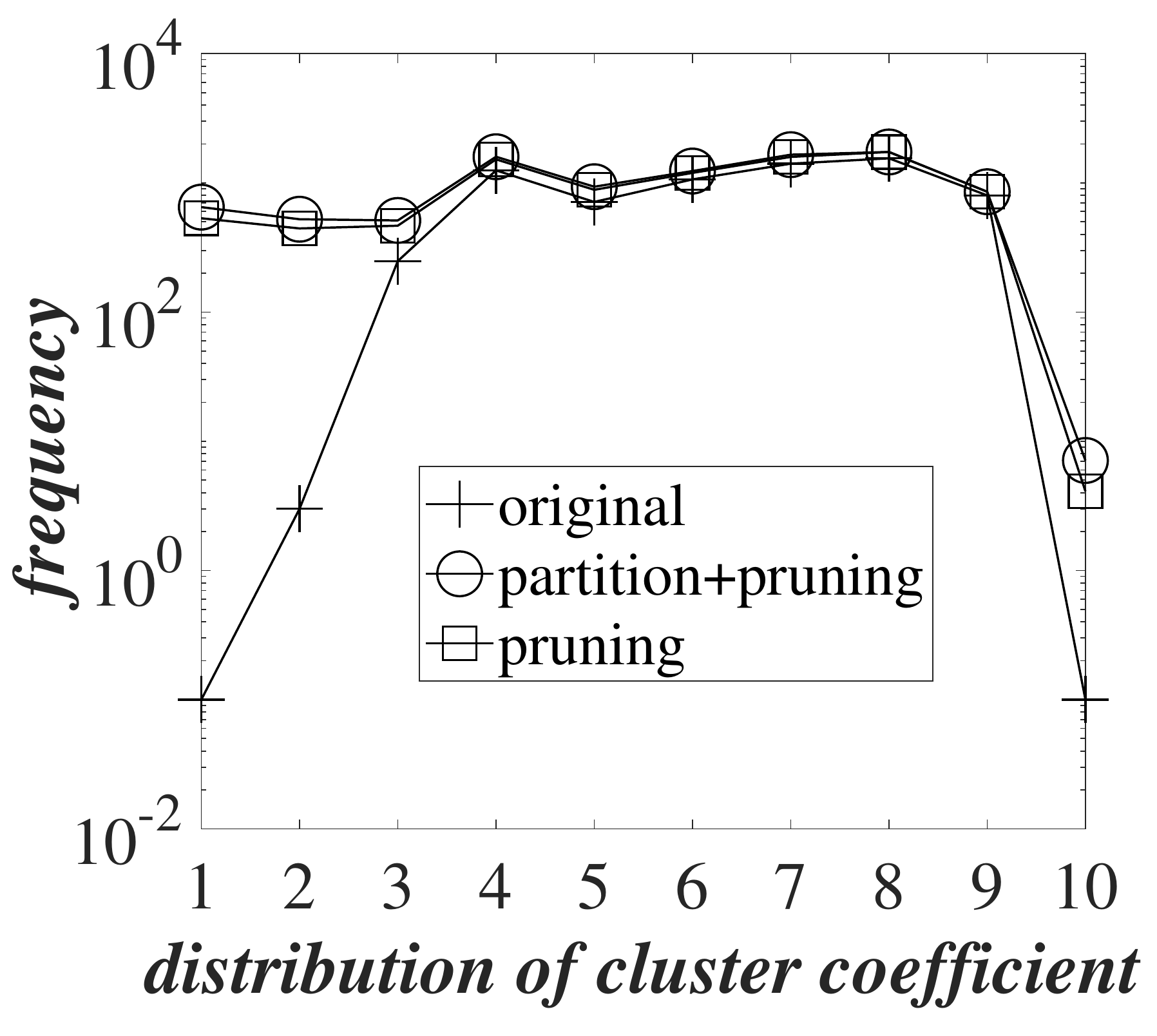}}
\label{subfig:zipf_vs_clc}
}
\caption{\small Distributions of Cluster Coefficient.} 
\label{exper:distribution_vs_cluster_coefficient}
\end{figure*}

\subsection{More Utility Evaluation}
\label{subsec:more_utility_eva}

\noindent {\bf Utility Evaluation vs. Vertex Degree Distribution.} Figure \ref{exper:more_distribution_vs_degree} evaluates the degree distributions of our proposed $kt$-safe graph generation approach over $Facebook$, $Wikipedia$, $Arxiv$, $Uniform$, $Gaussian$ and $Zipf$. Similar as Figure \ref{subfig:Epinions_vs_degree}, we can see that our proposed approaches can well reveal the degree distribution trends of original graph. Specifically, as shown in figures \ref{subfig:Facebook_vs_degree} and \ref{subfig:gaussian_vs_degree}, our methods may behave badly for degree ranges $[0, 5]$ on dense graphs, since we need to add more fake vertices to protect the privacy of each vertex. 

\noindent {\bf Utility Evaluation vs. Shortest Path Length Distribution.} Figure \ref{exper:more_distribution_vs_shortest_path} demonstrates the distributions of shortest path lengths among 1,000 randomly selected vertex pairs over $Facebook$, $Wikipedia$, $Arxiv$, $Uniform$, $Gaussian$ and $Zipf$. Similar as Figure \ref{subfig:Epinions_vs_spl}, we can observe that our proposed approaches vividly reveal the shortest path length distribution of the original graph, which indicates the utility performance of our proposed approaches.

\noindent {\bf Utility Evaluation vs. Cluster Coefficient Distribution.} Figure \ref{exper:distribution_vs_cluster_coefficient} evaluates the distributions of cluster coefficient over real/synthetic data. From figures, all approaches can well reproduce the distribution of cluster coefficient of the unanonymized graphs. The only exception is when the transitivity is 0.1 and/or 0.2. This is because, we add some fake nodes with low transitivity for meeting the requirements of both $kt\text{-}safety$. As a result, the cluster coefficients of $kt\text{-}safety$ deviate from that of the original graph. Moreover, the transitivity of the anonymized graphs via $partition+pruning$ have larger frequencies than that via $pruning$, since $partition+pruning$ needs to add more fake nodes into divided subgraphs to guarantee the privacy of each subgraph.

\noindent {\bf Degree Statistics of randomly selected vertices.} Table \ref{table:degree_statistics} reports the degree mean and standard derivation of 1,000 randomly selected vertices from real and synthetic data sets, where $partition+pruning$ and its 3 variants adopt the same selected vertices on the same data set. Specifically, we report these two statistics of the original graph as the standard. From the table, we can see that all our approaches can well reproduce the mean and standard derivation of the unanonymized graphs. The $pruning$ and $none$ methods behave slightly better than $partition+pruning$ and $partition$, since they do not consider the efficiency requirement and do not need to partition original graphs.

\noindent {\bf Shortest Path Length Statistics of randomly selected vertex pairs.} Table \ref{table:shortest_path_statistics} demonstrates the mean and standard derivation of the shortest path length for 1,000 randomly selected vertex pairs from all data sets, where $partition+pruning$ and its 3 variants apply the same vertex pairs on the same data set. Similar to degree statistics, $partition+pruning$ and $pruning$ have the same statistics as $partition$ and $none$ respectively. From the table, all our approaches can well reveal these two statistics on the 1,000 selected vertex pairs from all real and synthetic data sets.

\begin{figure}[H]
\centering
\scalebox{0.25}[0.25]{\includegraphics{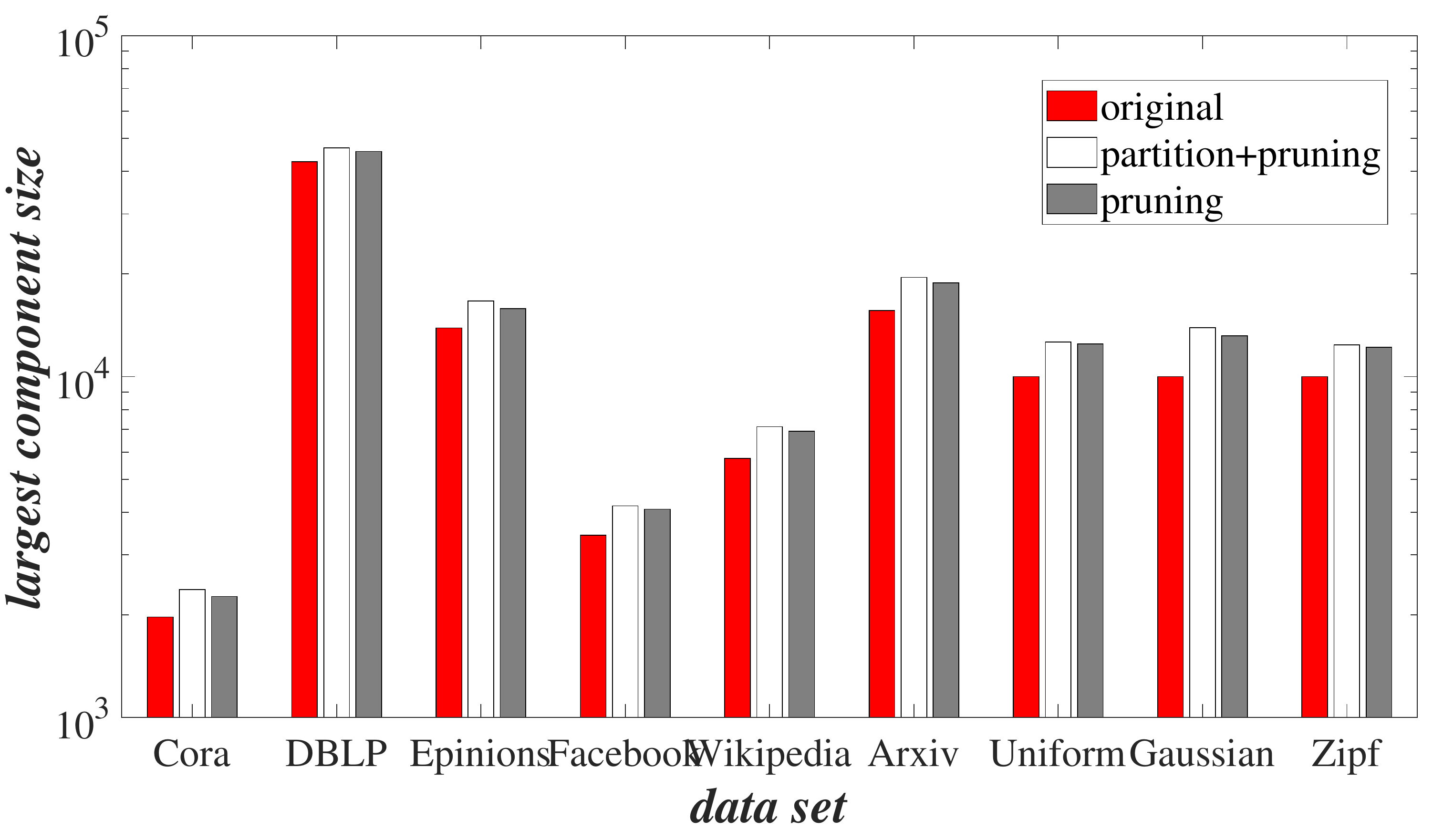}}
\caption{\small The largest component size vs. real/synthetic data sets.}
\label{fig:graph_analysis}
\end{figure}

\noindent {\bf Largest Component Size vs. Real/Synthetic Data Sets.} Figure \ref{fig:graph_analysis} reports the largest component sizes of the anonymized graphs via our approaches on real and synthetic data sets. In order to improve the readability, we omit the results of $partition$ and $none$, since they produce the same anonymized graphs as that of $partition+pruning$ and $pruning$, respectively. From the figures, all our approaches can well reveal the largest component sizes of the unanonymized graphs, which indicates that the anonymized graphs via our approaches can be well used for graph analysis tasks. Moreover, $pruning$ behaves slightly better than $partition+pruning$, since $pruning$ do not need to partition graphs for efficiency purpose.

\begin{table*}[t!]
\centering
\scriptsize
\caption{\small The degree statistics of 1,000 randomly selected vertices from data sets.} \label{table:degree_statistics}
\begin{tabular}{|c|c|c|c|c|c|c|}
\hline
\textbf{Data Sets} & \textbf{Statistics} & \textbf{original} & \textbf{partition+pruning} & \textbf{partition} & \textbf{pruning} &  \textbf{none} \\
\hline
Cora & mean & 4.141	& 4.141 & 4.141 & 4.141	& 4.141 \\
 & standard derivation & 213.951 & 213.951 & 213.951 & 213.951 & 213.951 \\\hline
DBLP & mean & 6.043 & 6.191 & 6.191 & 6.165 & 6.165 \\
 & standard derivation & 267.95 & 290.23 & 290.23 & 281.6 & 281.6 \\\hline
Epinions & mean & 30.803 & 30.945 & 30.945 & 30.933 & 30.933 \\
 & standard derivation & 2300.12 & 2348.31 & 2348.31 & 2340.84 & 2340.84\\\hline
Facebook & mean & 42.292 & 42.411 & 42.411 & 42.347 & 42.347 \\
 & standard derivation & 1719.05 & 1747.14 & 1747.14 & 1733.15 & 1733.15\\\hline
Wikipedia & mean & 33.103 & 33.31 & 33.31 & 33.23 & 33.23 \\
 & standard derivation & 2051.78 & 2068.21 & 2068.21 & 2063.51 & 2063.51\\\hline
Arxiv & mean & 8.267 & 8.293 & 8.293 & 8.281 & 8.281 \\
 & standard derivation & 335.099 & 347.516 & 347.516 & 342.526 & 342.526\\\hline
Uniform & mean & 19.25 & 19.38 & 19.38 & 19.32 & 19.32 \\
 & standard derivation & 358.438 & 367.275 & 367.275 & 363.541 & 363.541\\\hline
Gaussian & mean & 45.789 & 45.932 & 45.932 & 45.813 & 45.813 \\
 & standard derivation & 14543.6 & 14686.24 & 14686.24 & 14621.56 & 14621.56\\\hline
Zipf & mean & 11.195 & 11.211 & 11.211 & 11.204 & 11.204 \\
 & standard derivation & 356.428 & 363.645 & 363.645 & 360.153 & 360.153\\\hline
\end{tabular}
\end{table*}

\begin{table*}[t!]
\centering
\scriptsize
\caption{\small The shortest path length statistics of 1,000 randomly selected vertex pairs from data sets.} \label{table:shortest_path_statistics}
\begin{tabular}{|c|c|c|c|c|c|c|}
\hline
\textbf{Data Sets} & \textbf{Statistics} & \textbf{original} & \textbf{partition+pruning} & \textbf{partition} & \textbf{pruning} &  \textbf{none} \\
\hline
Cora & mean & 5.979 & 5.979 & 5.979 & 5.979 & 5.979 \\
 & standard derivation & 39.7815 & 39.7815 & 39.7815 & 39.7815 & 39.7815 \\\hline
DBLP & mean & 6.431 & 6.492 & 6.492 & 6.384 & 6.384 \\
 & standard derivation & 29.7529 & 30.351 & 30.351 & 28.849 & 28.849 \\\hline
Epinions & mean & 3.339 & 3.452 & 3.452 & 3.288 & 3.288 \\
 & standard derivation & 27.9657 & 29.034 & 29.034 & 27.425 & 27.425\\\hline
Facebook & mean & 3.717 & 3.816 & 3.816 & 3.761 & 3.761 \\
 & standard derivation & 36.2617 & 37.135 & 37.135 & 36.611 & 36.611\\\hline
Wikipedia & mean & 3.224 & 3.376 & 3.376 & 3.187 & 3.187 \\
 & standard derivation & 24.2037 & 24.417 & 24.417 & 24.0561 & 24.0561\\\hline
Arxiv & mean & 5.557 & 5.611 & 5.611 & 5.557 & 5.557 \\
 & standard derivation & 37.2392 & 37.7135 & 37.7135 & 37.2392 & 37.2392\\\hline
Uniform & mean & 6.888 & 6.889 & 6.889 & 6.888 & 6.888 \\
 & standard derivation & 21.0584 & 21.0592 & 21.0592 & 21.0584 & 21.0584\\\hline
Gaussian & mean & 2.134 & 2.161 & 2.161 & 2.134 & 2.134 \\
 & standard derivation & 23.5381 & 23.605 & 23.605 & 23.5381 & 23.5381\\\hline
Zipf & mean & 6.898 & 6.881 & 6.881 & 6.835 & 6.835 \\
 & standard derivation & 19.4833 & 19.4794 & 19.4794 & 18.847 & 18.847\\\hline
\end{tabular}
\end{table*}

\subsection{More Efficiency Evaluation}
\label{subsec:more_efficiency_eva}

\begin{figure}[t!]
\centering
\scalebox{0.27}[0.27]{\includegraphics{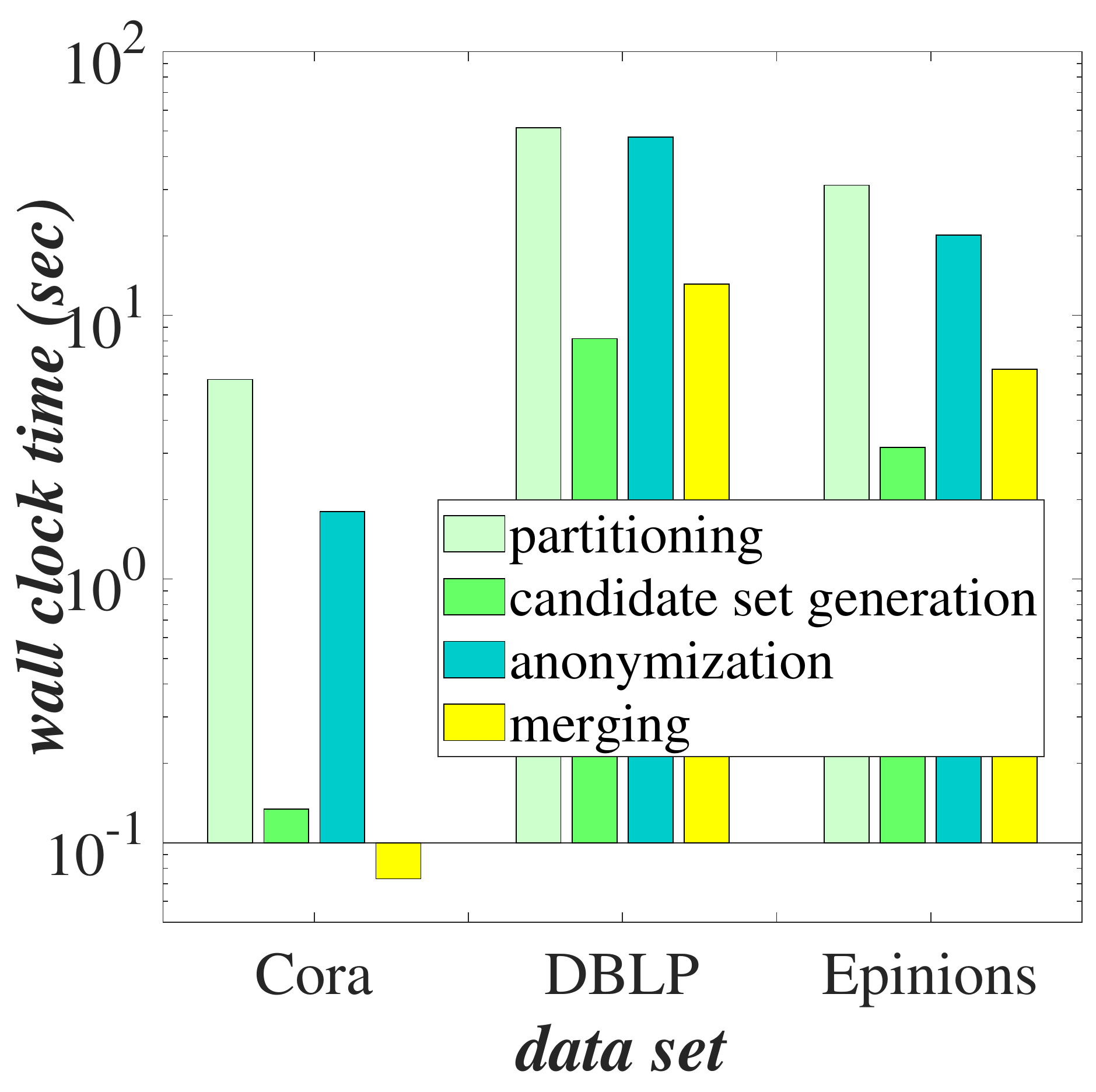}}
\caption{\small A break-up cost of $partition$$+$$pruning$ in Figure \ref{fig:exper:time_cost_vs_dataset}.}
\label{fig:break_up_cost}
\end{figure}

\noindent {\bf A break-up cost analysis of $partition+pruning$ in Figure \ref{fig:exper:time_cost_vs_dataset}.} Figure \ref{fig:break_up_cost} illustrates the break-up cost of our $partition+pruning$ approach over $Cora$, $DBLP$ and $Epinions$, which includes the partitioning cost (based on the cost model in Section \ref{subsec:cost_model}), candidate set generation cost (based on the pruning strategy in Section \ref{subsec:pruning_strategy} and the $kt$-tree in Section \ref{subsec:kt-tree}), anonymization cost (based on Algorithm \ref{alg:kt_safe_vertex}), and merging cost (based on Algorithm \ref{alg:graph_merge}). From the figure, we can see that the partitioning part of $partition+pruning$ is the most time-consuming, since it needs to cluster vertices into different subgraphs. However, as shown in Figure \ref{fig:exper:time_cost_vs_dataset}, $partition+pruning$ outperforms $pruning$ and $none$, in terms of the wall clock time, on large data sets, $DBLP$ and $Epinions$, by 1-2 order of magnitude, which confirms the necessity of the partitioning for large graphs. On the other hand, it indicates that we may not need to apply the partitioning strategy for small graphs (e.g., $Cora$). Moreover, the anonymization part is the second most time-consuming, as it needs to anonymize each vertex to a $kt$-safe one.

\begin{figure}[H]
\centering
\scalebox{0.22}[0.22]{\includegraphics{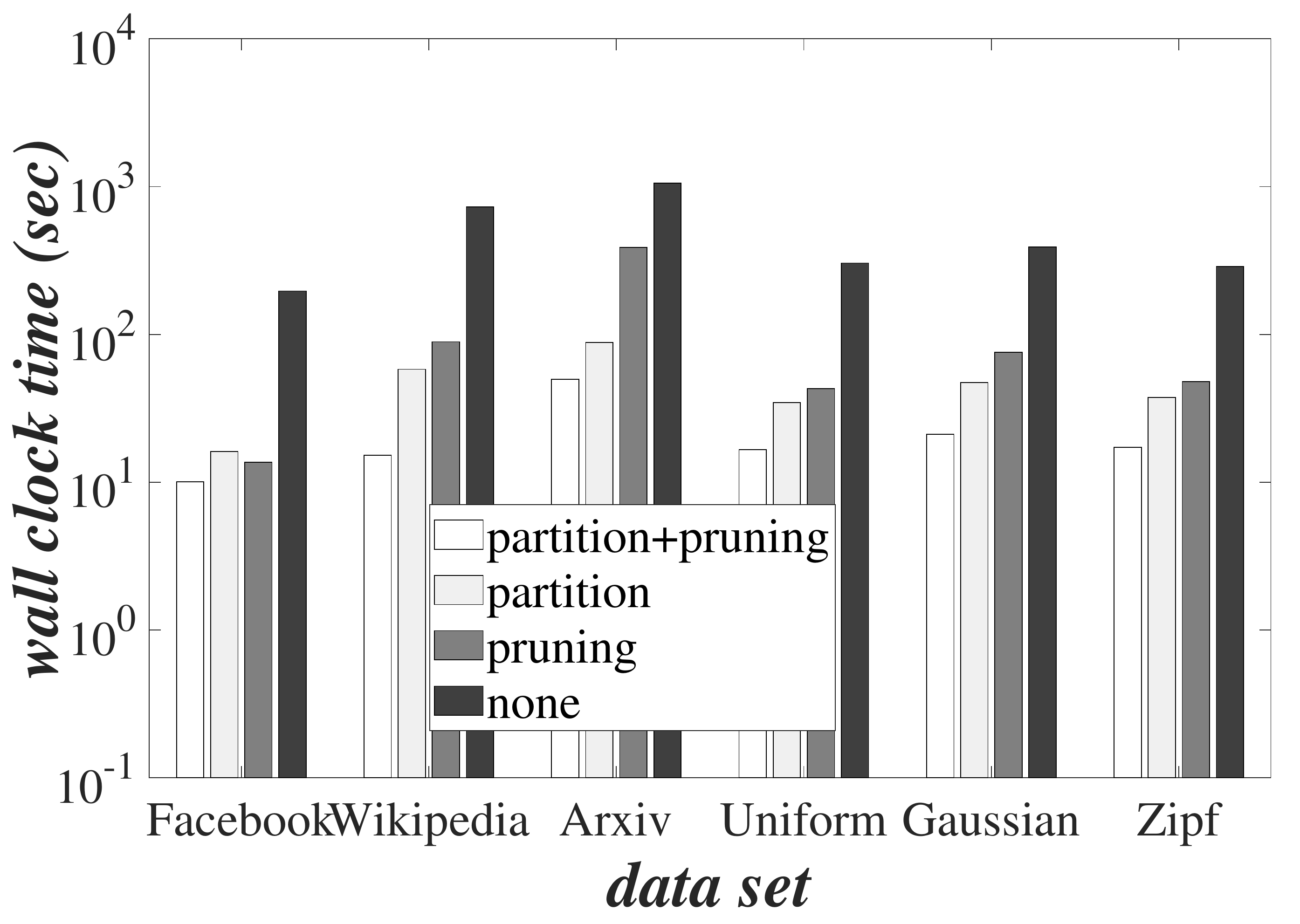}}
\caption{\small The time cost vs. more real/synthetic data sets.}
\label{fig:more_time_cost_vs_dataset}
\end{figure}

\noindent {\bf More Efficiency Evaluation vs. Wall Clock Time.} Figure \ref{fig:more_time_cost_vs_dataset} illustrates the wall clock time of anonymizing the graph for $Facebook$, $Wikipedia$, $Arxiv$, $Uniform$, $Gaussian$ and $Zipf$, where default values of parameters are used for synthetic data. From the figure, we can see that $partition+pruning$ needs the least wall clock time on all data sets, which confirms the efficiency of our graph anonymization approach. For the three variants of our methods, similar as Figure \ref{fig:exper:time_cost_vs_dataset}, $partition$ has the smallest wall clock time on all data sets except $Facebook$, followed by $pruning$ and $none$, respectively. Specifically, $partition$ is slower than $pruning$ on $Facebook$, which indicates that we may not need to adopt the partitioning strategy on small graph with less vertices.

\begin{figure}[H]
\centering
\scalebox{0.24}[0.24]{\includegraphics{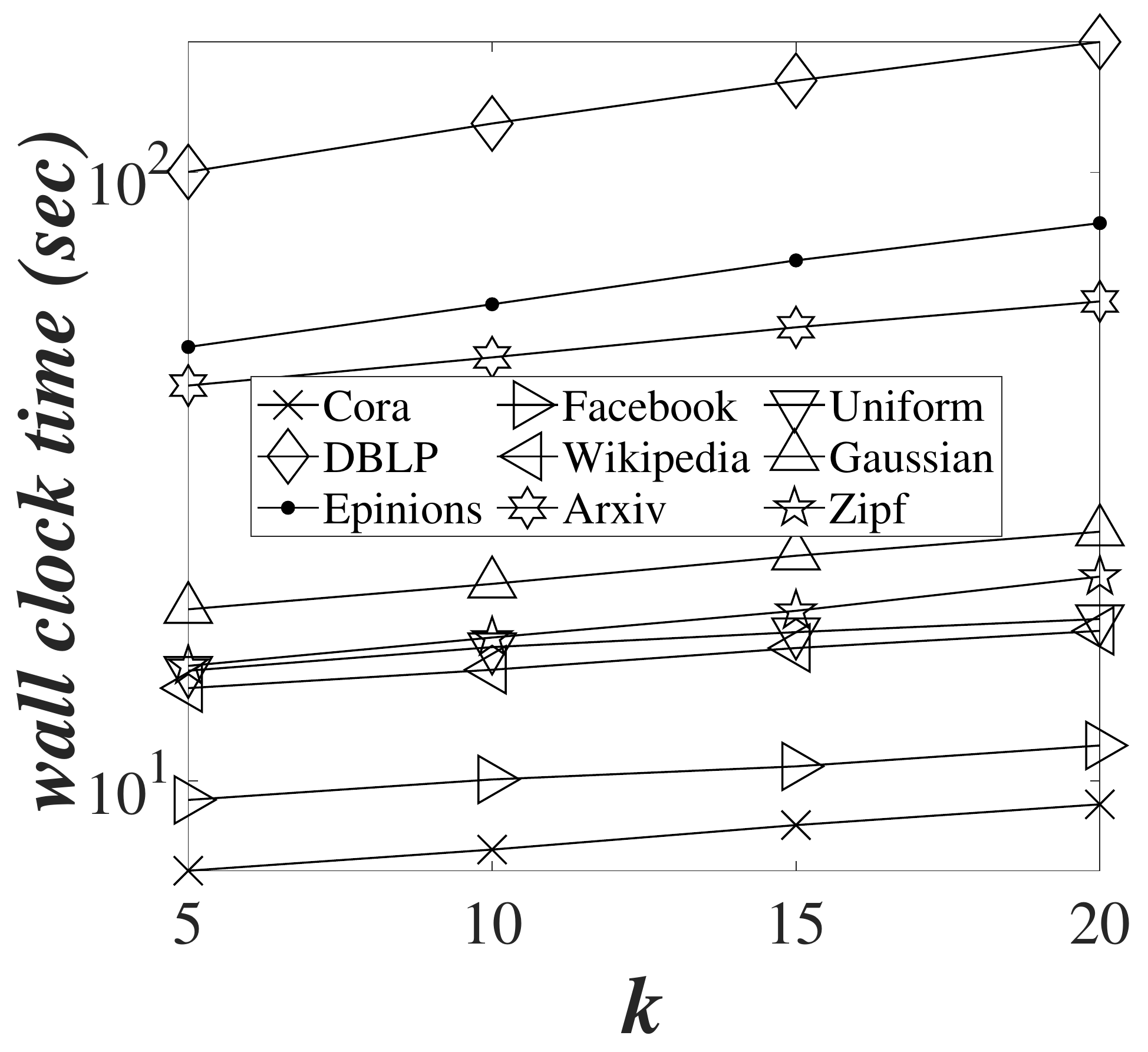}}
\caption{\small The wall clock time of $partition+pruning$ vs. privacy threshold $k$.}
\label{fig:exper:time_k}
\end{figure}

\noindent {\bf Efficiency vs. identity privacy threshold $k$.} Figure \ref{fig:exper:time_k} illustrates the performance of our $partition+pruning$ method over both real and synthetic data sets, where $k=5$, $10$, $15$ and $20$ and other parameters are set as their default values in Table \ref{table:exp_parameter_setting}. From the figure, with larger $k$, the wall clock time slightly increase over all data sets, since each node needs to find more similar candidates for a stronger privacy guarantee. Specifically, the $partition+pruning$ approach achieves the highest time cost on DBLP on all $k$ values, since DBLP has the largest number of vertices. Nevertheless, our $partition+pruning$ still needs low time cost (less than 163.78 $sec$ even when $k=20$).

\begin{figure}[H]
\centering
\scalebox{0.23}[0.23]{\hspace{0ex}\includegraphics{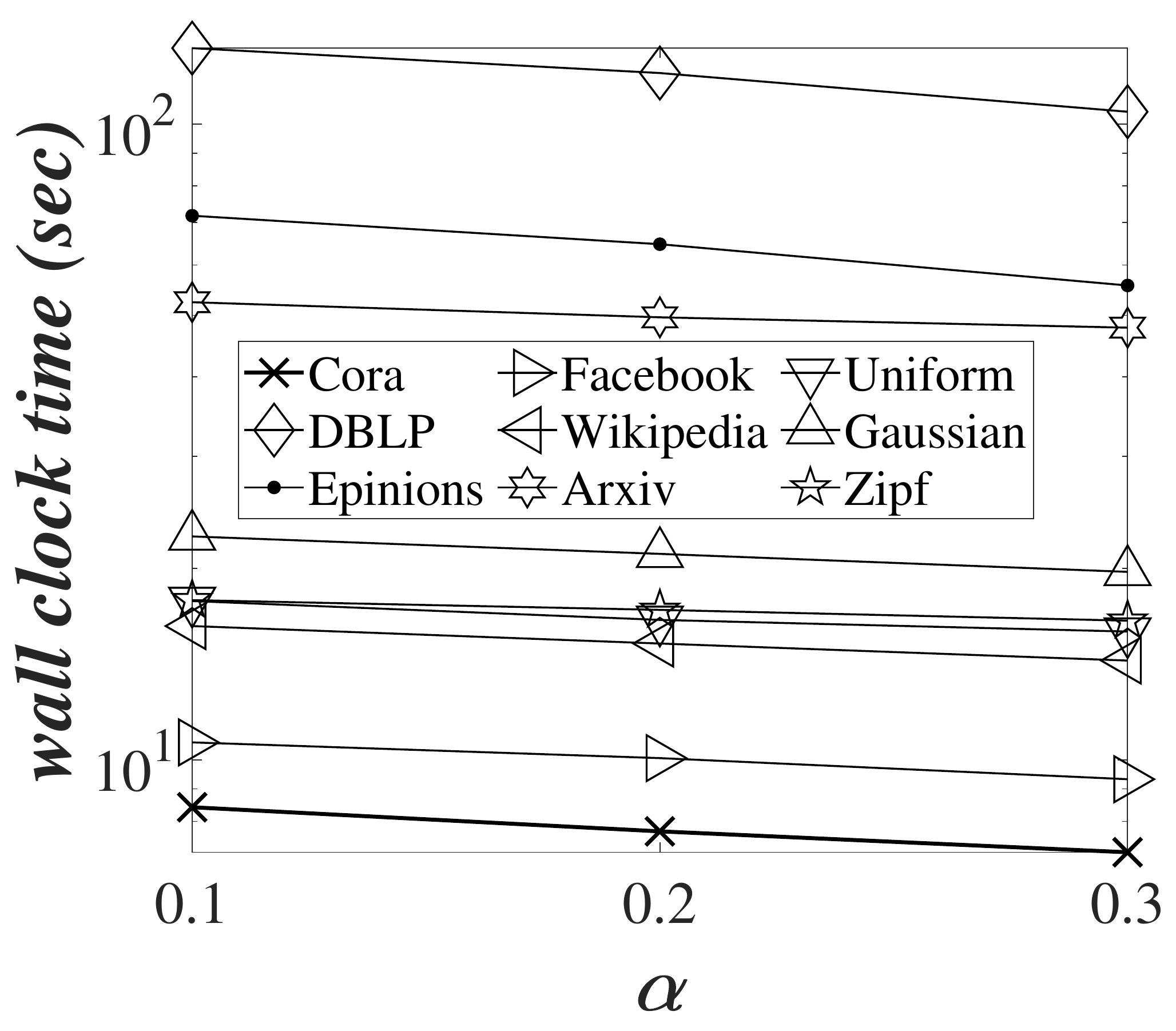}}
\caption{\small The wall clock time of $partition+pruning$ vs. sensitive privacy threshold $\alpha$.}
\label{fig:exper:time_alpha}
\end{figure}

\noindent {\bf Efficiency vs. sensitive privacy threshold $\alpha$.} Figure \ref{fig:exper:time_alpha} shows the effect of the sensitive privacy threshold on our $partition+pruning$ method by varying $\alpha$ from 0.1 to 0.3, where other parameters follow their defaults. From the figure, when $\alpha$ increases, the wall clock time slightly decreases, since there will be more tolerable sensitive attribute values in the protection set of each node (as defined in Definition \ref{def:KT_Safe_Vertex}). Nevertheless, $DBLP$ still has the highest time cost due to the large size of vertices.

\begin{figure}[H]
\centering
\scalebox{0.25}[0.25]{\hspace{0ex}\includegraphics{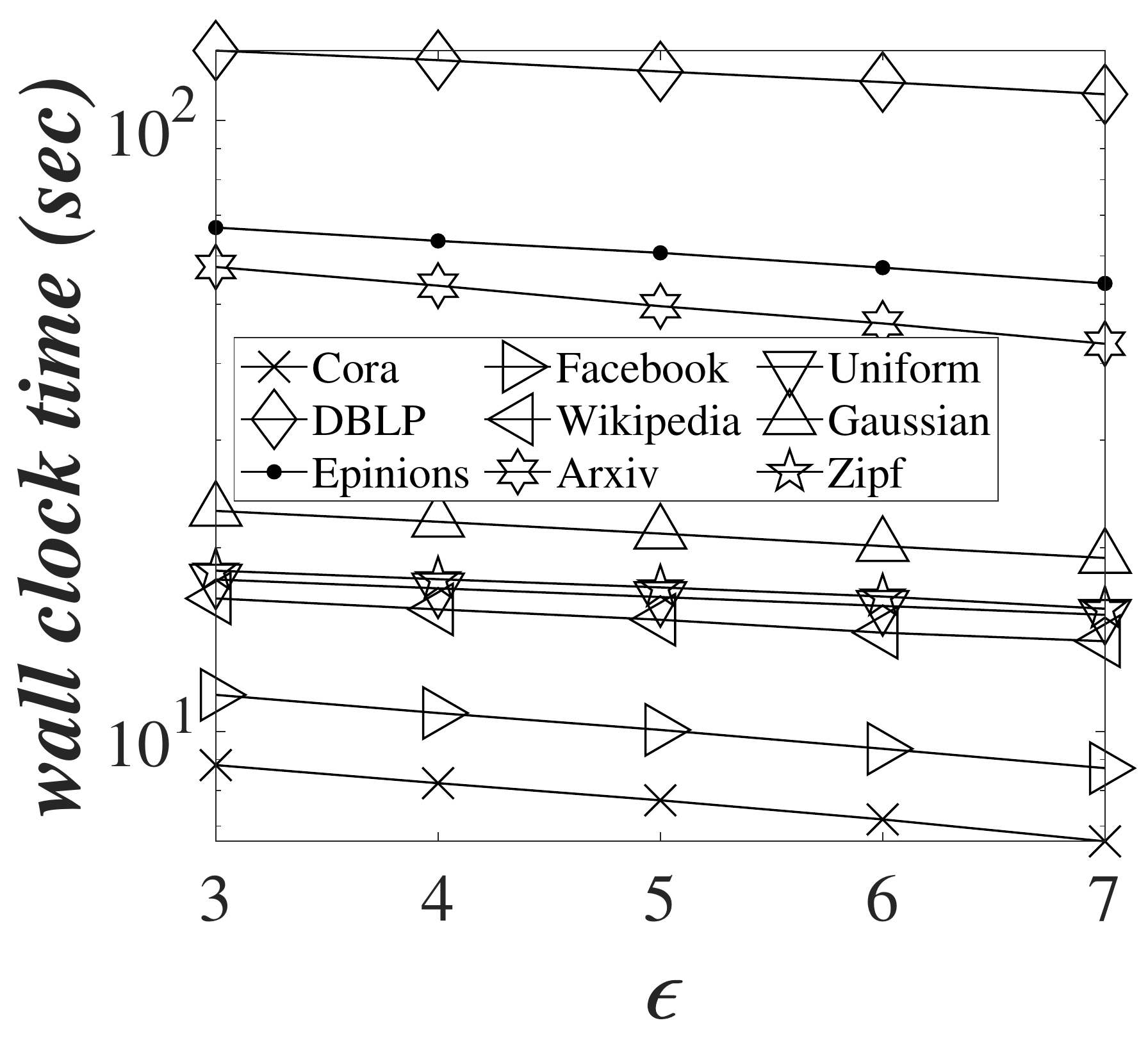}}
\caption{\small The wall clock time of $partition+pruning$ vs. graph edit distance threshold $\epsilon$.}
\label{fig:exper:time_epsilon}
\end{figure}

\noindent {\bf Efficiency vs. graph edit distance threshold $\epsilon$.} Figure \ref{fig:exper:time_epsilon} reports the performance of our $partition+pruning$ method, by varying $\epsilon$ from 3 to 7, where other parameters are set to their defaults. From the result, when $\epsilon$ increases, the wall clock time decreases. This is because, with larger $\epsilon$, there will be more initial detected candidates for each node (via Algorithm \ref{alg:init_cand}), which needs less time for obtaining the node's protection sets. 

\begin{figure}[H]
\centering
\scalebox{0.22}[0.22]{\hspace{0ex}\includegraphics{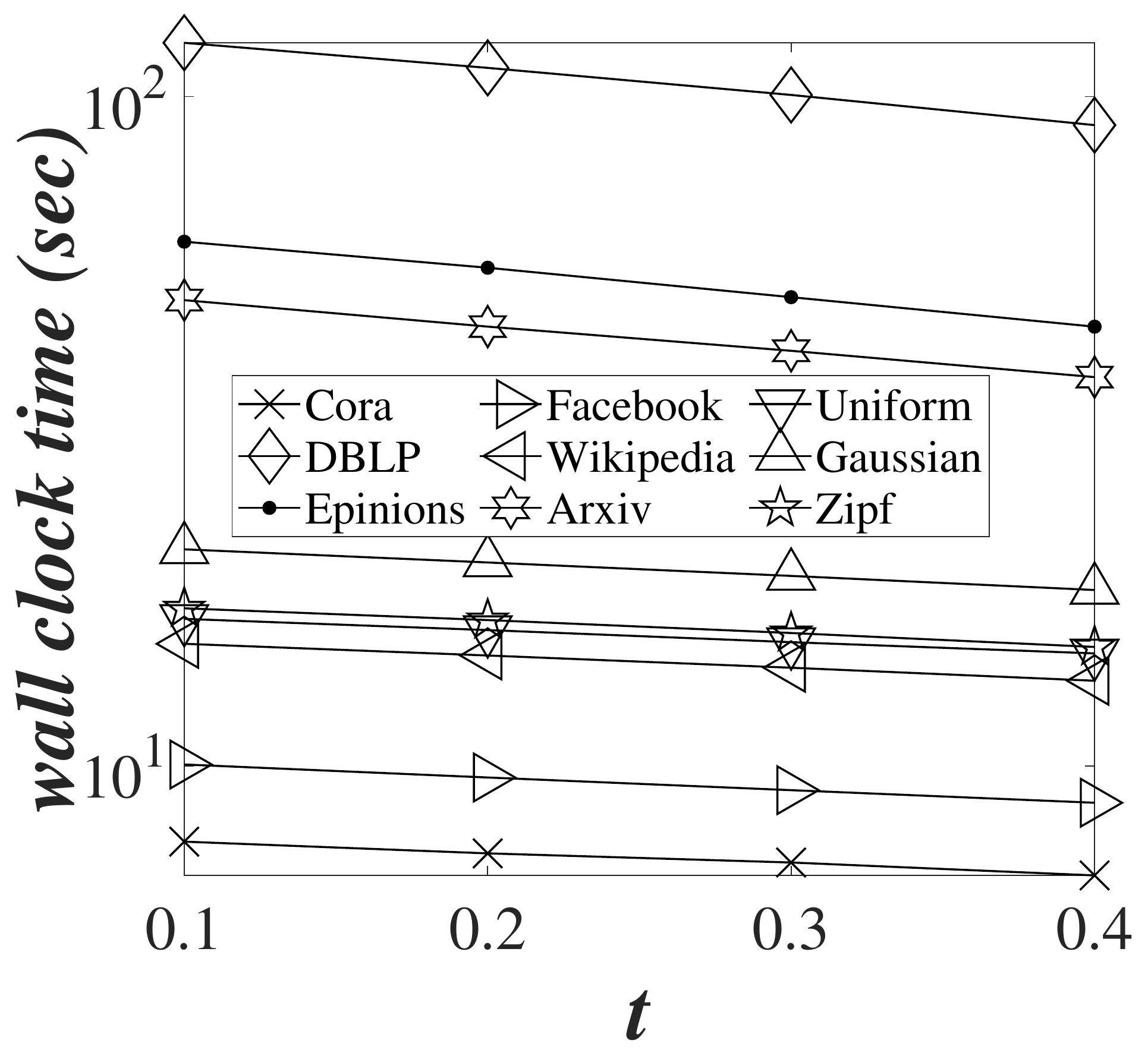}}
\caption{\small The wall clock time of $partition+pruning$ vs. distance threshold $t$.}
\label{fig:exper:time_t}
\end{figure}

\noindent {\bf Efficiency vs. distance threshold $t$.} Figure \ref{fig:exper:time_t} demonstrates the effect of the distance threshold $t$ on our $partition+pruning$ method on real and synthetic data sets, where $t=0.1$, $0.2$, $0.3$ and $0.4$ and other parameters are set as their defaults. From the figure, when $t$ increases, the time cost decrease on all data sets. Intuitively, with larger $t$, there will be more initial candidate nodes (obtained via Algorithm \ref{alg:init_cand}) satisfying the requirements of $kt$-safety model, due to the requirement loosing of distribution similarity of n-hop neighbors between graph nodes.

\begin{figure}[t!]
\centering
\scalebox{0.25}[0.25]{\includegraphics{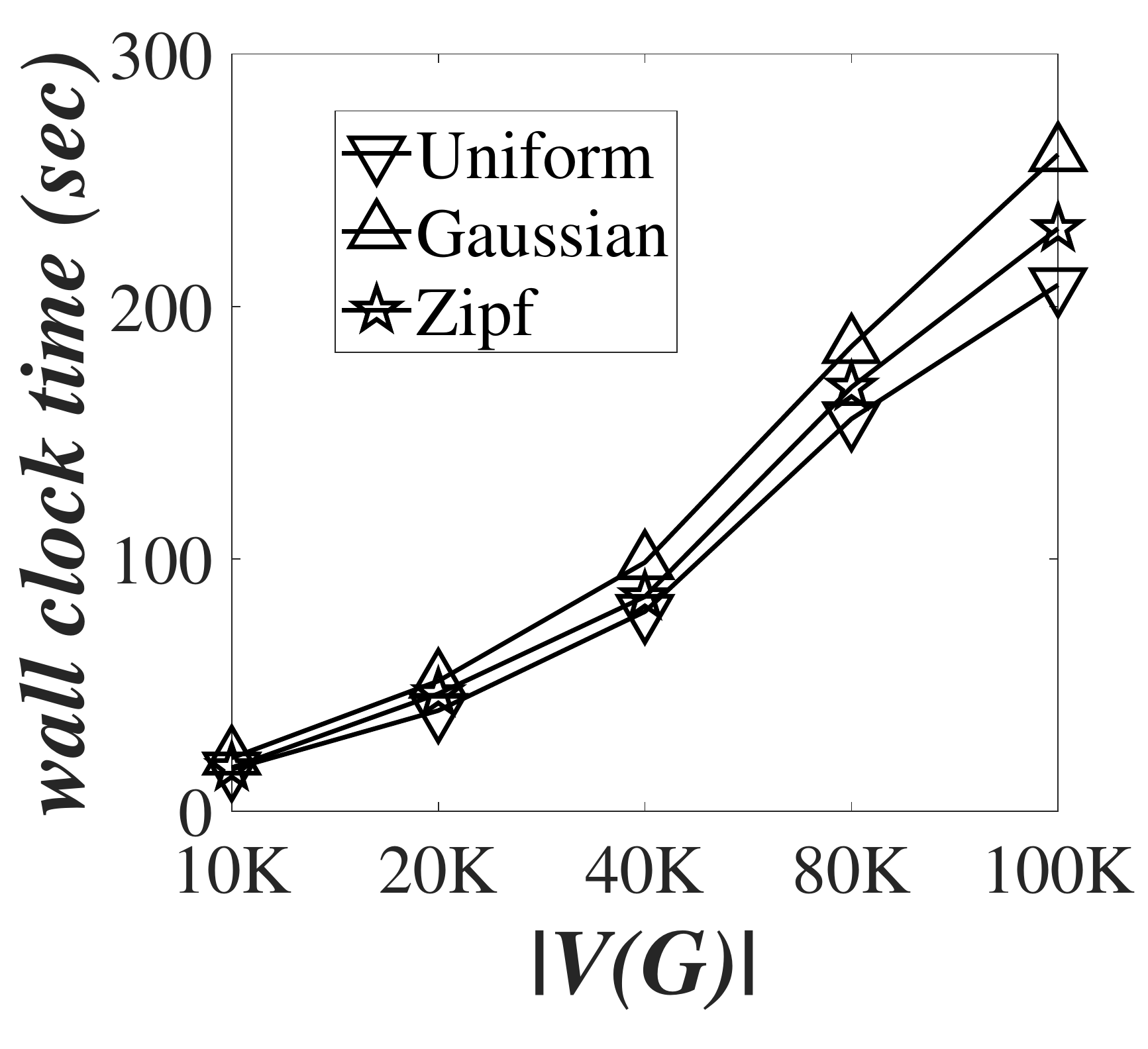}}
\caption{\small The time cost of $partition$$+$$pruning$ vs. graph size $|V(G)|$.}
\label{fig:exper:scalability_VG}
\end{figure}

\noindent {\bf Scalability Evaluation vs. Graph Size $|V(G)|$.} Figure \ref{fig:exper:scalability_VG} tests the scalability of our approach, $partition+pruning$, on synthetic graphs, where $|V(G)|$ varies from $10K$ to $100K$ and other parameters are set to their default values. From figures, we can see that the wall clock time of $partition+pruning$ increases for larger $|V(G)|$. Even when $|V(G)|=100K$, the time cost is still acceptable (i.e., 208.51 $sec\sim$ 260.17 $sec$). Furthermore, our $partition+pruning$ approach over $Uniform$ data can achieve the lowest time cost. This is because, $Gaussian$ has the largest average node degree, and $Zipf$ has more nodes with sensitive values (i.e., small attribute values), which require more time cost.

\begin{figure}[H]
\centering
\scalebox{0.22}[0.22]{\hspace{0ex}\includegraphics{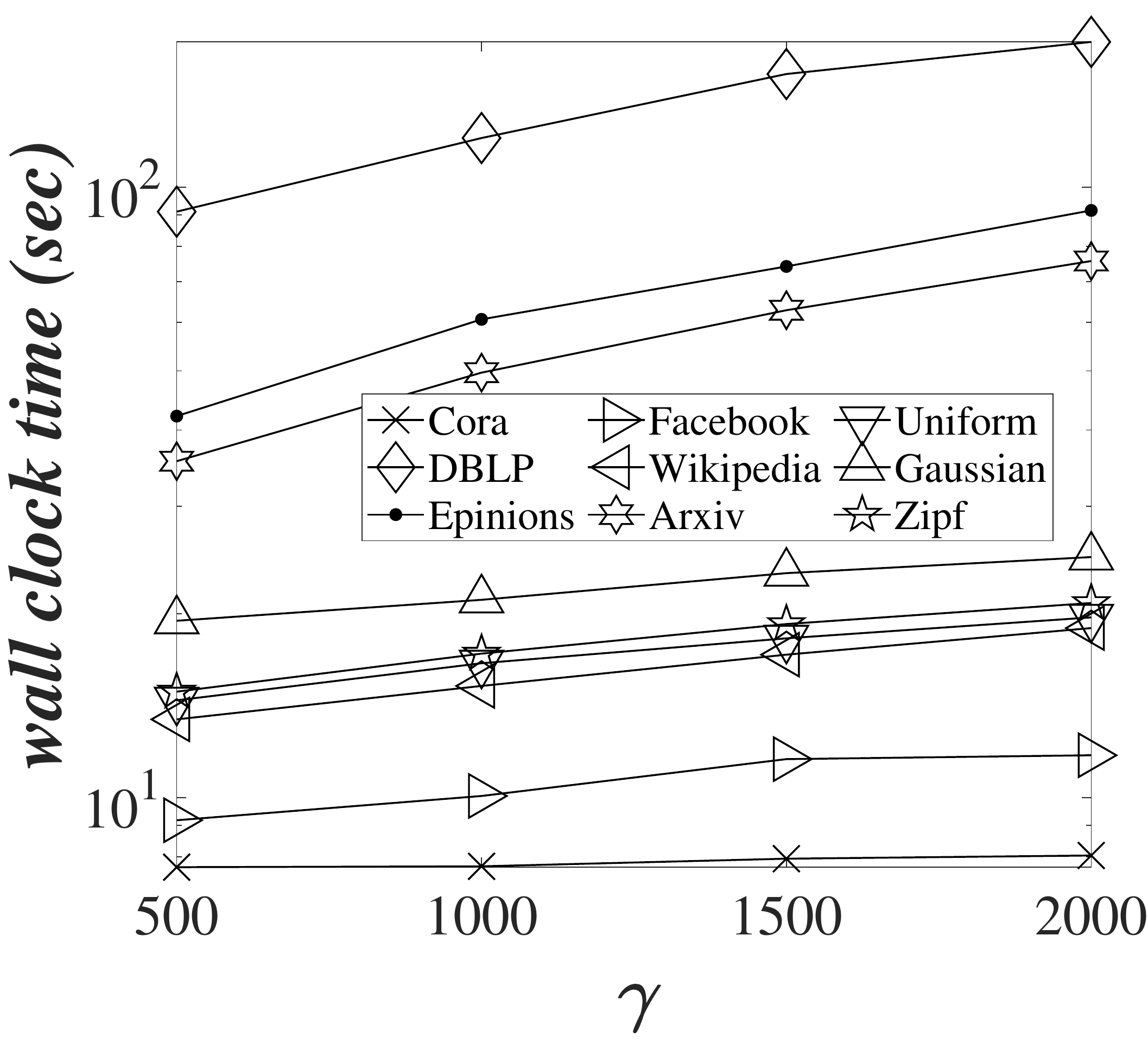}}
\caption{\small The wall clock time of $partition+pruning$ vs. the maximum size, $\gamma$, of subgraphs.}
\label{fig:exper:time_cost_gamma}
\end{figure}

\noindent {\bf Efficiency vs. the maximum size, $\gamma$, of subgraphs.} Figure \ref{fig:exper:time_cost_gamma} illustrates the performance of our $partition+pruning$ approach by varying $\gamma$ from $500$ to $2000$, where other parameters follow their defaults. From the figure, when $\gamma$ increases, the time cost increases. This is because, with larger $\gamma$, it needs to check more vertices for obtaining the protection set for each vertex in subgraphs. Nevertheless, the time cost of our $partition+pruning$ approach is still acceptable even when $\gamma=2,000$ (i.e., $172.86$ sec for $DBLP$). Specifically, the wall clock time of our approach on $Cora$ and $Facebook$ hardly increases when $\gamma\geq 1500$, since $partition+pruning$ needs to partition these two graphs only once given $s=4$ (by default).

\begin{figure}[H]
\centering
\scalebox{0.22}[0.22]{\hspace{0ex}\includegraphics{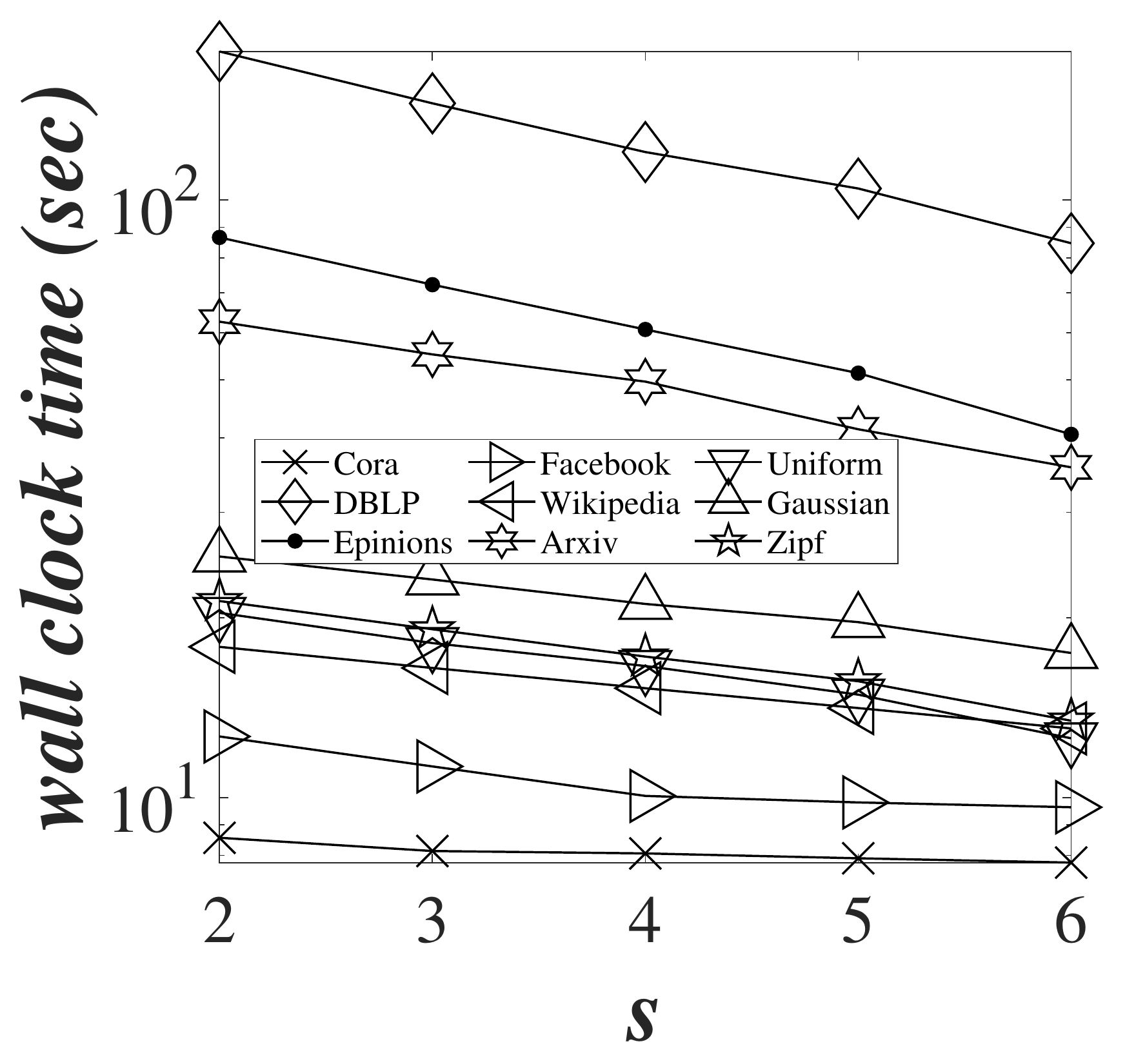}}
\caption{\small The wall clock time of $partition+pruning$ vs. the number, $s$, of graph partitions.}
\label{fig:exper:time_cost_s}
\end{figure}

\noindent {\bf Efficiency vs. the number, $s$, of graph partitions.} Figure \ref{fig:exper:time_cost_s} reports the wall clock time of our $partition+pruning$ approach, by varying $s$ from $2$ to $6$, where other parameters are set to their default values. From the figure, with larger $s$, our approach achieves lower time cost. The reason for this is that, with larger $s$, there will be more subgraphs with fewer nodes, each of which needs lower time to be anonymized.

\subsection{More Anonymization Cost Evaluation}
\label{subsec:more_anony_cost_eva}

\begin{figure}[H]
\centering
\scalebox{0.22}[0.22]{\includegraphics{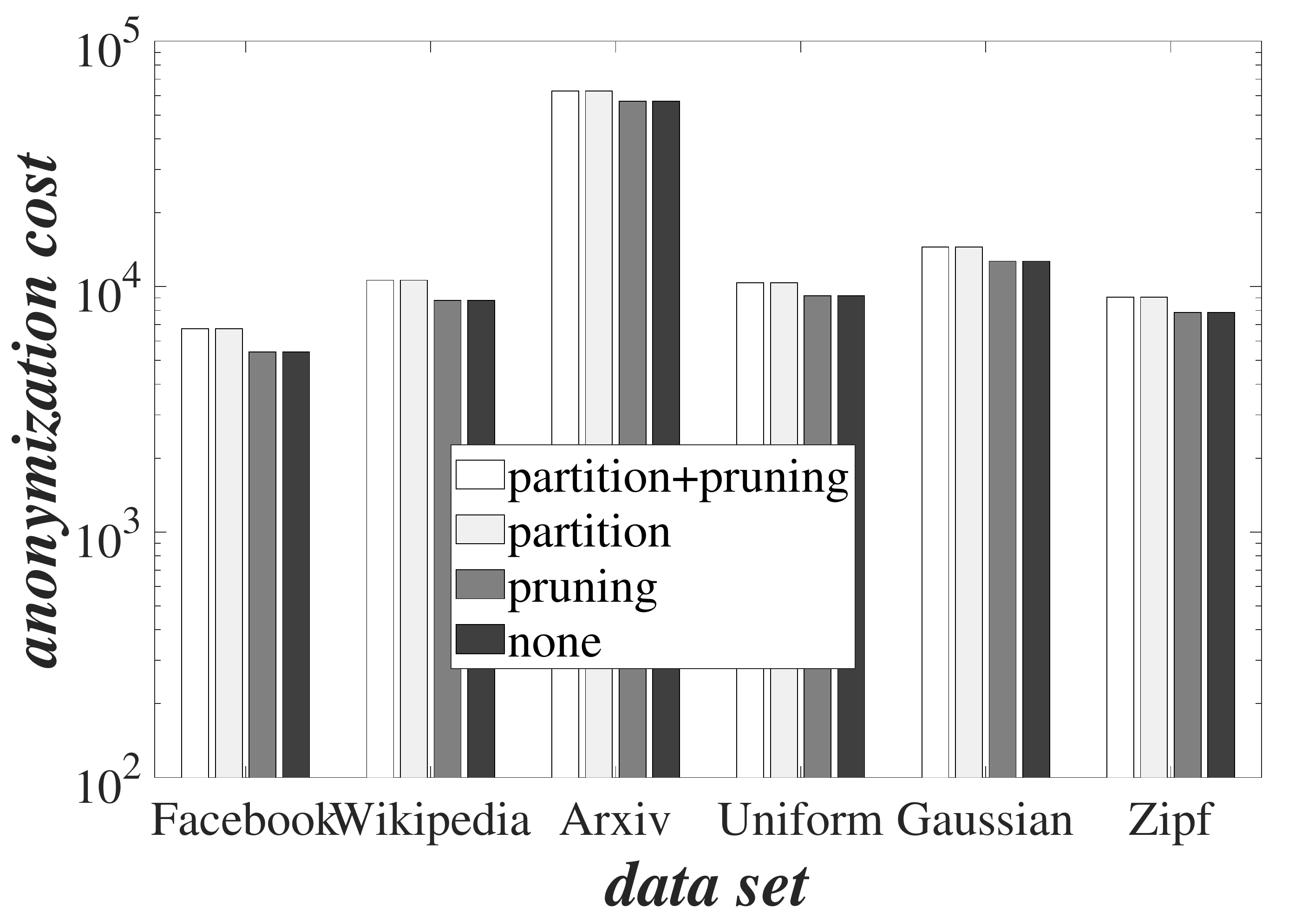}}
\caption{\small The anonymization cost vs. more real/synthetic data sets.}
\label{fig:more_anonymization_cost_vs_dataset}
\end{figure}

\noindent {\bf More Anonymization Cost Evaluation.} Figure \ref{fig:more_anonymization_cost_vs_dataset} illustrates the anonymization cost of all the approaches on $Facebook$, $Wikipedia$, $Arxiv$, $Uniform$, $Gaussian$ and $Zipf$. Similar as Figure \ref{fig:exper:anonymization_cost_vs_dataset}, $partition+pruning$ and $partition$ have slightly higher cost than $pruning$ and $none$, since our approach needs to add more fake nodes on the partitioned subgraphs than that of the original single graph to make each vertex as $kt$-safe. 

\begin{figure}[H]
\centering
\scalebox{0.23}[0.23]{\hspace{0ex}\includegraphics{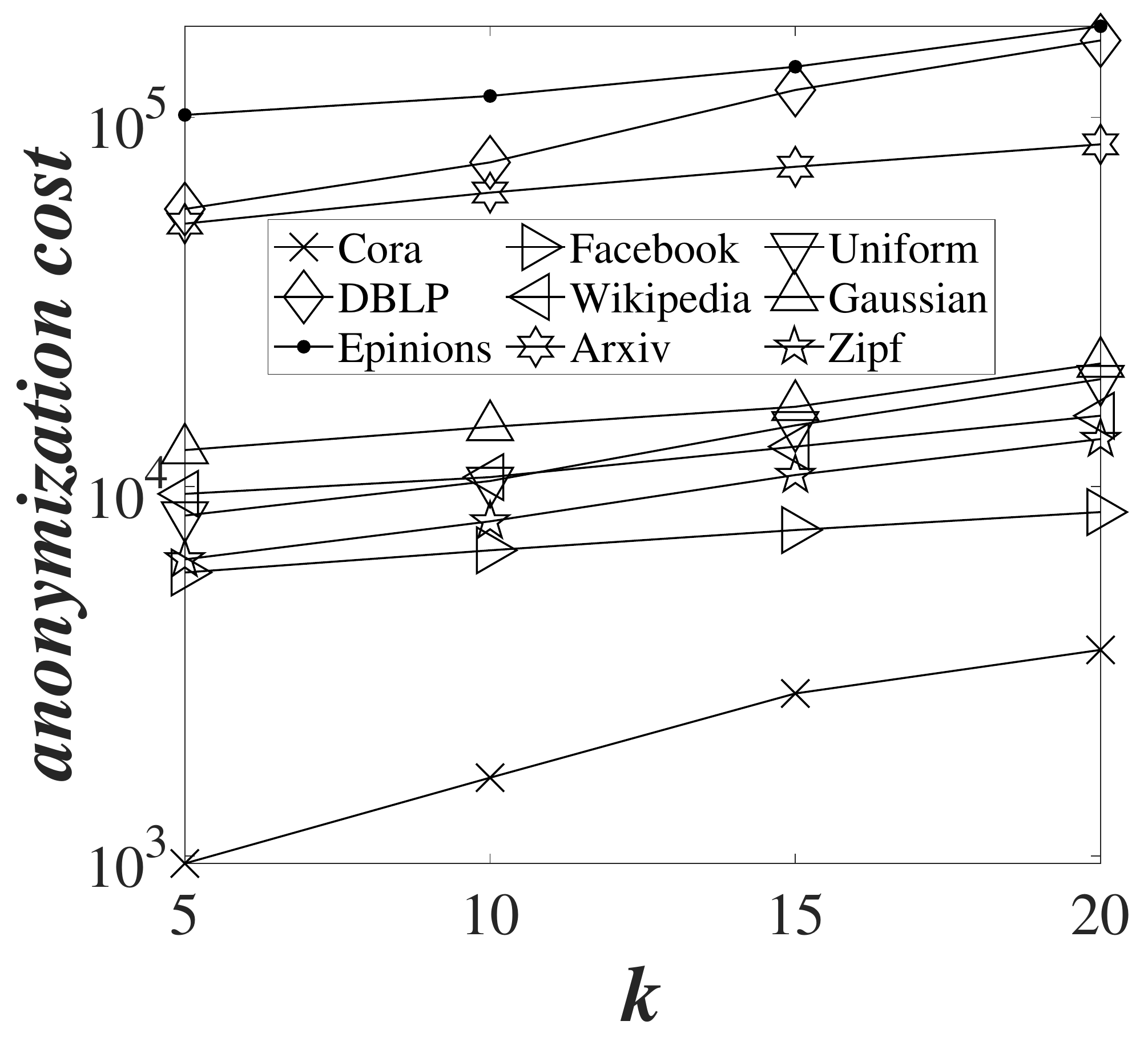}}
\caption{\small The anonymization cost of $partition+pruning$ vs. privacy threshold $k$.}
\label{fig:exper:cost_k}
\end{figure}

\noindent {\bf The anonymization cost vs. identity privacy threshold $k$.} Figure \ref{fig:exper:cost_k} shows the effect of the privacy threshold $k$ on our $partition+pruning$ approach over real and synthetic data sets, where $k$ varies from $5$ to $20$ and other parameters are set as their default (bold) values in Table \ref{table:exp_parameter_setting}. From the figure, when $k$ increases, the anonymization cost increases. This is reasonable, since larger $k$ indicates stronger privacy power and larger size of protection set for each graph node. Specifically, the $partition+pruning$ approach achieves the highest anonymization cost on $Epinions$ on all $k$ values. After we carefully examined $Epinions$, we found that it has the largest size of quasi-identifiers, which leads to less candidates in protection set (Definition \ref{def:anonymization_set}) for each vertex. As a result, our approach needs more anonymization cost to make each graph vertex as a $kt$-safe one. Nevertheless, the anonymization cost of our $partition+pruning$ method is still reasonable (less than 176,523 even when $k=20$).

\begin{figure}[H]
\centering
\scalebox{0.23}[0.23]{\hspace{0ex}\includegraphics{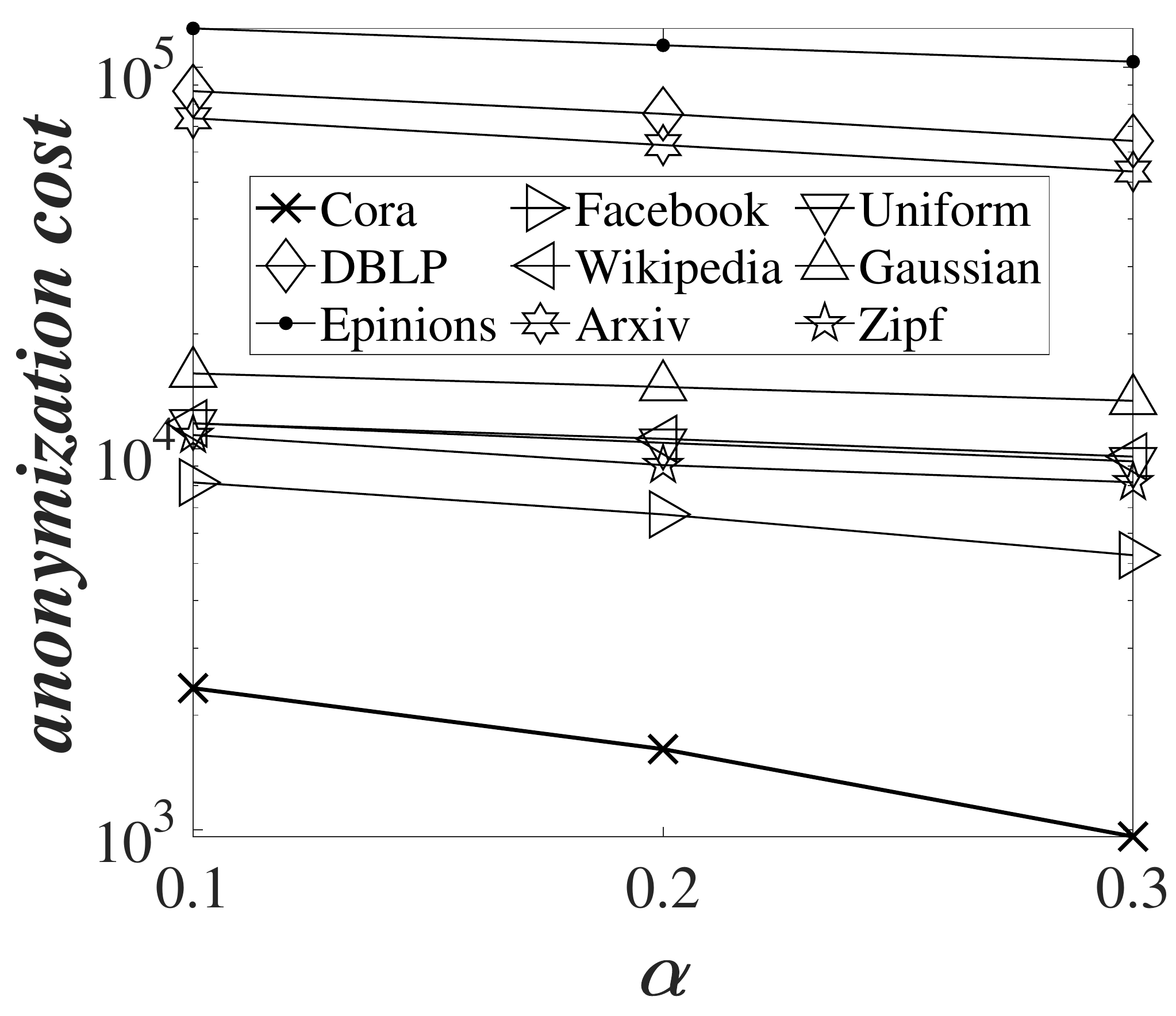}}
\caption{\small The anonymization cost of $partition+pruning$ vs. privacy threshold $\alpha$.}
\label{fig:exper:cost_alpha}
\end{figure}

\noindent {\bf The anonymization cost vs. sensitive privacy threshold $\alpha$.} Figure \ref{fig:exper:cost_alpha} reports the performance of our $partition+pruning$ method, by varing $\alpha$ from 0.1 to 0.3, where other parameters follow their defaults in Table \ref{table:exp_parameter_setting}. From the result, when $\alpha$ increases, the anonymization cost decreases, since larger $\alpha$ indicates more tolerance of sensitive values in protection set of graph nodes. Nevertheless, $Epinions$ still needs the highest anonymization cost due to the large size of quasi-identifiers.

\begin{figure}[H]
\centering
\scalebox{0.25}[0.25]{\hspace{0ex}\includegraphics{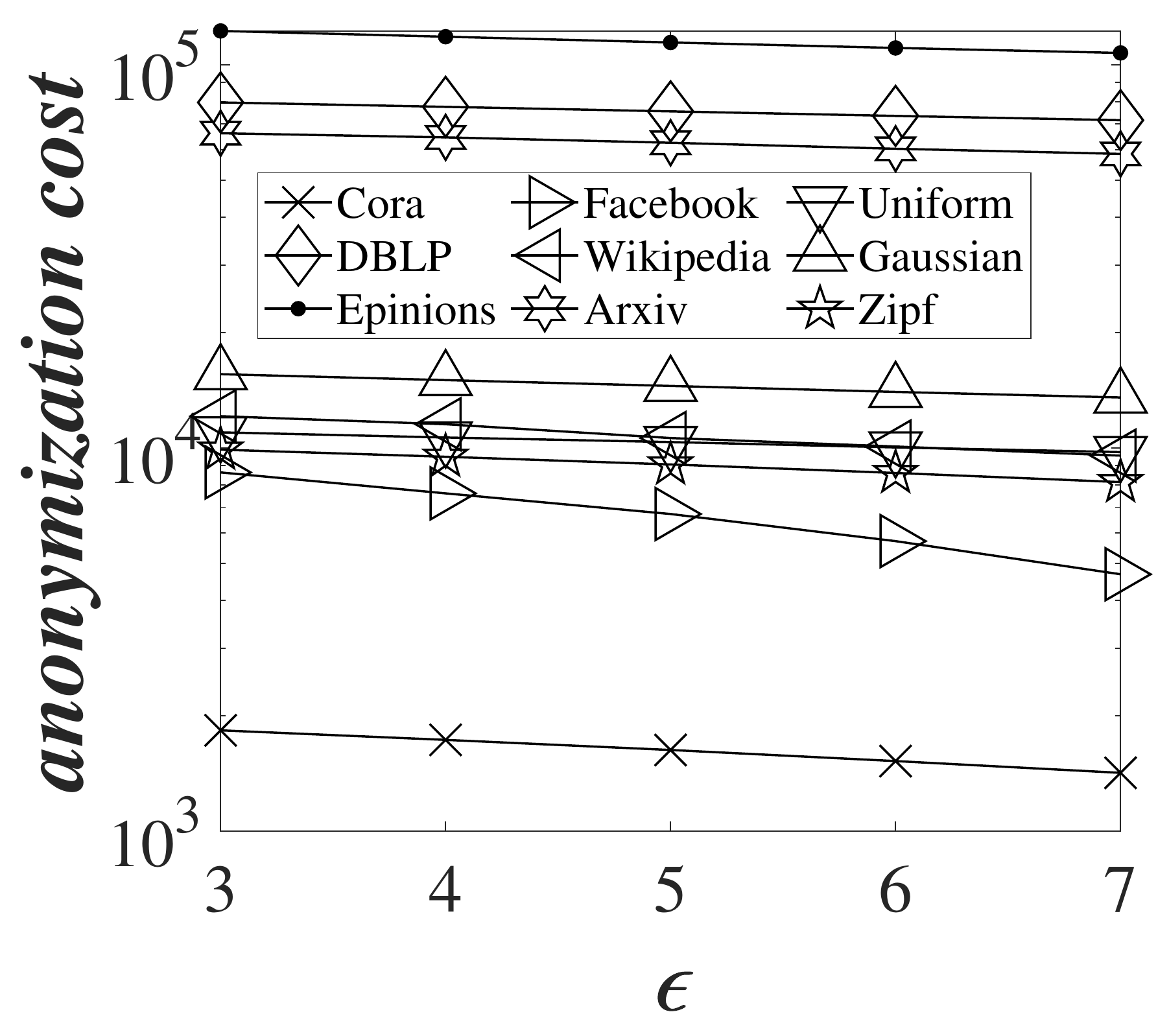}}
\caption{\small The anonymization cost of $partition+pruning$ vs. distance threshold $\epsilon$.}
\label{fig:exper:cost_epsilon}
\end{figure}

\noindent {\bf The anonymization cost vs. graph edit distance threshold $\epsilon$.} Figure \ref{fig:exper:cost_epsilon} illustrates the anonymization cost of our $partition+pruning$ method by varying $\epsilon$ from $3$ to $7$, where other parameters take their defaults in Table \ref{table:exp_parameter_setting}. From the figure, when $\epsilon$ increases, the anonymization cost smoothly decreases, since we can find more candidates for each node and need less graph operations. 

\begin{figure}[H]
\centering
\scalebox{0.22}[0.22]{\hspace{0ex}\includegraphics{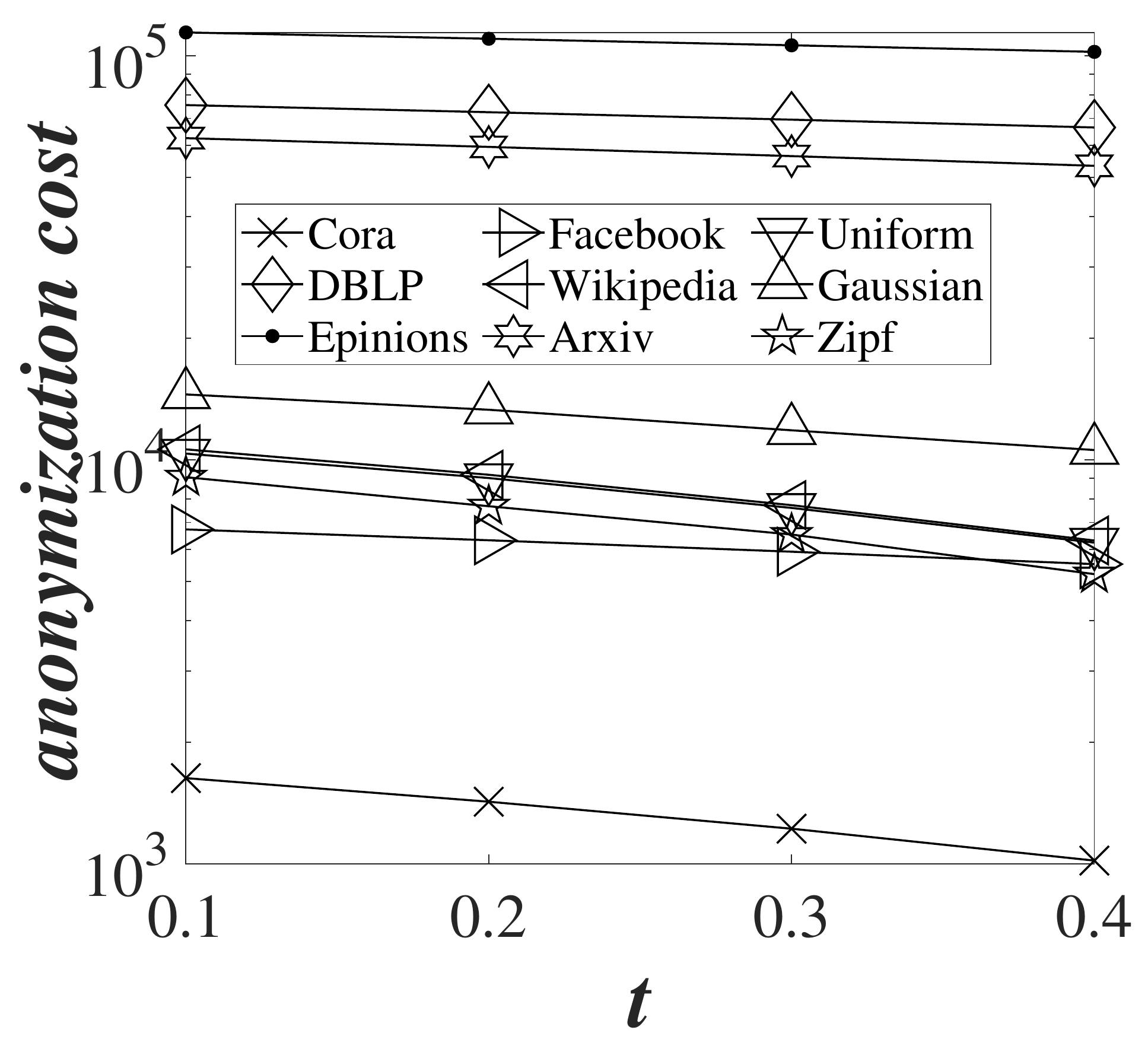}}
\caption{\small The anonymization cost of $partition+pruning$ vs. distance threshold $t$.}
\label{fig:exper:cost_t}
\end{figure}

\noindent {\bf The anonymization cost vs. distance threshold $t$.} Figure \ref{fig:exper:cost_t} shows the effect of the distance threshold $t$ on our $partition+pruning$ method on real/synthetic data sets, where $t=0.1$, $0.2$, $0.3$ and $0.4$ and other parameters follow their default values in Table \ref{table:exp_parameter_setting}. From the result, with larger $t$, the anonymization cost decreases, since it has a higher tolerance of distribution difference of n-hop neighbors between graph nodes. 

\begin{figure}[H]
\centering
\scalebox{0.25}[0.25]{\hspace{0ex}\includegraphics{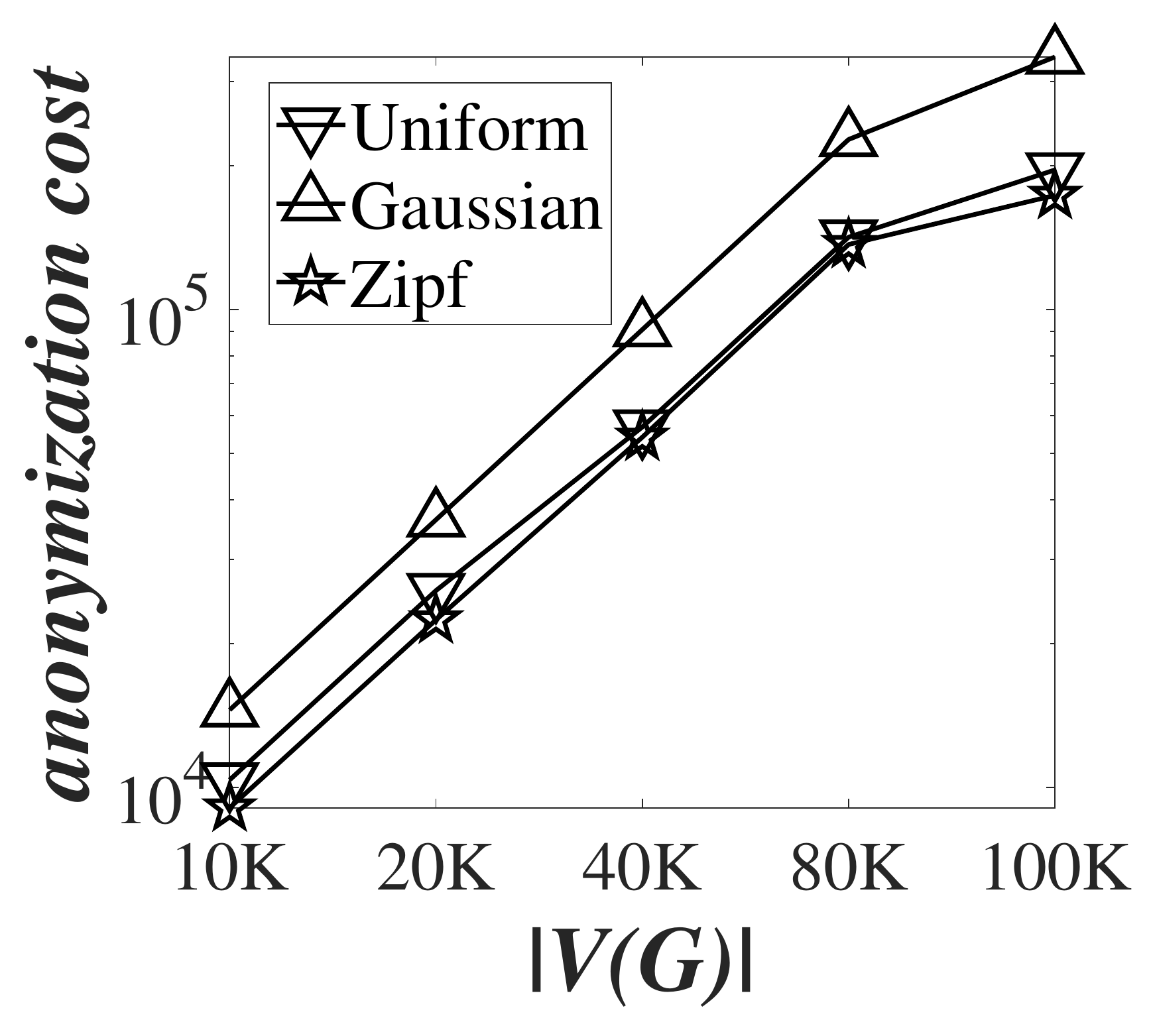}}
\caption{\small The anonymization cost of $partition+pruning$ vs. size, $|V(G)|$, of graph $G$.}
\label{fig:exper:cost_vg}
\end{figure}

\noindent {\bf The anonymization cost vs. size, $|V(G)|$, of graph $G$.} Figure \ref{fig:exper:cost_vg} demonstrates the performance of our $partition+pruning$ approach on synthetic data sets by varying graph size $|V(G)|$ from $10K$ to $100K$, and other parameters are set as their defaults in Table \ref{table:exp_parameter_setting}. Note that, since the real graphs have fixed size, we tested parameter $|V(G)|$ only on synthetic data. From the figure, when $|V(G)|$ increases, the anonymization cost increases, since the graph operation over a graph node will affect more other nodes. Nevertheless, our $partition+pruning$ method still achieves relative low anonymization cost (less than 337,493 even when $|V(G)|=100K$).

\begin{figure}[H]
\centering
\scalebox{0.22}[0.22]{\hspace{0ex}\includegraphics{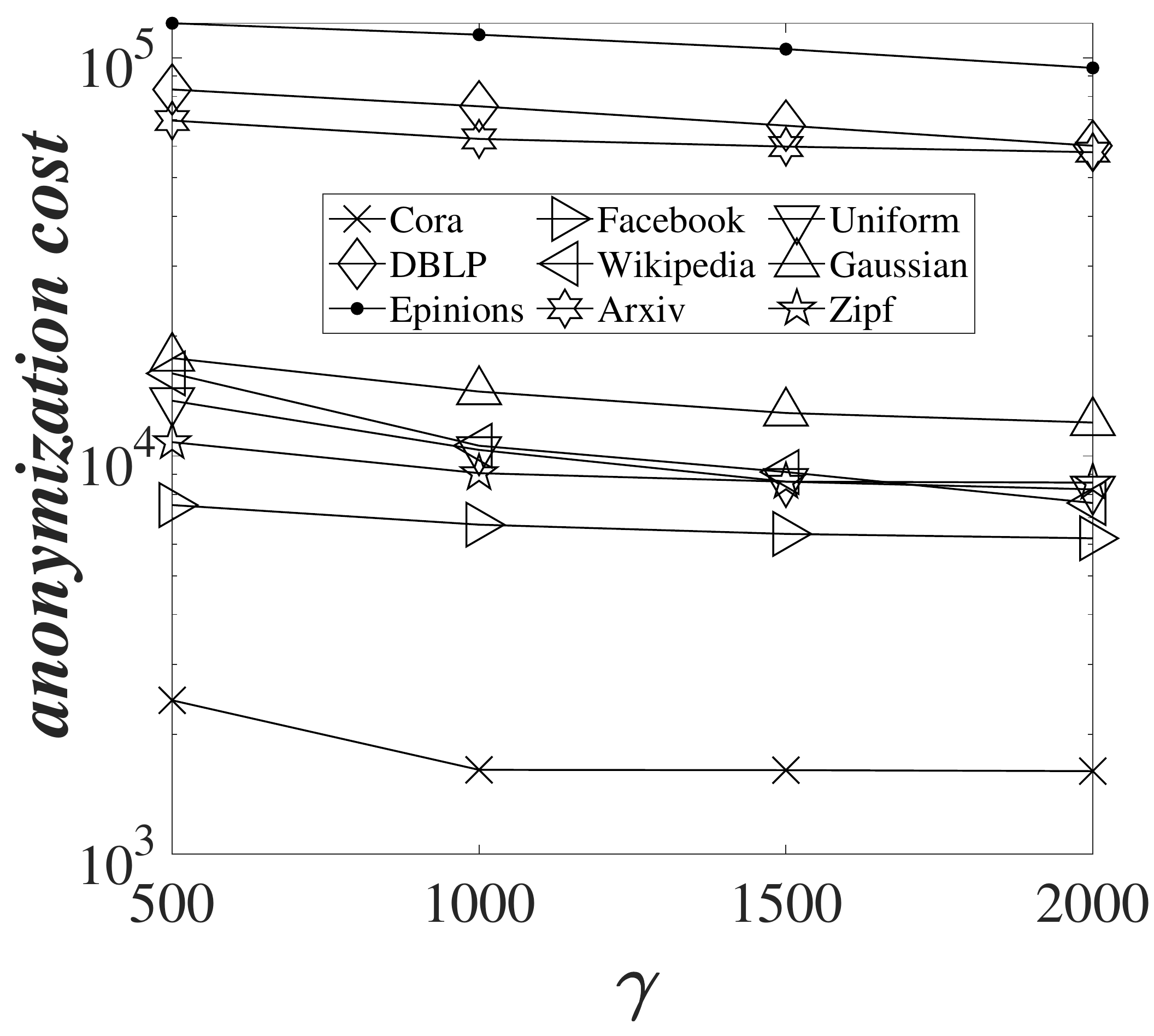}}
\caption{\small The anonymization cost of $partition+pruning$ vs. the maximum size, $\gamma$, of subgraphs.}
\label{fig:exper:cost_gamma}
\end{figure}

\noindent {\bf The anonymization cost vs. the maximum size, $\gamma$, of subgraphs.} Figure \ref{fig:exper:cost_gamma} reports the anonymization cost of our $partition+pruning$ approach, by setting $\gamma=500$, $1,000$, $1,500$ and $2,000$ and other parameters by default. From the figure, when $\gamma$ increases, the anomymization cost slightly decreases for all data sets except for $Cora$. The anonymization cost on Cora remains almost the same when $\gamma \geq$ 1,000, since our approach needs to partition the original graph only once when $\gamma \geq \frac{|V(G)|}{s}$, where $|V(G)|$ = 2,708 and $s=$ 4. 

\begin{figure}[H]
\centering
\scalebox{0.25}[0.25]{\hspace{0ex}\includegraphics{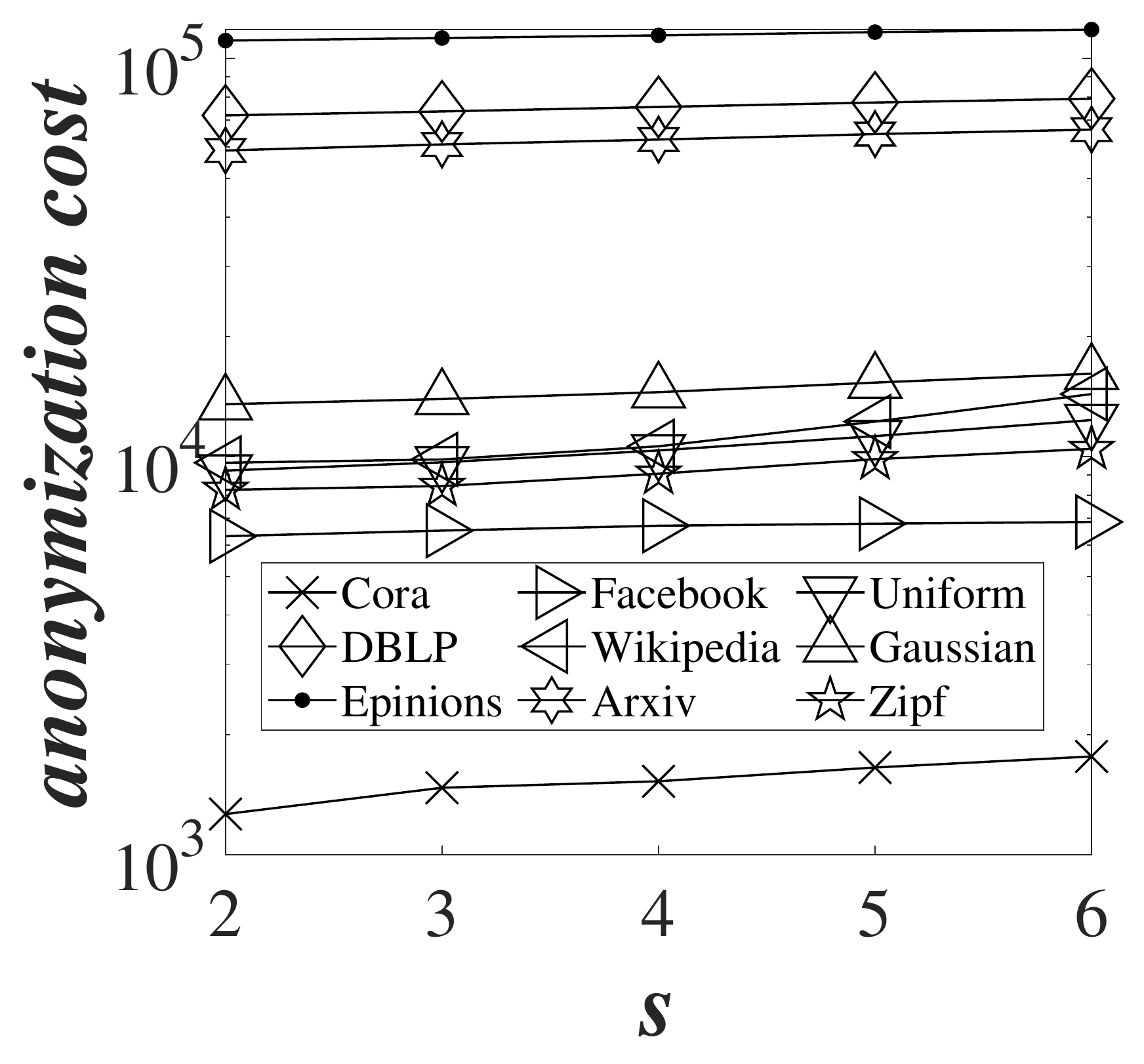}}
\caption{\small The anonymization cost of $partition+pruning$ vs. the number, $s$, of graph partitions.}
\label{fig:exper:cost_s}
\end{figure}

\noindent {\bf The anonymization cost vs. the number, $s$, of graph partitions.} Figure \ref{fig:exper:cost_s} shows the effect of the number $s$ of graph partitions on the anonymization cost of our $partition+pruning$ approach on real and synthetic data sets for $s$ from 2 to 6, where default values are used for other parameters. From the figure, when $s$ increases, the anonymization cost increases. This is because, with larger $s$, our approach will produce more subgraphs, which needs more cost to anonymize.

\end{document}